\documentclass[structabstract]{aa}
\usepackage{txfonts}
\usepackage {graphicx}
\usepackage{subfig}
\usepackage{color}
\usepackage{amssymb}
\usepackage{ulem}
\usepackage{rotating}
\usepackage{lscape}
\usepackage[]{natbib} 
\usepackage[]{longtable}
\usepackage[]{url}
\bibpunct{(}{)}{;}{a}{}{,}  

\begin{document}


\title{Analysis of  old very metal rich stars in the solar neighbourhood~\thanks{Observations collected at the European  Southern  Observatory, La Silla, Chile}$^{,}$~\thanks{Tables A.1 to A.4 are only available in electronic form at the CDS via anonymous ftp to cdsarc.u-strasbg.fr (130.79.128.5) 
    or via http://cdsweb.u-strasbg.fr/cgi-bin/qcat?J/A+A/}}

\titlerunning{Metal Rich Stars in the Solar Neighbourhood}

\author{
M. Trevisan\inst{1}
\and
B. Barbuy\inst{1}
\and
K. Eriksson\inst{2}
\and
B. Gustafsson\inst{2}
\and
M. Grenon\inst{3}
\and
L. Pomp\'eia\inst{4}
}
\offprints{M. Trevisan}
\institute{
 Universidade de S\~ao Paulo, Rua do Mat\~ao 1226, S\~ao Paulo 05508-900,
 Brazil\\
 e-mail: trevisan@astro.iag.usp.br, barbuy@astro.iag.usp.br
\and
Departament of Astronomy and Space Physics, Uppsala University,
Box 515, SE 751 20 Uppsala, Sweden \\
 e-mail: kjell.eriksson@fysast.uu.se, bengt.gustafsson@astro.uu.se
\and
Observatoire de Gen\`eve, 
51 Chemin des Maillettes, 1290 Sauverny, Switzerland \\
e-mail: michel.grenon@unige.ch
\and
Universidade do Vale do Para\'iba, Av. Shishima Hifumi 2911,
S\~ao Jos\'e dos Campos, 12244-000 S\~ao Paulo, Brazil \\
e-mail: pompeia@univap.br
}

\date{Accepted for publication in Astronomy \& Astrophysics, 2011 September 8}

\abstract
{A sample of mostly old metal-rich dwarf and turn-off stars with high eccentricity and low maximum height above the Galactic plane has been identified. From their kinematics, it was suggested that the inner disk is their most probable birthplace. Their chemical imprints may therefore reveal important information about the formation
 and evolution of the still poorly understood  inner~disk. }
{To probe the formation history of these stellar populations, a detailed analysis of a sample of very metal-rich stars is carried out. We derive the metallicities, abundances of $\alpha$ elements, ages, and Galactic orbits.}
{The analysis of 71 metal-rich  stars is  based on optical high-resolution \'echelle spectra obtained with the  FEROS spectrograph at the ESO 1.52-m Telescope at La Silla, Chile. The metallicities and abundances of C, O, Mg, Si, Ca, and Ti were derived based on  LTE detailed analysis, employing the MARCS model atmospheres.}
{We confirm the high metallicity of these stars reaching up to [\ion{Fe}{I}/H]~$= 0.58$, and the sample of metal-rich dwarfs can be kinematically subclassified in samples of thick disk, thin disk, and intermediate stellar populations. All sample stars show solar $\alpha$-Fe ratios, and most of them are old and still quite metal rich. The orbits suggest that the thin disk, thick disk and intermediate populations were formed at Galactocentric distances of $\sim 8$~kpc, $\sim 6$~kpc, and $\sim 7$~kpc, respectively. The mean maximum height of the thick disk subsample of  Z$_{\rm max}\sim 380$~pc, is lower than  for typical thick disk stars.A comparison of $\alpha$-element abundances of the sample stars with bulge stars shows that the oxygen is compatible with a bulge or inner thick disk origin. Our results suggest that models of radial mixing and dynamical effects of the bar  and bar/spiral arms might explain the presence of these old metal-rich dwarf stars in the solar neighbourhood.}
{}


\maketitle


\section{Introduction}

The formation and properties of the thin and thick disks of our Galaxy have been the subject of several studies, and several scenarios have been proposed to explain their formation  \citep[e.g.][and references therein]{Schonrich.Binney:2009a, Schonrich.Binney:2009b, Villalobos.etal:2010}. In each scenario,  typical signatures into the velocity and metallicity  distribution of stars are imprinted. For this reason there have been numerous studies devoted to determination of the thick disk velocity
ellipsoid and metallicity distribution, the study of the thin disk to the thick disk interface, abundance trends and correlations between abundance and kinematics, or the existence of gradients \citep[e.g. ][]{Ivezic.etal:2008, Katz.etal:2011}.

It is well-known that thick disk stars move on higher eccentricity orbits and present larger velocity dispersions than thin disk stars. The thick disk is a more slowly rotating stellar system than the thin disk, and as a whole it lags behind the local standard of rest by $\sim 50$~km~s$^{-1}$, while the thin disk component lags by only $\sim 12$~km~s$^{-1}$ \citep{Soubiran.etal:2003, Robin.etal:2003}. Thick disk stars also appear to be significantly older than thin disk ones \citep{Fuhrmann:1998}. 

On the other hand, the behaviour of the chemical abundance characteristics of these components still is a matter of debate. Some studies suggest that the thick disk component is composed mainly of metal-poor stars \citep[e.g.][]{Chiba.Beers:2000, Reddy.Lambert:2008}, while metal-rich stars appear to be restricted to the thin disk, with a transition occurring at [Fe/H] $\sim -0.3$ \citep{Mishenina.etal:2004, Reddy.etal:2006}. Previous results show that thick disk stars exhibit a larger abundance of $\alpha$-elements relative to iron than the thin disk members \citep{Fuhrmann:1998, Gratton.etal:2000, Ruchti.etal:2010}.  \citet{Bensby.etal:2003} and \citet{Feltzing.etal:2003} found that thick disk stars extend to solar metallicities, showing an inflexion in [$\alpha$/Fe] vs. [Fe/H] around [Fe/H] $\sim$ -0.5, reaching the solar ratios at [Fe/H] $\sim$ 0.0. Also, a new population has been identified in several studies:  \citet{Reddy.etal:2006} and \citet{Haywood:2008} identify a population having thick disk kinematics but thin disk abundances (TKTA subsample in Reddy et al.).  \citet{Mishenina.etal:2004} and  \citet{Soubiran.Girardi:2005} find  metal-rich stars with kinematics of the thick disk.

A detailed study of stars with these properties can clarify the origin of this population. Therefore, in this work we study a sample of  71 metal-rich stars in terms of kinematics and abundances.

The paper is organized as follows. In Sect.~\ref{Sec_Sample}, a description of the sample is presented, and the observations and reductions are described in Sect.~\ref{Sec_Obs}. The Galactic orbits are derived  in Sect.~\ref{Sec_kinematics}.  Derivations of stellar parameters effective temperatures, gravities, and metallicities are given in Sect.~\ref{Sec_Pars}. In Sect.~\ref{Sec_abonds}, the element abundances are derived. In Sect.~\ref{Sec_discussion} the present results are compared with other samples from the literature, and we briefly discuss the possible origins of the identified stellar populations in the context of the Galaxy formation. Finally,  results are summarized in Sect.~\ref{Sec_Summary}.
 

\section{Sample selection}
\label{Sec_Sample}

\citet{Grenon:1972, Grenon:1989, Grenon:1990, Grenon:1998, Grenon:2000} selected 7824 high proper motion stars from the New Luyten's Two Tenths catalogue (NLTT) (a catalogue of nearby stars with proper motions $\mu$ $>$ 0.18$\pm$0.02 arcsec yr$^{-1}$) that have been included in the HIPPARCOS programme. Among these,  radial velocities  and Geneva photometry were gathered for 5443 stars. Only stars with parallaxes larger than 10 mas were kept for the study presented in \citet{Raboud.etal:1998}. Among these, space velocities were measured for 4143 stars, and metallicities from Geneva photometry were gathered for 2619 of them.
From their kinematics, \citet{Raboud.etal:1998} found that the old disk stars in this sample appeared to show a positive mean U motion. In particular, an imbalance between positive and negative U velocities was found for old disk stars selected in the parallax range 10 to 40 mas (with U positive in the direction of the anti-centre), reaching up to 50 km s$^{-1}$. After corrections for local motions, the U anomaly is +29$\pm$2 km s$^{-1}$ with respect to the Sun, and +19$\pm$9 km s$^{-1}$ with respect to the Galactic centre. Raboud et al. suggested that the metal-rich stars within this sample appear to wander from inside the bar, reaching the solar neighbourhood.

A subsample of 202 of these stars was selected for this project by M. Grenon when gathering the oldest disk stars, with high metallicities and eccentricities, as well as thin disk very metal-rich stars. We were able to obtain high-resolution spectra for 100 of them using the FEROS spectrograph at the 1.5m ESO telescope at La Silla, during an IAG/ON and ON/ESO agreement in 1999-2002. 

The Geneva photometry was used by \citet{Grenon:1978} to derive the effective temperatures, absolute magnitudes, and metallicities, with internal errors of $20-40$~K on
effective temperature $T_{\rm eff}$, $0.03-0.05$~dex on  metallicity [M/H] and $0.15$ on V magnitudes. 

For the present analysis, we selected the 71 most metal-rich stars of the sample of 100 observed stars, indicated by Geneva photometry to have [Fe/H] $> 0.00$, hereafter called {\it sample stars}. A study of $\alpha$-elements vs. [Fe/H] in their full metallicity range, for 36 among the 100 observed such stars, covering -0.8 $<$ [Fe/H] $<$ +0.4, was  presented by \citet{Pompeia.etal:2003}. There are 12 stars in common between the present sample and   \citet{Pompeia.etal:2003}, where another 24 stars with metallicities below solar, were also analysed, with the aim  of identifying the downturn knee of [$\alpha$/Fe] vs. [Fe/H], as discussed in Sect. \ref{Sec_discussion}.
All the stars in the sample have parallaxes larger than 10 mas, with errors of 6\% in average. 

Table \ref{Tab_logbook} shows the log of spectroscopic observations (available electronically only).

\section{Observations and reductions}
\label{Sec_Obs}

Optical spectra were obtained using the Fiber Fed Extended Range Optical Spectrograph (FEROS) \citep{Kaufer.etal:2000} at the 1.52~m telescope at ESO, La Silla. The total wavelength coverage is 3560-9200 {\rm \AA} with a resolving power of (R=$\lambda/\Delta\lambda$) = 48,000. Two fibres, with entrance aperture of 2.7~arcsec, simultaneously recorded star light and sky background. The detector is a back-illuminated CCD with 2948~x~4096 pixels of 15~$\mu$m size.

 Reductions were carried out through a pipeline package for reductions (DRS) of FEROS data, in {\tt MIDAS} environment.  The pipeline performs the subtraction of bias and scattered light in the CCD, orders extraction, flatfielding and wavelength calibration with a ThAr calibration frame. The data reduction proceeded in the {\tt IRAF} environment as follows. The spectra were cut into parts of 500 \AA \ each using the {\tt SCOPY} task, and the normalization was carried out with the {\tt CONTINUUM} task. Spectra  of rapidly rotating hot B stars at similar airmasses as the target were also observed,  in order to correct for telluric lines using the {\tt TELLURIC} task.  The radial and heliocentric velocities, v$_{\rm r}$ and v$_{\rm Helio}$,  were determined  using the {\tt RVCORRECT} task. The standard errors of the  velocities are $\sim$~0.2~km~s$^{-1}$.  Typical signal-to-noise ratios of  the spectra were obtained considering average values at different wavelengths. The mean signal-to-noise ratio for the sample stars is $\sim 100$, as reported in Table \ref{Tab_logbook}.

\begin{table*}
\scriptsize
\caption{Basic stellar data} 
\label{Tab_basic}      
\centering          
\begin{tabular}{ccccccccc}    
\hline\hline    
Star & $T_{\rm Gen}$ (K) & V & J  & K$_{\rm S}$ &$\mathcal{M}_{\rm Bol}$& BC$_{\rm V}$   & $\pi$ (mas) & d (pc)\\
\hline
G   161-029 & 4869 & 10.33  & 9.52 $\pm$ 0.02 & 8.90 $\pm$ 0.02 & ...  & -0.41 & ...		  & ... 	 \\ 
BD-02   180 & 4917 & 10.09  & 8.44 $\pm$ 0.02 & 7.89 $\pm$ 0.02 & 5.59 & -0.29 & 14.41 $\pm$ 1.62 & 69 $\pm$  8  \\ 
BD-05  5798 & 4875 & 10.38  & 8.64 $\pm$ 0.03 & 8.07 $\pm$ 0.04 & 5.68 & -0.35 & 13.48 $\pm$ 1.94 & 74 $\pm$ 11  \\ 
BD-17  6035 & 4830 & 10.30  & 8.53 $\pm$ 0.03 & 7.98 $\pm$ 0.02 & 5.71 & -0.35 & 14.23 $\pm$ 2.41 & 70 $\pm$ 12  \\ 
CD-32  0327 & 5001 & 10.41  & 8.75 $\pm$ 0.03 & 8.16 $\pm$ 0.03 & 5.88 & -0.32 & 14.41 $\pm$ 1.77 & 69 $\pm$  9  \\ 
CD-40 15036 & 5341 & 10.08  & 8.69 $\pm$ 0.02 & 8.27 $\pm$ 0.02 & 5.08 & -0.16 & 10.75 $\pm$ 1.70 & 93 $\pm$ 15  \\ 
HD     8389 & 5135 &  7.85  & 6.39 $\pm$ 0.02 & 5.92 $\pm$ 0.02 & 5.27 & -0.18 & 33.09 $\pm$ 0.99 & 30 $\pm$  1  \\ 
HD     9174 & 5459 &  8.40  & 7.13 $\pm$ 0.02 & 6.74 $\pm$ 0.03 & 4.03 & -0.10 & 14.04 $\pm$ 1.13 & 71 $\pm$  6  \\ 
HD     9424 & 5332 &  9.17  & 7.81 $\pm$ 0.03 & 7.38 $\pm$ 0.02 & 5.17 & -0.15 & 16.94 $\pm$ 0.99 & 59 $\pm$  3  \\ 
HD    10576 & 5883 &  8.51  & 7.41 $\pm$ 0.03 & 7.07 $\pm$ 0.03 & 3.77 & -0.08 & 11.66 $\pm$ 0.73 & 86 $\pm$  5  \\ 
HD    11608 & 4917 &  9.31  & 7.61 $\pm$ 0.02 & 7.07 $\pm$ 0.03 & 5.91 & -0.31 & 24.12 $\pm$ 1.34 & 41 $\pm$  2  \\ 
HD    12789 & 5706 &  8.89  & 7.74 $\pm$ 0.02 & 7.38 $\pm$ 0.02 & 4.00 & -0.08 & 10.92 $\pm$ 1.00 & 92 $\pm$  8  \\ 
HD    13386 & 5131 &  8.91  & 7.39 $\pm$ 0.02 & 6.97 $\pm$ 0.02 & 5.46 & -0.19 & 22.27 $\pm$ 1.09 & 45 $\pm$  2  \\ 
HD    15133 & 5113 &  9.36  & 7.85 $\pm$ 0.03 & 7.38 $\pm$ 0.02 & 5.33 & -0.20 & 17.18 $\pm$ 1.38 & 58 $\pm$  5  \\ 
HD    15555 & 4793 &  7.34  & 5.60 $\pm$ 0.02 & 4.99 $\pm$ 0.02 & 3.42 & -0.37 & 19.45 $\pm$ 1.06 & 51 $\pm$  3  \\ 
HD    16905 & 4821 &  9.44  & 7.65 $\pm$ 0.02 & 7.09 $\pm$ 0.02 & 6.06 & -0.37 & 24.94 $\pm$ 0.93 & 40 $\pm$  1  \\ 
HD    25061 & 5247 &  9.27  & 7.84 $\pm$ 0.02 & 7.36 $\pm$ 0.02 & 5.35 & -0.18 & 17.86 $\pm$ 0.82 & 56 $\pm$  3  \\ 
HD    26151 & 5285 &  8.49  & 7.08 $\pm$ 0.02 & 6.65 $\pm$ 0.03 & 5.02 & -0.16 & 21.79 $\pm$ 1.12 & 46 $\pm$  2  \\ 
HD    26794 & 4932 &  8.78  & 7.07 $\pm$ 0.02 & 6.49 $\pm$ 0.02 & 5.83 & -0.34 & 30.02 $\pm$ 1.68 & 33 $\pm$  2  \\ 
HD    27894 & 4879 &  9.36  & 7.64 $\pm$ 0.02 & 7.07 $\pm$ 0.03 & 5.89 & -0.34 & 23.60 $\pm$ 0.91 & 42 $\pm$  2  \\ 
HD    30295 & 5350 &  8.86  & 7.49 $\pm$ 0.02 & 7.04 $\pm$ 0.02 & 4.88 & -0.15 & 17.14 $\pm$ 0.77 & 58 $\pm$  3  \\ 
HD    31452 & 5206 &  8.43  & 6.94 $\pm$ 0.03 & 6.47 $\pm$ 0.02 & 5.26 & -0.20 & 25.50 $\pm$ 1.27 & 39 $\pm$  2  \\ 
HD    31827 & 5463 &  8.26  & 7.00 $\pm$ 0.02 & 6.61 $\pm$ 0.02 & 4.56 & -0.10 & 19.05 $\pm$ 0.69 & 52 $\pm$  2  \\ 
HD    35854 & 4943 &  7.70  & 6.01 $\pm$ 0.03 & 5.39 $\pm$ 0.02 & 6.08 & -0.35 & 55.76 $\pm$ 0.76 & 18 $\pm$  0  \\ 
HD    37986 & 5455 &  7.36  & 6.06 $\pm$ 0.02 & 5.62 $\pm$ 0.02 & 5.02 & -0.13 & 36.05 $\pm$ 0.92 & 28 $\pm$  1  \\ 
HD    39213 & 5337 &  8.96  & 7.61 $\pm$ 0.02 & 7.20 $\pm$ 0.02 & 4.82 & -0.13 & 15.75 $\pm$ 0.91 & 63 $\pm$  4  \\ 
HD    39715 & 4781 &  8.84  & 6.99 $\pm$ 0.03 & 6.35 $\pm$ 0.02 & 6.28 & -0.44 & 37.57 $\pm$ 1.26 & 27 $\pm$  1  \\ 
HD    43848 & 5091 &  8.65  & 7.12 $\pm$ 0.03 & 6.61 $\pm$ 0.02 & 5.58 & -0.22 & 26.99 $\pm$ 0.83 & 37 $\pm$  1  \\ 
HD    77338 & 5283 &  8.63  & 7.22 $\pm$ 0.02 & 6.76 $\pm$ 0.02 & 5.39 & -0.16 & 24.23 $\pm$ 1.18 & 41 $\pm$  2  \\ 
HD    81767 & 4943 &  9.45  & 7.77 $\pm$ 0.02 & 7.21 $\pm$ 0.02 & 5.74 & -0.31 & 20.89 $\pm$ 1.49 & 48 $\pm$  3  \\ 
HD    82943 & 5849 &  6.54  & 5.51 $\pm$ 0.02 & 5.11 $\pm$ 0.02 & 4.28 & -0.07 & 36.42 $\pm$ 0.84 & 27 $\pm$  1  \\ 
HD    86065 & 4963 &  9.36  & 7.65 $\pm$ 0.03 & 7.09 $\pm$ 0.02 & 6.14 & -0.33 & 26.42 $\pm$ 1.25 & 38 $\pm$  2  \\ 
HD    86249 & 4935 &  8.99  & 7.32 $\pm$ 0.02 & 6.74 $\pm$ 0.02 & 6.02 & -0.32 & 29.57 $\pm$ 1.16 & 34 $\pm$  1  \\ 
HD    87007 & 5302 &  8.82  & 7.35 $\pm$ 0.02 & 6.89 $\pm$ 0.04 & 5.46 & -0.19 & 23.23 $\pm$ 1.41 & 43 $\pm$  3  \\ 
HD    90054 & 5986 &  7.87  & 6.85 $\pm$ 0.02 & 6.52 $\pm$ 0.02 & 3.63 & -0.05 & 14.52 $\pm$ 0.97 & 69 $\pm$  5  \\ 
HD    91585 & 5077 &  9.43  & 7.90 $\pm$ 0.03 & 7.37 $\pm$ 0.03 & 5.66 & -0.24 & 19.67 $\pm$ 1.33 & 51 $\pm$  3  \\ 
HD    91669 & 5175 &  9.70  & 8.26 $\pm$ 0.02 & 7.77 $\pm$ 0.02 & 4.95 & -0.18 & 12.19 $\pm$ 1.52 & 82 $\pm$ 10  \\ 
HD    93800 & 5129 &  9.12  & 7.58 $\pm$ 0.02 & 7.10 $\pm$ 0.02 & 5.32 & -0.22 & 19.18 $\pm$ 1.42 & 52 $\pm$  4  \\ 
HD    94374 & 4825 &  9.27  & 7.42 $\pm$ 0.02 & 6.79 $\pm$ 0.02 & 6.13 & -0.43 & 28.70 $\pm$ 1.29 & 35 $\pm$  2  \\ 
HD    95338 & 5144 &  8.62  & 7.10 $\pm$ 0.02 & 6.59 $\pm$ 0.02 & 5.56 & -0.23 & 27.14 $\pm$ 0.91 & 37 $\pm$  1  \\ 
HD   104212 & 5694 &  8.38  & 7.24 $\pm$ 0.02 & 6.88 $\pm$ 0.03 & 3.62 & -0.08 & 11.59 $\pm$ 1.09 & 86 $\pm$  8  \\ 
HD   107509 & 5944 &  7.91  & 6.90 $\pm$ 0.02 & 6.58 $\pm$ 0.03 & 3.68 & -0.06 & 14.65 $\pm$ 0.82 & 68 $\pm$  4  \\ 
HD   120329 & 5511 &  8.34  & 7.09 $\pm$ 0.02 & 6.69 $\pm$ 0.02 & 4.00 & -0.11 & 14.24 $\pm$ 1.05 & 70 $\pm$  5  \\ 
HD   143102 & 5432 &  7.88  & 6.59 $\pm$ 0.02 & 6.17 $\pm$ 0.02 & 3.51 & -0.12 & 14.13 $\pm$ 0.99 & 71 $\pm$  5  \\ 
HD   148530 & 5346 &  8.81  & 7.42 $\pm$ 0.03 & 6.97 $\pm$ 0.02 & 5.31 & -0.17 & 21.50 $\pm$ 1.27 & 47 $\pm$  3  \\ 
HD   149256 & 5271 &  8.42  & 7.04 $\pm$ 0.02 & 6.60 $\pm$ 0.02 & 3.89 & -0.15 & 13.32 $\pm$ 1.21 & 75 $\pm$  7  \\ 
HD   149606 & 4936 &  8.95  & 7.25 $\pm$ 0.02 & 6.72 $\pm$ 0.02 & 6.09 & -0.31 & 30.89 $\pm$ 1.37 & 32 $\pm$  1  \\ 
HD   149933 & 5424 &  8.05  & 6.72 $\pm$ 0.02 & 6.29 $\pm$ 0.03 & 5.03 & -0.14 & 26.56 $\pm$ 1.22 & 38 $\pm$  2  \\ 
HD   165920 & 5261 &  7.91  & 6.50 $\pm$ 0.03 & 6.03 $\pm$ 0.02 & 5.22 & -0.17 & 31.27 $\pm$ 1.12 & 32 $\pm$  1  \\ 
HD   168714 & 5552 &  8.90  & 7.67 $\pm$ 0.02 & 7.31 $\pm$ 0.02 & 4.33 & -0.09 & 12.67 $\pm$ 1.51 & 79 $\pm$  9  \\ 
HD   171999 & 5257 &  8.34  & 6.89 $\pm$ 0.02 & 6.43 $\pm$ 0.02 & 5.31 & -0.18 & 26.97 $\pm$ 1.12 & 37 $\pm$  2  \\ 
HD   177374 & 5011 &  9.40  & 7.75 $\pm$ 0.03 & 7.24 $\pm$ 0.02 & 5.44 & -0.28 & 18.35 $\pm$ 1.65 & 54 $\pm$  5  \\ 
HD   179764 & 5374 &  9.01  & 7.58 $\pm$ 0.02 & 7.11 $\pm$ 0.02 & 4.84 & -0.19 & 15.98 $\pm$ 1.30 & 63 $\pm$  5  \\ 
HD   180865 & 5132 &  8.97  & 7.45 $\pm$ 0.02 & 6.98 $\pm$ 0.02 & 5.54 & -0.21 & 22.66 $\pm$ 1.32 & 44 $\pm$  3  \\ 
HD   181234 & 5220 &  8.59  & 7.15 $\pm$ 0.03 & 6.69 $\pm$ 0.02 & 4.97 & -0.17 & 20.49 $\pm$ 1.19 & 49 $\pm$  3  \\ 
HD   181433 & 4866 &  8.40  & 6.66 $\pm$ 0.02 & 6.09 $\pm$ 0.02 & 5.97 & -0.35 & 38.24 $\pm$ 1.15 & 26 $\pm$  1  \\ 
HD   182572 & 5461 &  5.17  & 3.55 $\pm$ 0.21 & 3.04 $\pm$ 0.32 & 4.01 & -0.26 & 66.01 $\pm$ 0.77 & 15 $\pm$  1  \\ 
HD   196397 & 5267 &  8.95  & 7.59 $\pm$ 0.02 & 7.13 $\pm$ 0.03 & 5.20 & -0.15 & 19.01 $\pm$ 1.65 & 53 $\pm$  5  \\ 
HD   196794 & 5075 &  8.52  & 6.94 $\pm$ 0.03 & 6.41 $\pm$ 0.02 & 5.99 & -0.26 & 35.22 $\pm$ 1.14 & 28 $\pm$  1  \\ 
HD   197921 & 4866 &  9.25  & 7.49 $\pm$ 0.02 & 6.90 $\pm$ 0.02 & 5.82 & -0.37 & 24.45 $\pm$ 1.58 & 41 $\pm$  3  \\ 
HD   201237 & 4886 & 10.10  & 8.31 $\pm$ 0.02 & 7.71 $\pm$ 0.02 & 4.97 & -0.39 & 11.23 $\pm$ 2.09 & 89 $\pm$ 17  \\ 
HD   209721 & 5388 &  9.51  & 8.18 $\pm$ 0.02 & 7.77 $\pm$ 0.02 & 4.59 & -0.13 & 11.00 $\pm$ 1.26 & 91 $\pm$ 10  \\ 
HD   211706 & 5830 &  8.90  & 7.84 $\pm$ 0.02 & 7.52 $\pm$ 0.02 & 4.19 & -0.07 & 11.78 $\pm$ 1.40 & 85 $\pm$ 10  \\ 
HD   213996 & 5203 &  8.66  & 7.21 $\pm$ 0.03 & 6.76 $\pm$ 0.03 & 5.30 & -0.18 & 23.10 $\pm$ 1.14 & 43 $\pm$  2  \\ 
HD   214463 & 4958 &  9.67  & 8.10 $\pm$ 0.02 & 7.59 $\pm$ 0.02 & 5.29 & -0.24 & 14.90 $\pm$ 1.77 & 67 $\pm$  8  \\ 
HD   218566 & 4834 &  8.59  & 6.82 $\pm$ 0.02 & 6.22 $\pm$ 0.02 & 5.83 & -0.38 & 33.40 $\pm$ 1.19 & 30 $\pm$  1  \\ 
HD   218750 & 5122 &  9.25  & 7.71 $\pm$ 0.02 & 7.18 $\pm$ 0.03 & 5.34 & -0.24 & 18.45 $\pm$ 1.50 & 54 $\pm$  4  \\ 
HD   221313 & 5075 &  9.90  & 8.38 $\pm$ 0.02 & 7.85 $\pm$ 0.02 & 5.16 & -0.23 & 12.52 $\pm$ 1.79 & 80 $\pm$ 11  \\ 
HD   221974 & 5109 &  9.31  & 7.80 $\pm$ 0.02 & 7.32 $\pm$ 0.02 & 5.68 & -0.20 & 20.61 $\pm$ 1.53 & 49 $\pm$  4  \\ 
HD   224230 & 4900 &  9.97  & 8.24 $\pm$ 0.03 & 7.63 $\pm$ 0.03 & 6.16 & -0.36 & 20.43 $\pm$ 1.22 & 49 $\pm$  3  \\ 
HD   224383 & 5689 &  7.89  & 6.74 $\pm$ 0.03 & 6.33 $\pm$ 0.02 & 4.40 & -0.10 & 20.98 $\pm$ 1.24 & 48 $\pm$  3  \\ 
\hline
\end{tabular}
\end{table*}

\section{Kinematics}
\label{Sec_kinematics}

Grenon (1999) found that the high metallicity stars in the sample have low maximum height from the Galactic plane $Z_{\rm max}$, and their turn-off location indicated  an age of $\sim$ 10~Gyr. The identification of an old population with such a high metallicity and low $Z_{\rm max}$ is unexpected. In order to investigate the kinematical properties of the sample stars, we derive the Galactic orbits in Sect. \ref{Sec_orbits}, using the {\tt GRINTON} integrator \citep{Carraro.etal:2002, Bedin.etal:2006}.
In Sect. \ref{Sec_disk_member}, we separate the sample into thin disk and thick disk stars, based on kinematical criteria. We assigned a probability of each star belonging to either the thin or the thick disk, assuming that the space velocities of each population follow a Gaussian distribution as defined by \citet{Soubiran.etal:2003}.

\subsection{Galactic orbits}
\label{Sec_orbits}

\citet{Grenon:1999} derived U, V, W space velocities for all the sample stars. U, V, W are defined in a right-handed Galactic system with U pointing outwards the Galactic centre, V in the direction of rotation and W towards the north Galactic pole. We used the {\tt GRINTON} integrator to calculate the Galactic orbits, with these velocities and the HIPPARCOS parallaxes.
 This code integrates the orbits back in time for several Galactic revolutions 
and returns the minimum and maximum distances from the Galactic centre ($R_{\rm min},\ R_{\rm max}$),  maximum height from the Galactic plane ($Z_{\rm max}$) and the eccentricity $e$ of the orbit. Before using the observed space motions, these were transformed to the local standard of rest. We used a solar motion of (-10.0, 5.3, 7.2)~km~s$^{-1}$ \citep{Dehnen.Binney:1998}. The gravitational  potential used in the orbit integration  is a simple one \citep{Allen.Santillan:1991}, for which a circular rotation speed of 220~km~s$^{-1}$ and a disk volume density of $0.15 M_{\odot}$~pc$^{-3}$ are adopted and a solar Galactocentric distance $R_{\odot}$ = $8.5$~kpc is assumed. 
 
Uncertainties in the orbital parameters were obtained using the bootstrapping technique, as follows. We integrated the orbit of each star 500 times. At each integration, the input parameters, U, V, W velocities and the parallax $\pi$, were varied following a normal distribution with mean $X$ and standard deviation of $\sigma_X$, where
$X$ is the parameter value and $\sigma_X$ the error associated with it. The final orbital parameters $R_{\rm min},\ R_{\rm max}$, $Z_{\rm max}$, and eccentricity, and their errors were then computed  as the mean and standard deviation of the output values of these 500 realizations. Uncertainties in $R_{\rm min}$, $R_{\rm max}$, and $Z_{\rm max}$ are typically $\approx 0.30$ kpc, $0.60$ kpc, and $0.05$ kpc, respectively. The derived orbital parameters are listed in Table \ref{Tab_kinematics}. Our sample contains 17 stars in common with the Geneva-Copenhagen survey \citep{Holmberg.etal:2009}, hereafter referred to as GCS, as listed in Table \ref{Tab_uvw_literature}. We compared the orbital parameters derived here with the values from the GCS survey. We found that our $R_{\rm min}$ distances are $\sim 6\%$ lower and  $R_{\rm max}$ are $\sim 5\%$ higher on average. For the orbit  eccentricities, we derived values which are $\sim 16\%$ higher than eccentricities from GCS. The maximum height from the Galactic plane from GCS are $\thicksim 80$~pc lower ($\sim 30\%$) than our sample, on average. 

It is important to stress that the gravitational potential used in the orbit integration does not take the Galactic bar into account. The bar potential could affect 
the orbits of our stars, since $R_{\rm min}$ are as close as 3-4~kpc from the Galactic centre. 

\begin{table*}
\centering
 \caption{Sample stars in common with the Geneva-Copenhagen survey \citep{Holmberg.etal:2009}.}
 \label{Tab_uvw_literature}
 \begin{tabular}{lcccccccc}
 \hline\hline
 Star       & $R_{\rm min}$ & $\Delta R_{\rm min}$ & $R_{\rm max}$ & $\Delta R_{\rm max}$ & $e$ & $\Delta e$ & $Z_{\rm max}$ & $\Delta Z_{\rm max}$ \\
            &  (kpc) & (kpc) &  (kpc) & (kpc) &  &  & (kpc) & (kpc)  \\
 \hline
HD 9424   &  3.44 & -0.35 &  8.28 & 0.47 & 0.41 &  0.07 & 0.08 &  0.03\\ 
HD 13386  &  4.74 & -0.41 &  8.02 & 0.55 & 0.26 &  0.07 & 0.10 &  0.15\\
HD 25061  &  4.94 & -0.23 &  9.20 & 0.44 & 0.30 &  0.04 & 0.04 &  0.15\\
HD 26151  &  6.39 & -0.19 &  9.27 & 0.41 & 0.18 &  0.04 & 0.07 &  0.00\\
HD 35854  &  6.54 & -0.36 &  8.02 & 0.53 & 0.10 &  0.06 & 0.17 & -0.13\\
HD 82943  &  6.96 & -0.25 &  8.23 & 0.41 & 0.08 &  0.05 & 0.03 &  0.16\\
HD 86249  &  8.00 &  0.51 &  8.78 & 0.21 & 0.05 & -0.02 & 0.24 &  0.19\\
HD 90054$^a$  &  3.39 & -0.33 &  8.02 & 0.50 & 0.41 &  0.06 & 0.18 &  0.13\\
HD 95338  &  2.98 & -0.30 &  8.02 & 0.53 & 0.46 &  0.06 & 0.74 &  0.14\\
HD 104212 &  3.81 & -0.63 &  7.98 & 0.53 & 0.35 &  0.11 & 0.24 &  0.36\\
HD 107509 &  3.43 & -0.42 &  8.03 & 0.53 & 0.40 &  0.08 & 0.25 &  0.19\\
HD 148530 &  3.76 & -0.20 &  9.12 & 0.59 & 0.42 &  0.04 & 0.51 &  0.17\\
HD 171999$^b$ &  3.94 & -0.45 &  7.97 & 0.52 & 0.34 &  0.08 & 0.06 &  0.16\\
HD 180865 &  4.41 & -0.30 &  8.02 & 0.50 & 0.29 &  0.06 & 0.11 &  0.15\\
HD 181433 &  6.72 &  0.07 & 10.06 & 0.41 & 0.20 &  0.01 & 0.27 & -0.24\\
HD 218750 &  4.52 & -0.54 &  8.16 & 0.53 & 0.29 &  0.08 & 0.30 & -0.22\\
HD 224383 &  3.84 & -0.38 &  8.59 & 0.44 & 0.38 &  0.07 & 0.09 &  0.02\\
\hline
Average   &       & $-0.28 \pm 0.26$ &  & $0.48 \pm 0.09$ &  & $0.06 \pm 0.03$ & & $0.08 \pm 0.16$ \\

 \hline
\end{tabular}
\tablefoot{
$\Delta = {\rm Our} - {\rm GCS}$. 
\tablefoottext{a}{Binary.}
\tablefoottext{b}{Spectroscopic binary.}
}
\end{table*}

\subsection{Thin and thick disk membership probabilities}
\label{Sec_disk_member}

Identifying stellar populations in velocity space is not straightforward. Thus, before discussing whether the sample stars belong to the thin or thick disk, we must analyse the criteria used to define the membership probabilities. 

The separation between thin and thick disks can be done either by selecting stars based on their kinematics, using chemical composition criteria, or a combination of both. Usually, separation based only on kinematics or abundances are not equivalent: thick (thin) disk samples selected on kinematical criteria can contain stars with thin (thick) chemical abundances \citep[e.g. ][]{Mishenina.etal:2004, Reddy.etal:2006}. Some authors argue that, since the chemical composition of a star does not change with time, while kinematics may change, the selection based on abundances is more reliable. On the other hand, if we want to trace the formation of the disk components through the study of the chemical abundances of their stars, the abundances must not be used to define these components. Therefore, here we assign the probability of each star belonging to either the thin disk or the thick disk by adopting the kinematical approach used in previous studies by \citet{Bensby.etal:2004}, \citet{Mishenina.etal:2004}, and \citet{Reddy.etal:2006}. The procedure relies on the assumption that the space velocities of each population follow a Gaussian distribution, with given mean values and dispersions $\sigma_{\rm U}$, $\sigma_{\rm V}$, $\sigma_{\rm W}$. The equations determining the probabilities are

\begin{eqnarray}
 p_{\rm thin} = f_1 \frac{p_1}{p},\ \ p_{\rm thick} = f_2 \frac{p_2}{p},\ \ p_{\rm halo} = f_3 \frac{p_3}{p}
\label{Eq_membership1}
\end{eqnarray}

\noindent where $p_{\rm thin}$, $p_{\rm thick}$, $p_{\rm halo}$ correspond to the probability that the star belongs to either the thin disk, thick disk or halo, respectively. Then, $p$ and $p_i$ are given by

\begin{eqnarray}
 p = f_1 p_1 + f_2 p_2 + f_3 p_3 \nonumber
\end{eqnarray}

\noindent and

\begin{eqnarray}
p_i = \frac{1}{(2 \pi)^{3/2} \sigma_{U_i} \sigma_{V_i} \sigma_{W_i}} \exp \left[ -\frac{U^2}{2 \sigma_{U_i}^2} 
 -\frac{(V - V_{\rm lag})^2}{2 \sigma_{V_i}^2}  -\frac{W^2}{2 \sigma_{W_i}^2} \right].
\label{Eq_membership2}
\end{eqnarray}

The parameters $f_i$ are the relative densities of thin disk, thick disk, and halo stars in the solar neighbourhood. Since there is an overlap of the Gaussian distributions in velocity space, the definition of the thin and thick disk populations is very sensitive to the choice of parameters defining the Gaussian distributions and the population fractions.There are several studies devoted to determining of the velocity ellipsoids of the thin disk,  thick disk, and  halo components, as well as the population fractions in the solar neighbourhood. Here we compared studies by \citet{Soubiran.etal:2003} and \citet{Robin.etal:2003}. We determined the probabilities using the values given in Table \ref{Tab_sigmas}, where the velocity ellipsoids for the thin and thick disks were taken from \citet{Soubiran.etal:2003}, and values from \citet{Robin.etal:2003} were used for the halo component. We also applied the same procedure to the thin disk, thick disk, and halo velocity dispersions and fractions from \citet{Robin.etal:2003}: ($\sigma_{\rm U}$,~$\sigma_{\rm V}$,~$\sigma_{\rm W})_{\rm thin} = (43, 28, 18)$~km~s$^{-1}$ and ($\sigma_{\rm U}$,~$\sigma_{\rm V}$,~$\sigma_{\rm W})_{\rm thick} = (67, 51, 42)$~km~s$^{-1}$, $f_{\rm thin} = 0.93$ and  $f_{\rm thick} = 0.07$. 

\begin{figure*}
\centering
\begin{tabular}{ccc}  \\
 \resizebox{0.32\hsize}{!}{\includegraphics{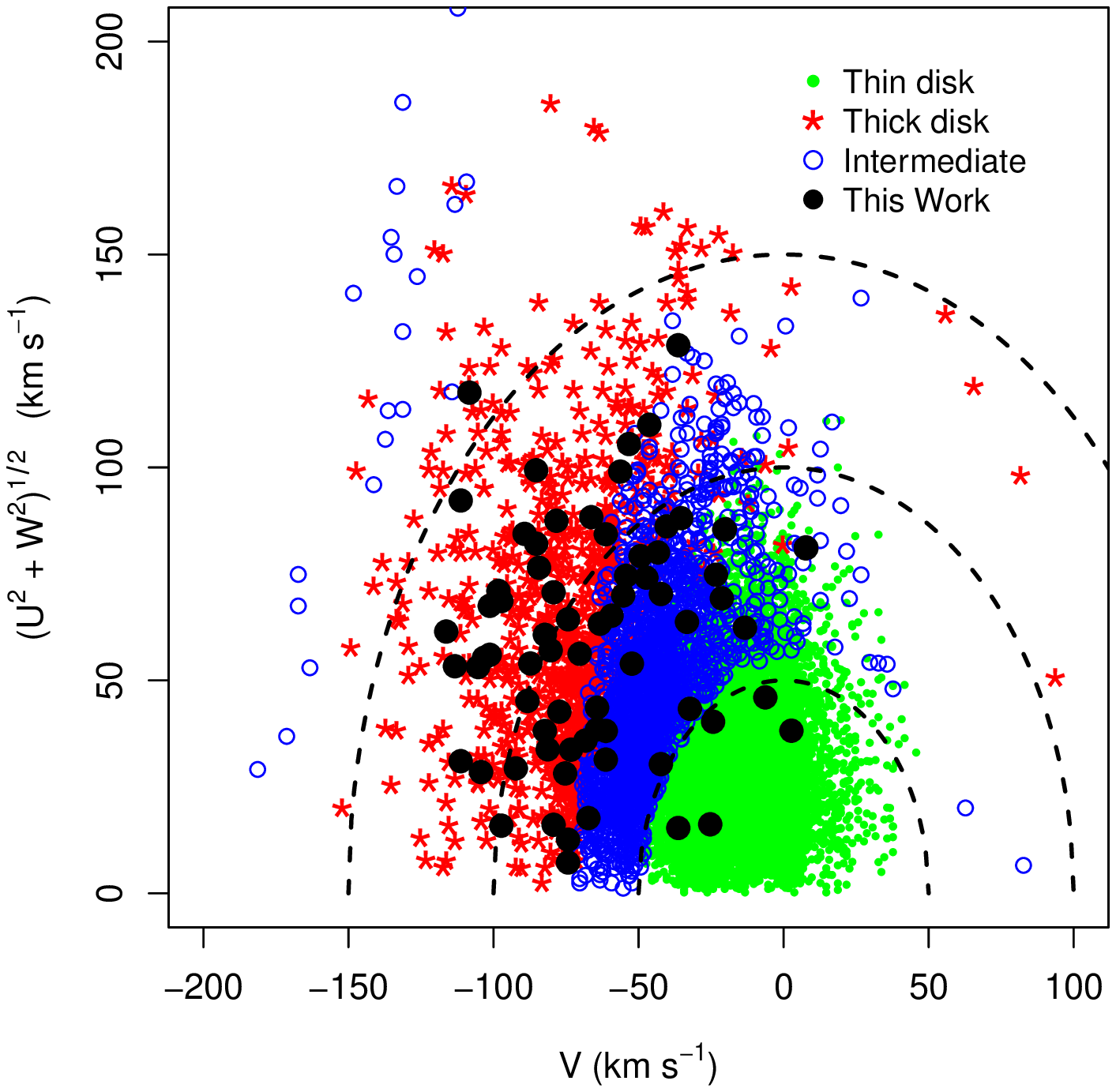}} &
 \resizebox{0.32\hsize}{!}{\includegraphics{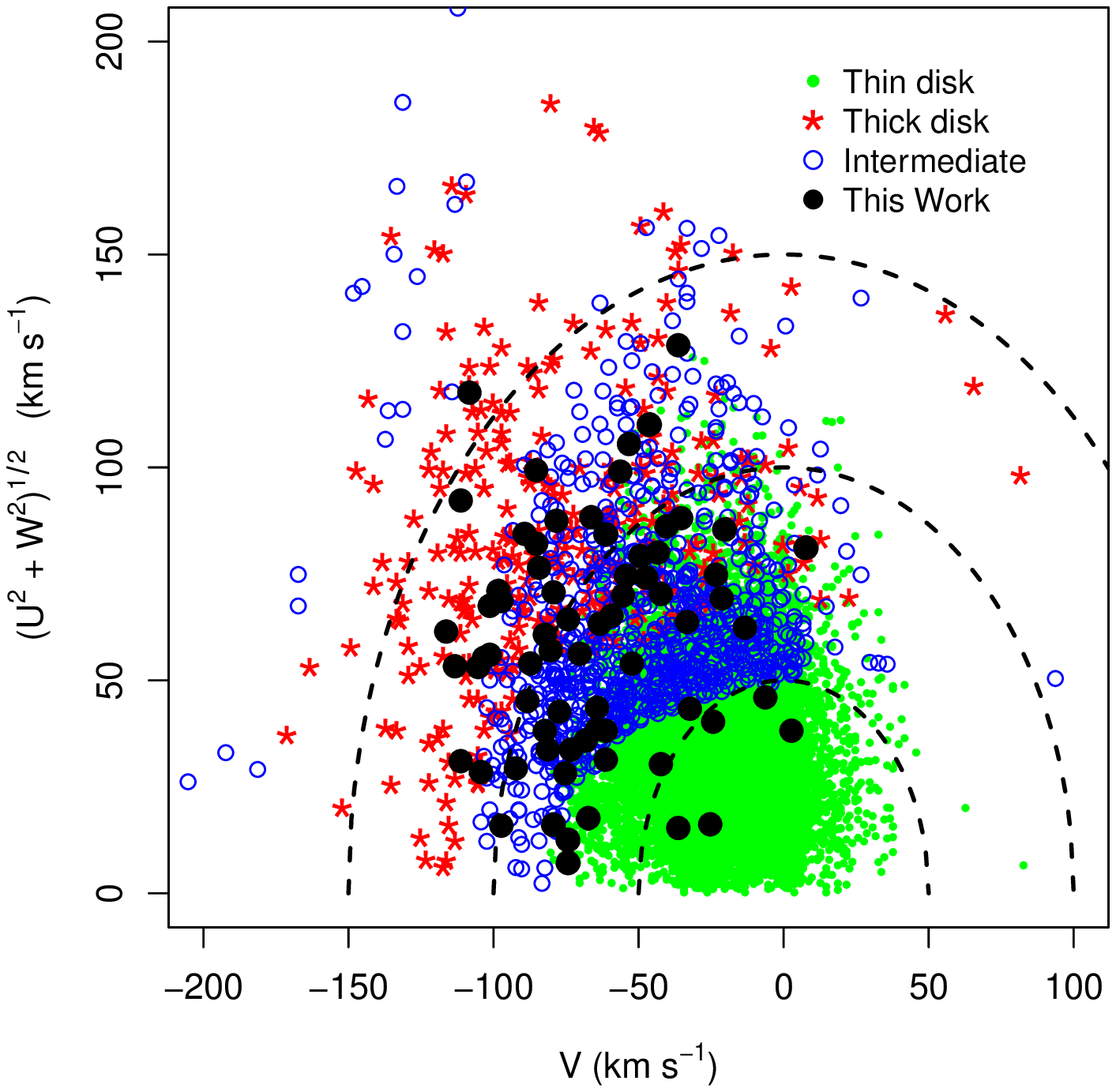}} &
 \resizebox{0.32\hsize}{!}{\includegraphics{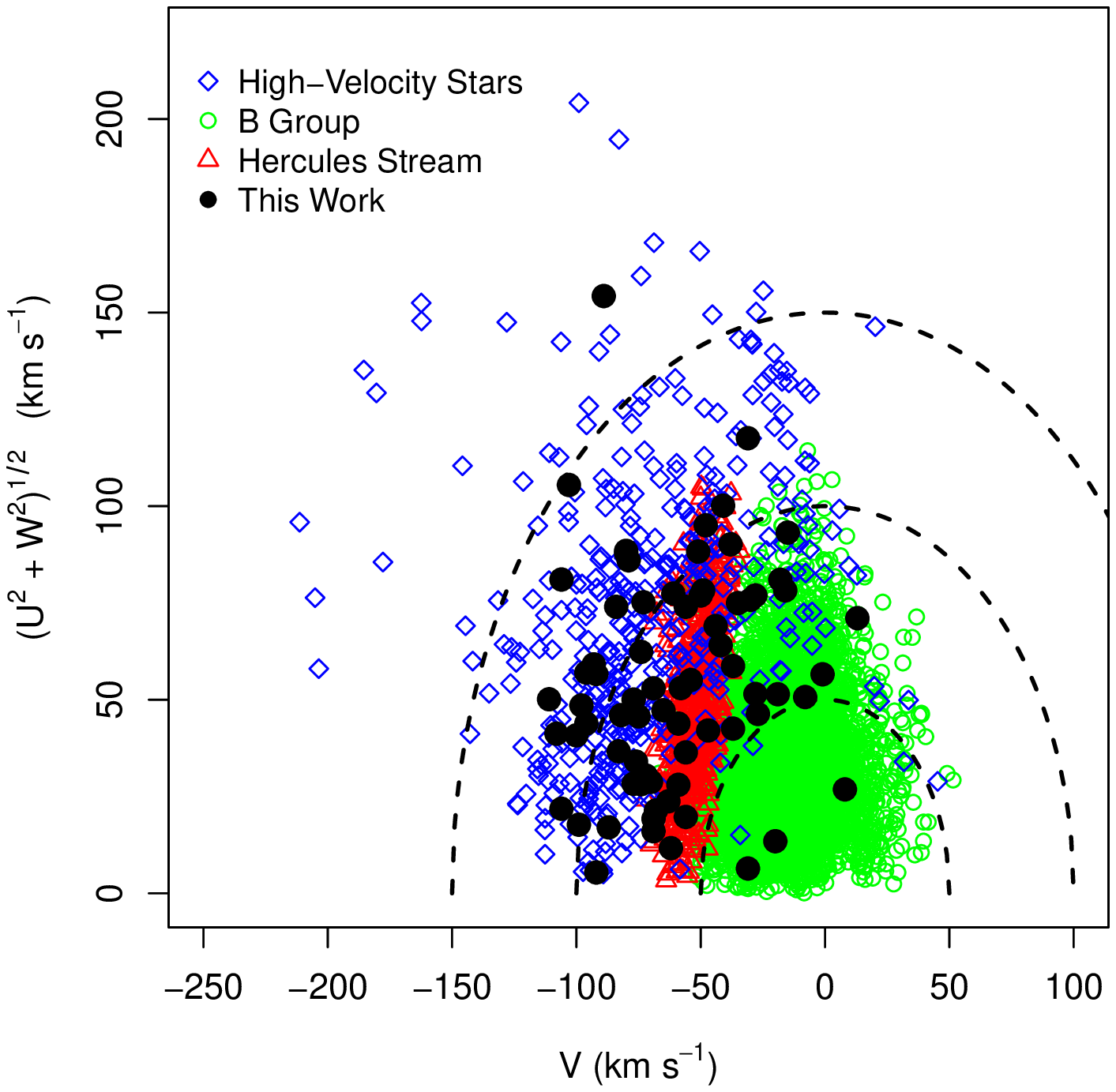}} 
 \end{tabular}
\caption {Toomre diagram of GCS stars. {\it Left and middle panels:} The thin/thick disk decomposition of GCS stars was performed using the velocity dispersions and star fractions from \citet{Soubiran.etal:2003} ({\it left}) and from \citet{Robin.etal:2003} ({\it middle}). The thin, thick, and intermediate stars of the GCS are represented by green dots, red stars, and open blue circles, respectively. The dashed lines indicate the total space velocity, $v_{\rm total} = \sqrt{U^2 + V^2 + W^2}$, in steps of 50~km~s$^{-1}$. {\it Right panel:}  Toomre diagram showing the groups identified by \citet{Famaey.etal:2005}: the Hercules stream (red open triangles), high-velocity stars (blue open diamonds), and B-group stars (green open circles). In all panels the present sample is indicated by black dots.}
\label{Fig_toomre_gcs}
\end{figure*}

As a test, we applied the procedure, using \citet{Soubiran.etal:2003} and \citet{Robin.etal:2003}, to the GCS stars, and the results are shown in Fig. \ref{Fig_toomre_gcs}. The decomposition of a larger sample into thin/thick disk makes the differences between \citet{Soubiran.etal:2003} and \citet{Robin.etal:2003} clearer. We considered that, if the probability of a star belonging to either the thin or thick disk is higher than $80\%$, then the star can be assigned to that component. If both $p_{\rm thin}$ and $p_{\rm thick}$ are lower than $80\%$, the star is classified as member of the intermediate population. Using velocity ellipsoids defined by \citet{Soubiran.etal:2003}, we found that 81\% of the GCS stars belong to the thin disk,  5\% are thick disk stars, and  14\% cannot be assigned to either of the components. The thin disk stars are restricted to $V > -50$~km~s$^{-1}$. Using \citet{Robin.etal:2003}, the following fractions were found: 92\%,  2\%, and 6\%  are thin, thick, and intermediate stars, respectively, and the thin disk stars can rotate as slowly as $V \sim -80$~km~s$^{-1}$. 

We then classified our 71 sample stars using velocity ellipsoids from \citet{Soubiran.etal:2003}, and we found that 42 stars in the sample can be assigned to the thick disk, and 11 are more likely to be thin disk stars. The other 17 stars in the sample are intermediate between thin and  thick disk components. Using \citet{Robin.etal:2003}, we found  that 16 stars in the sample belong to the thick disk, and 29 are more likely to be thin disk stars. The other 26 stars in the sample are not clearly members of either the thin or the thick disk components. 

The same procedure (i.e., equations \ref{Eq_membership1} and \ref{Eq_membership2}) was applied to the groups and stellar streams identified by \citet{Famaey.etal:2005}. They applied a maximum-likelihood method to the kinematical data of 6691 K and M giants in the solar neighbourhood. They identified six kinematical groups: {\it i)} group Y, containing stars with ``young'' kinematics; {\it ii)} group HV, composed of high-velocity stars, which are probably mostly halo or thick disk stars; {\it iii)} group HyPl, the  Hyades-Pleiades supercluster; {\it iv)} group Si, the Sirius moving group; {\it v)} group He, the Hercules stream; and {\it vi)} group B, which is composed of a ``smooth'' background in the UV plane, that are mostly thin disk stars. We obtained the probabilities of 
our sample stars belonging to these groups following the same procedure as above, assuming that the space velocities of each group follow a Gaussian distribution, with mean values, dispersions, and population fractions taken from Famaey et al. 

As expected, neither of the sample stars belong to groups Y, HyPl, and Si, with probabilities below 1\%. Only two stars have $p_{\rm Y} > 1\%$: HD~35854 has $p_{\rm Y} = 6\%$ and HD~82943 has $p_{\rm Y} = 17\%$, respectively. Despite 11 stars in the sample
 having a probability of belonging to the Hercules stream larger than 50\%, neither of them satisfied our criteria $p_{\rm He} > 80\%$ to be assigned to this group. Six stars in our sample appear to belong to the B group ($p_{\rm B} > 80\%$), and 37 have kinematics compatible with the HV group ($p_{\rm HV} > 80\%$). The other 28 stars in the sample cannot be assigned clearly to either of these groups. 

The right-hand panel in Fig. \ref{Fig_toomre_gcs} presents the high velocity group, the B group and Hercules Stream stars, as classified by \citet{Famaey.etal:2005}, and our sample star data overplotted in the Toomre diagram. In Table \ref{Tab_vels} the kinematics of our sample are compared with the data for 6030 stars from \citet{Famaey.etal:2005}, where all the samples analysed  show low maximum height above the plane. It appears that our sample is compatible with the high-velocity group (HV), and might be identified with that subpopulation. This group is probably composed of halo or thick-disk stars, and represents about 10\% of the whole sample analysed by \citet{Famaey.etal:2005}.
 Most stars assigned to the thick disk component, following the \citet{Soubiran.etal:2003} velocity distributions, are also members of the HV group. 

The probabilities assumed for classifying the present 71 sample stars  are those obtained with \citet{Soubiran.etal:2003} velocity ellipsoids, and not those suggested by \citet{Robin.etal:2003} (that give a low fraction of thick disk stars $f_{\rm thick} = 0.07$). A higher fraction of nearby thick disk stars  is supported by recent studies of the SDSS data \citep{Juric.etal:2008}. Moreover, membership probabilities obtained with \citet{Soubiran.etal:2003} criteria are in better agreement with the more detailed analysis of the velocity space by \citet{Famaey.etal:2005}. The final probabilities $p_{\rm thin}$ and $p_{\rm thick}$ are reported in Table \ref{Tab_kinematics}. In Table \ref{Tab_populations}, we report the mean ages (Sect. 5.5), metallicities (Sect. 5.3), [$\alpha$/Fe] (Sect. 6), space velocities, eccentricities, and maximum height above the Galactic plane for each of the populations: thin disk, thick disk, and intermediate.  The stars with intermediate properties are so classified for having intermediate eccentricity, V velocity, Z$_{\rm max}$, and R$_{\rm m}$ relative to the thick and thin disks.

%
\begin{table}
\centering
\caption{Velocity ellipsoids for the thin disk, thick disk, and halo.}
 \begin{tabular}{lccccr}
\hline\hline
 Component & $\sigma_{\rm U}$ &  $\sigma_{\rm V}$ &  $\sigma_{\rm W}$ & V$_{\rm lag}$ & Fraction \\
\hline
  Thin disk$^a$    &  39 &  20 & 20 &   -7 &  0.85  \\
  Thick disk$^a$   &  63 &  39 & 39 &  -46 &  0.15  \\
  Halo$^b$         & 131 & 106 & 85 & -220 & 0.006  \\
\hline
 \end{tabular}
\tablefoot{ 
\tablefoottext{a}{\citet{Soubiran.etal:2003}.}
\tablefoottext{b}{\citet{Robin.etal:2003}.}
\tablefoottext{c}{\citet{Famaey.etal:2005}.}
}
\label{Tab_sigmas}
\end{table}

\begin{table*}
\centering
\caption{Mean data for the thin, thick, and intermediate populations.}
 \begin{tabular}{cccccccccc}
\hline\hline
Group  &  Age  &  [Fe/H]  &  [$\alpha$/Fe]  & $R_m^a$  &  U$_{\rm LSR}^b$   &  V$_{\rm LSR}^b$ &  W$_{\rm LSR}^b$  &  $e$  &  $Z_{\rm max}$ \\
       & (Gyr) &   (dex)  &  (dex)  & (kpc)  & (km s$^{-1}$) & (km s$^{-1}$) & (km s$^{-1}$) & & (kpc) \\
\hline
Thin disk    &  $7.8 \pm 3.5$ &  $0.20 \pm 0.22$ &  $-0.01 \pm 0.05$ &  $8.2 \pm 0.6$ &  $-10 \pm 51$ &  $-19 \pm 16$ &  $-11 \pm 15$ &  $0.20 \pm 0.08$ &  $0.21 \pm 0.16$ \\
Thick disk   &  $7.5 \pm 3.1$ &  $0.22 \pm 0.17$ &  $ 0.00 \pm 0.04$ &  $6.3 \pm 0.4$ &  $ 36 \pm 43$ &  $-84 \pm 17$ &  $-21 \pm 23$ &  $0.40 \pm 0.07$ &  $0.38 \pm 0.40$ \\
Intermediate &  $6.8 \pm 2.9$ &  $0.29 \pm 0.17$ &  $-0.02 \pm 0.03$ &  $7.3 \pm 0.5$ &  $ 37 \pm 61$ &  $-48 \pm 13$ &  $-20 \pm 14$ &  $0.29 \pm 0.05$ &  $0.28 \pm 0.22$ \\
\hline
 \end{tabular}
\tablefoot{ 
\tablefoottext{a}{Mean Galactocentric distance, $R_{\rm m} = (R_{\rm max} + R_{\rm min}) / 2$.}
\tablefoottext{b}{Space velocities with respect to the local standard of rest. }
}
\label{Tab_populations}
\end{table*}

%

\begin{table*}
\centering
\caption{Mean kinematical data for our sample, compared with kinematic groups studied by \citet{Famaey.etal:2005}.}
 \begin{tabular}{cccccc}
\hline\hline
\hbox{Kinematic group} & $<$U(km s$^{-1}$)$>$ &  $<$V(km s$^{-1}$)$>$ &  $<$W(km s$^{-1}$)$>$ &  $<$Z$_{\rm max}$(pc)$>$ &  \\
\hline
present sample    & 17.6 (55.0) & -60.1 (29.7) & -10.6 (21.0) & 330.0    \\
HV Stars          & 18.5 (62.6) & -53.3 (37.2) &  -6.6 (45.9) & 208.1  \\
B Group           &  2.9 (31.8) & -15.2 (17.6) &  -8.2 (16.3) & 196.1  \\
Hercules Stream   & 42.1 (25.3) & -51.6 ( 7.1) &  -8.1 (15.4) & 132.9 \\
\hline
 \end{tabular}
\tablefoot{ 
\tablefoottext{}{HV refers to high-velocity stars. Standard deviations are presented in parenthesis.}
}
\label{Tab_vels}
\end{table*}

\section{Stellar parameters}
\label{Sec_Pars}

A common method for deriving of stellar parameters relies on the abundances derived from \ion{Fe}{I} and \ion{Fe}{II} lines, by requiring excitation and ionization equilibria. In this work it was not assumed {\it a priori} that the absolute equilibria are reached, since this procedure could hide non-LTE effects, leading to misleading parameters. This choice is justified by previous work, which suggests that deviations from LTE are present in metal-rich dwarf stars \citep[e.g.][]{Feltzing.Gustafsson:1998, Melendez.Ramirez:2005}. Therefore, the stellar temperatures and surface gravities were obtained without recourse to the \ion{Fe}{I} and \ion{Fe}{II} lines as follows.

\begin{description}
\item {\it i)} The effective temperatures were calculated from the (V-K$_S$) colour using the \citet[][hereafter CRM10]{Casagrande.etal:2010} colour-temperature relations, as described in Sect.~\ref{Sec_Photo_Temp}. 

\item {\it ii)} Then, $\log g$ is derived from masses and parallaxes. Using the HIPPARCOS parallaxes and the stellar masses from the Yonsei-Yale evolutionary tracks \citep{Demarque.etal:2004}, hereafter $Y^2$, the surface gravities were derived following the procedure presented in Sect.~\ref{Sec_grav}. 

\item {\it iii)} We used the metallicities from Geneva photometry  as first guesses in steps {\it i} and {\it ii}. 

\item{\it iv)} Then, fixing $T_{\rm eff}$ and $\log g$, the iron abundance and microturbulence velocity were derived from \ion{Fe}{I} and \ion{Fe}{II} lines, through local thermodynamic equilibrium (LTE) analysis, using the MARCS model atmospheres \citep{Gustafsson.etal:2008}. The microturbulence velocity was obtained by imposing constant iron abundance as a function of equivalent width (except for the cooler stars). The iron abundance and microturbulence velocity determinations are described in detail in Sect.~\ref{Sec_Met}. 

\item{\it v)} The Geneva metallicity was then replaced by the new iron abundance, to go through the whole iteration process

\item{\it vi)} the procedure was repeated until there were no significant changes in ($T_{\rm eff}$, $\log g$, [Fe/H]). 

\end{description}

The changes on temperatures and gravities due to changes on metallicities are small, but not zero. Therefore, the procedure includes iteration
to keep internal consistency. The final photometric parameters were tested against excitation and ionization equilibria,
 and $T_{\rm eff}$ and $\log g$ were further adjusted if necessary, as shown in Sect. \ref{Sec_equilibria}.
 Only two stars in the sample required adjustments to be in satisfactory spectroscopic equilibrium.

We estimated ages for the sample stars, using the $Y^2$ isochrones. Details about the determination of
 stellar masses and ages are given in Sect.~\ref{Sec_masses_ages}. Final considerations about the parameter
 determinations and comparison
 with other studies are presented in Sect.~\ref{Sec_Final_pars}.

\subsection{Temperatures}
\label{Sec_Photo_Temp}

The basic photometric data used in temperature determinations are presented in Table \ref{Tab_basic}: photometric temperatures
from Geneva photometry;
 V magnitudes \citep{ESA:1997}; J and K$_{\rm S}$ magnitudes from 2MASS \citep{Skrutskie.etal:2006}; Bolometric correction BC$_{\rm V}$ from \citet{Alonso.etal:1995} (see Sect. \ref{Sec_grav}); and HIPPARCOS parallaxes $\pi$ \citep{ESA:1997}. Errors of about $0.02$~mag apply to V magnitudes; the errors on the other magnitudes are reported in Table \ref{Tab_basic}.
The sample stars are all within $90$~pc of the Sun and since interstellar reddening is usually zero for stars lying within $100$~pc of the Sun \citep{Schuster.Nissen:1989}, no reddening corrections were applied. We checked this assumption using the extinction law by \citet{Chen.etal:1998}, and we verified that the maximum reddening correction would be 0.1 mag for HD~104212. This level of extinction would raise the temperature by $130$~K. Even so, excitation equilibrium was reached for this star 
(see Sect. \ref{Sec_equilibria}). Thus, we adopted $A_{\rm V} = 0$ for all the sample stars. 

We derived temperatures from  CRM10's colour-temperature calibrations. The results were compared with those determined with the widely adopted relations from \citet[][hereafter AAM96]{Alonso.etal:1996} and with temperatures from Geneva photometry, which are also available for all the sample stars.  

CRM10 provide colour-temperature relations for Johnson V and 2MASS J and K$_{\rm S}$ magnitudes, so no magnitude system transformations are needed. To determine the photometric $T_{\rm (J-K)}$ and $T_{\rm (V-K)}$ temperatures from the colour-temperature calibrations described in AAM96, the following photometric system transformations were used. The J, K$_{\rm S}$ magnitudes and colours were  transformed from the 2MASS system to CIT (California Institute of  Technology) system, and from the latter to TCS (Telescopio Carlos S\'anchez) system, with  the relations established by \citet{Carpenter:2001}  and  \citet{Alonso.etal:1994}. The transformations between the Johnson and TCS systems used the relations presented in \citet{Alonso.etal:1994}.  

\begin{figure}
\centering
 \resizebox{\hsize}{!}{\includegraphics{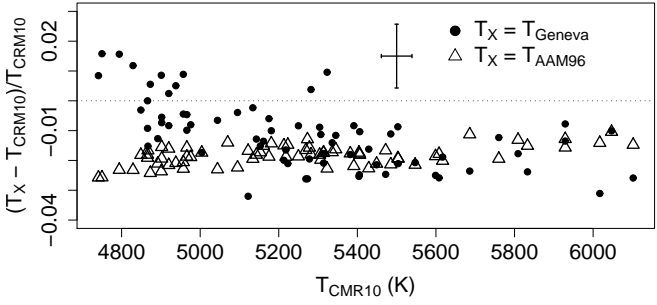}}
\caption{Photometric temperatures. The temperatures from CRM10 calibrations are compared with those from AAM96 relations (open triangles) and from Geneva photometry (filled circles). CRM10 temperatures are about 2\% ($\sim 90$~K) hotter than AAM96 ones.}
\label{Fig_teffs}
\end{figure}

Figure \ref{Fig_teffs} presents the comparison between photometric temperatures.
 AAM96 relations give temperatures about $2$\% ($\sim 90$~K) lower than CMR10 ones, in agreement with differences found by CRM10 between these two calibrations. As discussed in CRM10, the main source of differences between  photometric $T_{\rm eff}$ scales is the absolute calibration of the photometric systems, which is essential when setting the zero point of the scale.  The estimated zero point of the CRM10 scale is defined by a sample of solar twins, resulting  in zero-point uncertainties of $\sim 15$~K.  For AAM96 calibrations, this uncertainty is $\sim 100$~K \citep{Casagrande.etal:2006, Casagrande.etal:2010}.  Moreover, AAM96 calibrations require photometric system transformations, which can introduce  unnecessary errors. Therefore, the CRM10 calibrations were chosen for the effective temperature, and the ${({\rm V} - {\rm K}_{\rm S})}$ colour calibration was preferred over the other colours, owing the extended base line, a confirmed  lower $\sigma(T_{\rm eff}) \sim 25$~K, and  smaller dependence with [Fe/H]. 

The internal errors in temperatures were computed considering the uncertainties in magnitudes and metallicities:

\begin{eqnarray}
 \sigma_{T_{\rm eff}} = \frac{5040}{\theta^2} \left[\left(\frac{\partial\theta}{\partial {\rm [Fe/H]}} \sigma_{\rm [Fe/H]}\right)^2
 + \left(\frac{\partial\theta}{\partial {\rm (V-K)}} \sigma_{\rm (V-K)}\right)^2  \right]^{1/2}
\end{eqnarray}

\noindent where $\sigma_{\rm (V-K)}$ is the quadratic sum of errors in V and K$_{\rm S}$ magnitudes, and $\theta\ (= 5040/{T_{\rm eff}})$ is a function of $({\rm V} - {\rm K})$ and [Fe/H], as given in CRM10. Uncertainties in the zero point of the scale ($15$~K) and the calibration deviations ($25$~K, as given in Table 4 of CRM10) were added quadratically to the internal error. The final temperatures and errors are presented in Table \ref{Tab_final}.

\subsection{Surface gravities}

\label{Sec_grav}

The trigonometric surface gravities, $\log g$, were derived from HIPPARCOS parallaxes, $\pi$, through the 
standard formula

\begin{eqnarray}
\log\left({\frac{g_{\star}}{ g_{\odot}}}\right) = 4\log \left(\frac{T_{\rm eff, \star}}{T_{\rm eff, \odot}}\right) 
+ 0.4(\mathcal{M}_{\rm Bol, \star} - \mathcal{M}_{\rm Bol,\odot}) + 
\log \left(\frac{\rm M_\star}{\rm M_{\odot}}\right) ,
\label{Eq_gravity}
\end{eqnarray}

\noindent where $T_{\rm eff, \star}$ and M$_{\star}$ are the stellar temperature and mass, respectively, and the bolometric magnitude, $\mathcal{M}_{\rm Bol,{\star}}$, is given by 

\begin{displaymath}
\mathcal{M}_{\rm Bol, \star} = {\rm V} - {\rm A}_{\rm V} + {\rm BC}_{\rm V} + 5 \log \pi + 5 .
\end{displaymath}

\noindent The following values were adopted for the Sun: $T_{\rm eff,\odot} = 5777$~K, $\mathcal{M}_{\rm Bol,{\odot}} = 4.75$ \citep{Barbuy:2007} and $\log g_{\odot} = 4.44$. We used bolometric corrections ${\rm BC}_{\rm V}$ from \citet{Alonso.etal:1995}, 
where (V$-$K)$_{\odot} = 1.486$, BC$_{\rm V,\odot} = -0.08$ were adopted.

Errors in gravities were calculated using the error propagation equation, derived from equation \ref{Eq_gravity}:

\begin{eqnarray}
 \sigma_{\log g}^2 =  \sigma_{\rm M}^2 \cdot k^2 \left( \frac{\rm M_{\odot}}{\rm M} \right) ^2 + 
 \sigma_{T_{\rm eff}}^2 \cdot k^2  \left(\frac{4}{T_{\rm eff}} \right) ^2 + 
 \sigma_{\pi}^2 \cdot k^2  \left(\frac{5}{\pi} \right) ^2 + \nonumber\\ +
 2\ \sigma_{{\rm M}, T_{\rm eff}}^2 \cdot k^2 \left( \frac{{\rm M}_{\odot}}{\rm M} \right) 
\left(\frac{4}{T_{\rm eff}} \right)
\end{eqnarray}

\noindent where $k = \log {\rm e} = 0.4343$, and errors in BC$_{\rm V}$ and magnitudes were considered to be small. The variables $\sigma_{\rm M}$, $\sigma_{T_{\rm eff}}$, and $\sigma_{\pi}$ are errors in masses, temperatures, and parallaxes, respectively. Since we used the temperatures to get the masses from isochrones, the covariance between $T_{\rm eff}$ and mass, $\sigma_{{\rm M}, T_{\rm eff}}$, does not vanish. For each star, we took all the possible solutions [M$_i$, ${T}_{{\rm eff}, i}$] within the error bars from the evolutionary tracks. Given that  $\overline{\rm M}$, $\overline{{T}_{\rm eff}}$ are the respective mean values, the covariance can be obtained through

\begin{eqnarray}
 \sigma_{{\rm M}, T_{\rm eff}} = \left\langle({\rm M}_i - \overline{\rm M}) \cdot ({T}_{{\rm eff}, i} - \overline{{T}_{\rm eff}})\right\rangle .
\end{eqnarray}

Table \ref{Tab_loggs_deltas} presents the variations in $\log g$ with temperature, masses, E(B-V), and [Fe/H]. A temperature change of $-2$\% ($\sim100$~K), which corresponds to the difference between CRM10 and AAM96 calibrations, would lead to lower  stellar masses by 4\% in average. The overall change in gravities is $-0.05$ dex on average. Despite the uncertainties involved in the determination of stellar masses, the effect on gravity is not significant: a 5\% change in the masses would change gravities by only $0.02$ dex. As described in Sect. \ref{Sec_masses_ages}, the masses estimated in this work have internal errors of about
 3\% and accuracy of 4\%, resulting in a total uncertainty of about 5\%. Effects of reddening corrections and changes in [Fe/H] are not significant ($0.01$ and $< 0.005$ dex, respectively). 

\begin{figure}
\centering
 \resizebox{\hsize}{!}{\includegraphics{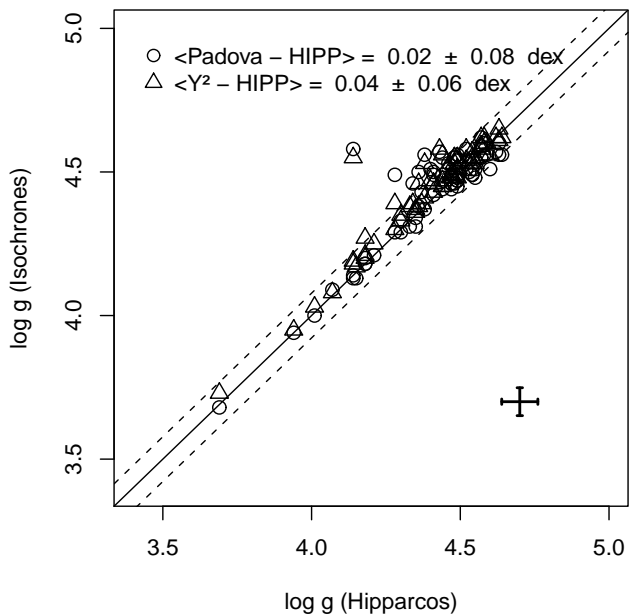}}
\caption {Comparison between trigonometric gravities from HIPPARCOS parallaxes and isochrone gravities. The open triangles represent the gravities obtained from $Y^2$  evolutionary tracks, and gravities from Padova isochrones are indicated by open circles. Solid and dashed lines show the very good agreement,
with variations within $\pm 0.05$ dex.}
\label{Fig_loggs}
\end{figure}

We checked the consistency between trigonometric gravities and evolutionary gravities. For this comparison, we used both $Y^2$ and Padova isochrones \citep{Girardi.etal:2000}.  Padova gravities were obtained using the tool PARAM\footnote{\url{http://stev.oapd.inaf.it/cgi-bin/param}} \citep{daSilva.etal:2006}. The $Y^2$ gravities are the mean values of all the solutions within $T_{\rm eff}$ $\pm$ $\sigma_{T_{\rm eff}}$ and $\mathcal{M}_{\rm abs}$ $\pm$ $\sigma_{\mathcal{M}_{\rm abs}}$, where $\sigma_{T_{\rm eff}}$ and $\sigma_{\mathcal{M}_{\rm abs}}$ are the errors in temperatures and absolute magnitudes. The agreement between both evolutionary ($Y^2$, Padova) gravities and trigonometric gravities is excellent, as shown in Fig. \ref{Fig_loggs}. A  large
difference of $\sim 0.4$~dex between HIPPARCOS and isochrones gravities was found for HD~201237. 
The photometry and/or Galactic extinction should not be the source of this discrepancy. This star has good quality photometry from the 2MASS catalogue; i.e., the photometric quality flags are set to A, and the errors on the magnitudes are $\sim 0.02$ mag. The reddening correction for this star is only E(B-V) $\sim 0.04$ mag, following the law by \citet{Chen.etal:1998}, and this level of correction should not affect the stellar parameter determinations significantly. On the other hand, the HIPPARCOS parallax has a large error ($\sim 19$\%), which leads to an uncertainty of $\sim 0.16$~dex on gravity. Therefore, the parallax error is the most probable source of the discrepancy between HIPPARCOS and isochrones gravities.
The stellar parameters found by \citet{Pompeia.etal:2002a} for this star,
($T_{\rm eff}$, $\log g$, [Fe/H]) $=$ ($4950$~K, $4.10$, $-0.05$), are in good agreement
with those found in this work, ($4829$~K, $4.14$, $0.00$). 

\begin{figure}
\centering
 \resizebox{\hsize}{!}{\includegraphics{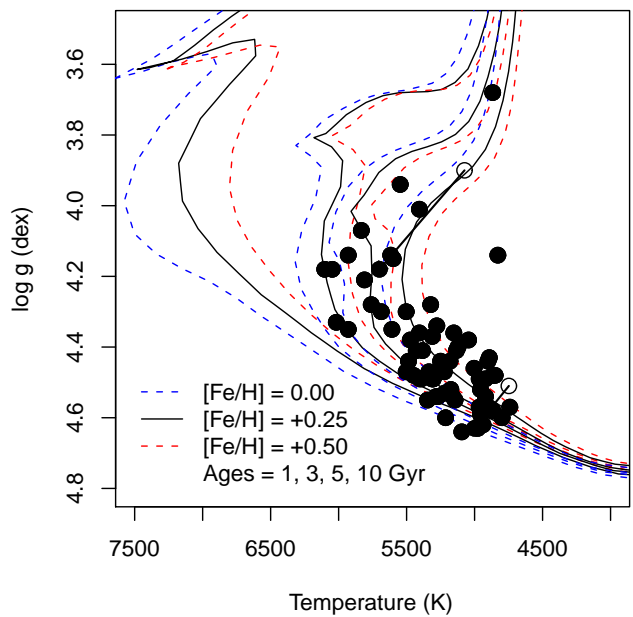}}
\caption {Temperatures {\it vs.} gravities. $Y^2$ isochrones representing different ages and metallicities are shown. The lines connecting an open and a filled circle depict changes in temperature and gravity of HD~94374 and HD~182572 to achieve excitation equilibrium (see text). The open circles indicate photometric temperatures obtained as described in Sect. \ref{Sec_Photo_Temp}; the filled circles correspond to the excitation-equilibrium values, which were adopted for these two stars.}
\label{Fig_teff_loggs}
\end{figure}

The final $T_{\rm eff}$ and $\log g$ parameters are presented in Fig. \ref{Fig_teff_loggs}, where $Y^2$ evolutionary tracks are shown for comparison.

\begin{table}
\centering
\caption{Gravity variations with stellar parameters}
 \begin{tabular}{ccccc}
\hline\hline
  & $\Delta {\rm T}_{\rm eff}$ & $\Delta$Mass & $\Delta$E(B - V)  & $\Delta$[Fe/H]  \\
  &  ($- 2$ \%)                &  ($- 5$ \%)  & ($+ 0.05$ mag)     &  ($ -0.30$ dex)  \\
\hline
$\Delta \log g$ (dex) & -0.05 &  -0.02      &    0.01           &     $<$ 0.01        \\
\hline
 \end{tabular}
\label{Tab_loggs_deltas}
\end{table}
\normalsize

\subsection{Metallicities}
\label{Sec_Met}

We performed an LTE analysis to obtain the iron abundances from the measured equivalent widths. The calculations were carried out using the Meudon code ABON2 \citep[][and updates implemented since then]{Spite:1967}. We used the MARCS 1D hydrostatic model atmospheres \citep{Gustafsson.etal:2008}, obtained by interpolation for the appropriate parameters of the sample stars.

The list of neutral and ionized iron lines used in this work were based on the lists
from \citet{Castro.etal:1997}, \citet{Bensby.etal:2003} and  \citet{Melendez.etal:2009}. The oscillator strengths, $\log gf$, adopted in this work were fitted in order to reproduce the solar iron abundance (\ion{Fe}{I}/H)$_{\odot} = 7.5$\footnote{(X/H) = $\log (N_{\rm X}/N_{\rm H}) - 12$} \citep{Grevesse.Sauval:1998}, using $T_{\rm eff_{\odot}}=5777$~K, $\log g_{\odot}=4.44$,  and $\xi_{{\odot}}=0.9$~km~s$^{-1}$. The damping constants were computed when possible, using  the collisional broadening theory of \citet{Barklem.etal:1998, Barklem.etal:2000} and \citet[][and references therein]{Barklem.AspelundJohansson:2005}. 

The final iron line list comprises only lines that

\begin{figure}
\centering
\begin{tabular}{c}
 \resizebox{\hsize}{!}{\includegraphics{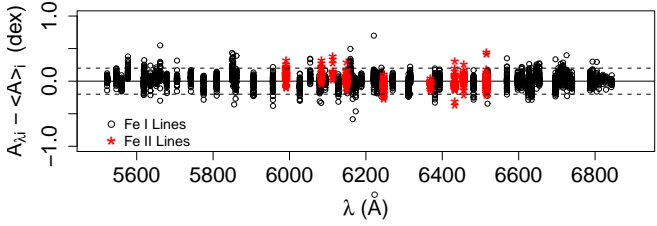}} \\ 
\end{tabular}
\caption {Process of selection for stable \ion{Fe}{I} and \ion{Fe}{II} lines.
The iron abundance of each star, $\left\langle A \right\rangle_i$, is the average of abundances derived from individual lines, $A_{\lambda i}$. $A_{\lambda i} - \left\langle A \right\rangle_i$ values indicate deviations from the mean abundance value for each star. \ion{Fe}{I} and \ion{Fe}{II} lines are indicated as open circles and red stars, respectively.}
\label{Fig_lines}
\end{figure}

\begin{description}
 \item $i)$ are free of blends. We used an atlas of the solar photospheric spectrum \citep{Wallace.etal:1998} and the VALD line lists \citep{Kupka.etal:1999} to check for possible blends, and blended lines were discarded;
 
 \item $ii)$ have solar W$_{\lambda}<100$~m\AA. For this range of W$_{\lambda}$, the astrophysical $\log gf$ values obtained from the solar equivalent widths are more reliable (for stronger lines oscillator strengths and broadening of wings have competing effects); and

 \item $iii)$ give systematically reliable abundances for the sample stars.
 Given that $\left\langle A \right\rangle_i$ is the iron abundance of each star,
 and $A_{\lambda i}$ is the abundance derived from an individual line,
 we checked the ``quality'' of the line by computing $A_{\lambda i} - \left\langle A \right\rangle_ii$ for all the sample stars (Fig. \ref{Fig_lines}). This approach allows us to detect and exclude lines that give abundances 
systematically higher or lower than the abundance of the star by 0.15~dex and lines that seem to lead to inaccurate
 abundance values (deviation in $A_{\lambda i} - \left\langle A \right\rangle_i$ larger than $0.12$~dex). 
\end{description}

Atomic data adopted for the final iron line list are given in Table \ref{Tab_ironlinelist}.
The equivalent widths of iron lines were measured using the automatic code {\tt ARES}\footnote{\url{http://www.astro.up.pt/~sousasag/ares/}} developed by \citet{Sousa.etal:2007}. Given a reference line list and a list of input configuration parameters, {\tt ARES} fits a continuum and measures W$_{\lambda}$ by fitting a Gaussian profile. 

We tested the dependence of $W_{\lambda}$ on the choice of the {\tt ARES}'s input parameters by comparing measurements made with different values, in particular the parameter required for the continuum definition ({\it rejt}) and the parameter that defines the wavelength interval around the line where the computation will be conducted ({\it space}). By considering different values for these parameters in the intervals $0.993<rejt<0.999$ and $2<space<5$~\AA, we obtained $\sigma_{W_{\lambda}}$, which is the standard deviation of the measurements. Despite the majority of the lines giving a stable result under different configurations, we found that some lines are very sensitive to the choice of these parameters. The differences between  measurements made with different values of {\it rejt} and {\it space} can be as high as $\sigma_{W_{\lambda}}\sim15-20$\%. To avoid the effect of these lines in computing of the final metallicity, for each line with $W_{\lambda}\pm\sigma_{W_{\lambda}}$, we computed the iron abundance [Fe/H]$\pm\sigma_{\rm [Fe/H]}$. The final [Fe/H] of each star was then considered as the weighted mean of the abundances, where $w=(1/\sigma_{\rm [Fe/H]})$ were used as weights for each line.

\begin{figure*}
\centering
\begin{tabular}{cc}
 \resizebox{0.49\hsize}{!}{\includegraphics{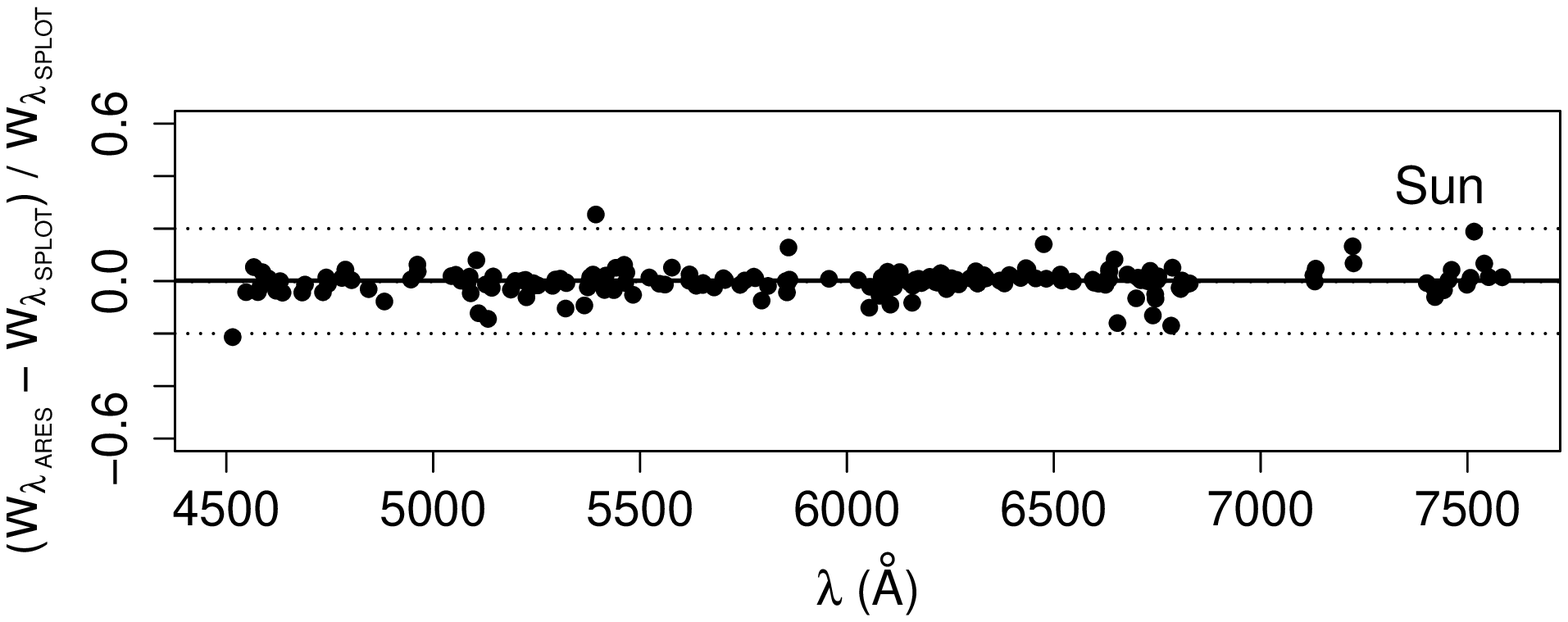}} &
 \resizebox{0.49\hsize}{!}{\includegraphics{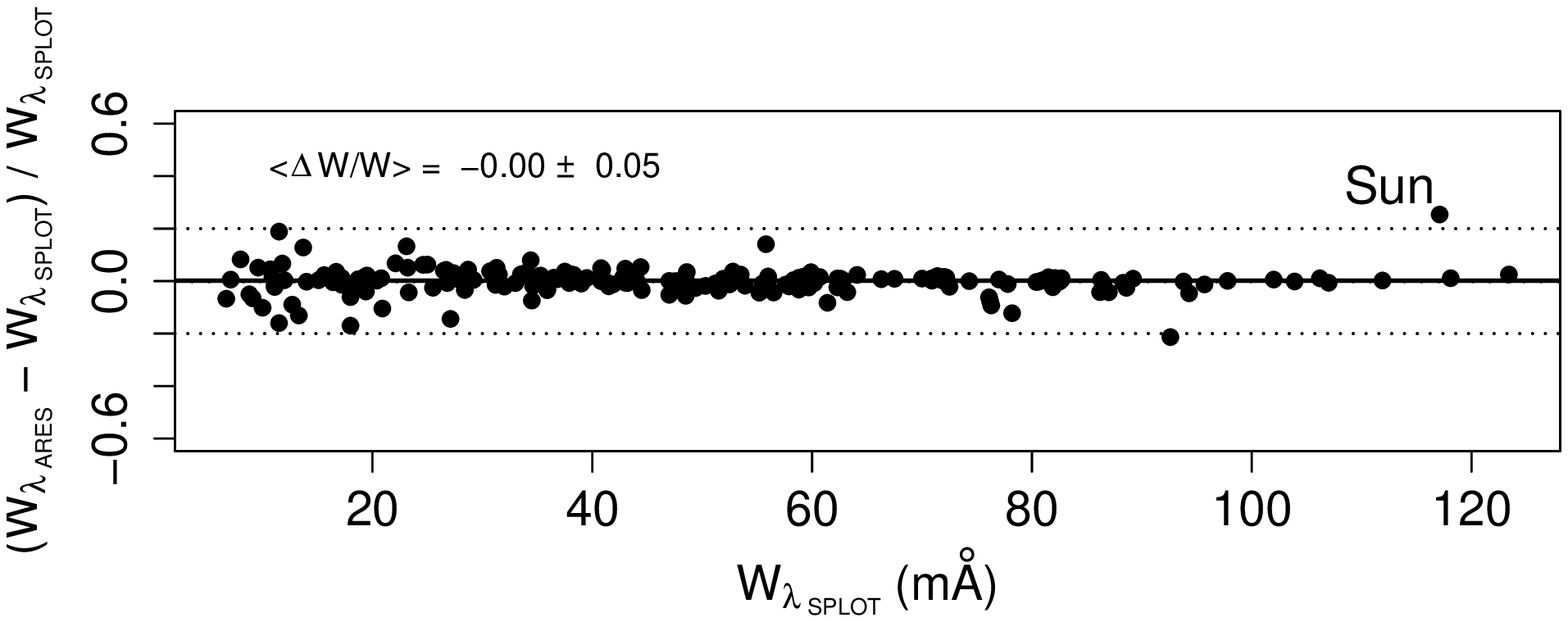}} \\
 \end{tabular}
\caption {Equivalent widths measurements from ARES and IRAF for the solar spectrum. The relative differences, $(W_{\lambda_{\rm ARES}} - W_{\lambda_{\rm IRAF}}) / W_{\lambda_{\rm IRAF}}$, are given as a function of wavelength ({\it left}) and $W_{\lambda_{\rm IRAF}}$ ({\it right}).}
\label{Fig_ares_iraf_sun}
\end{figure*}

\begin{figure*}
\centering
\begin{tabular}{cc}
\vspace{-2cm}
 \resizebox{0.49\hsize}{!}{\includegraphics{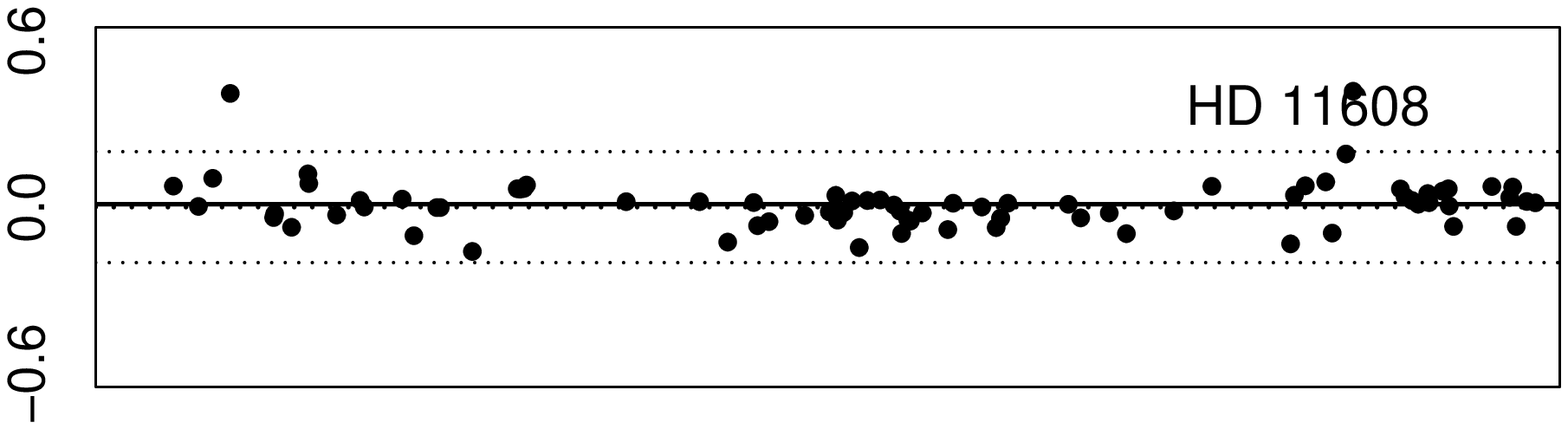}} &
 \resizebox{0.49\hsize}{!}{\includegraphics{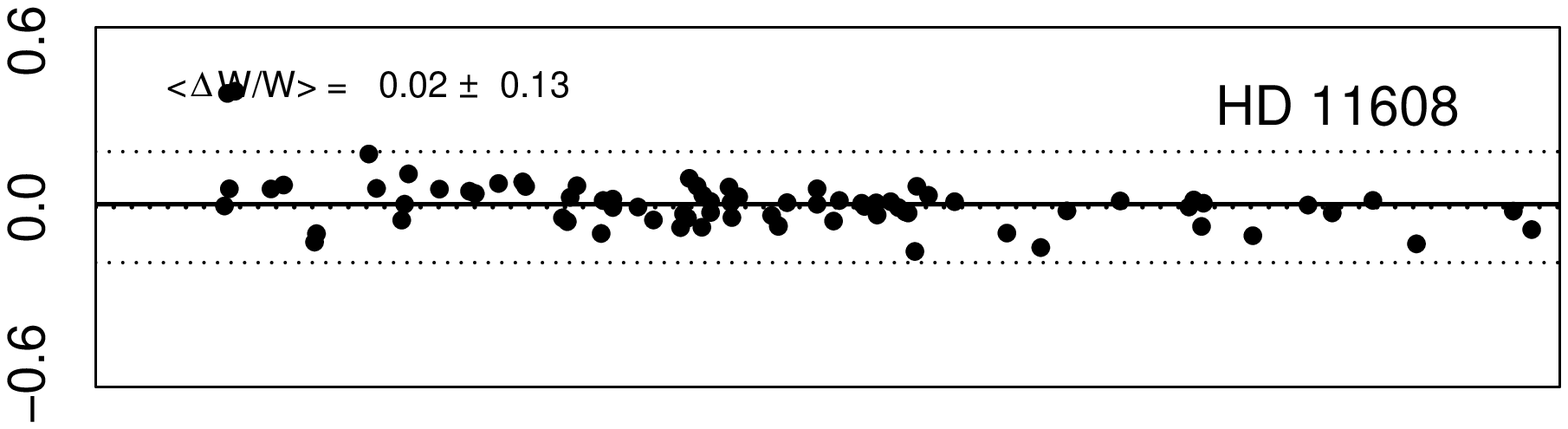}}\\
\vspace{-2cm}
 \resizebox{0.49\hsize}{!}{\includegraphics{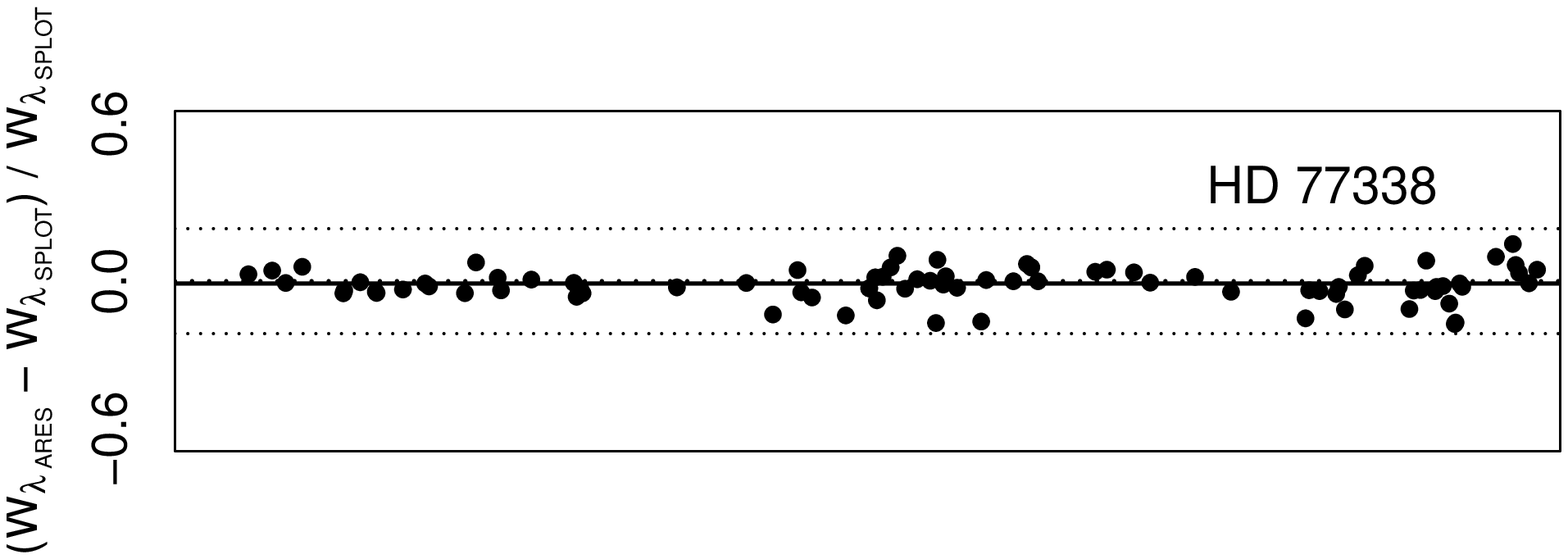}} &
 \resizebox{0.49\hsize}{!}{\includegraphics{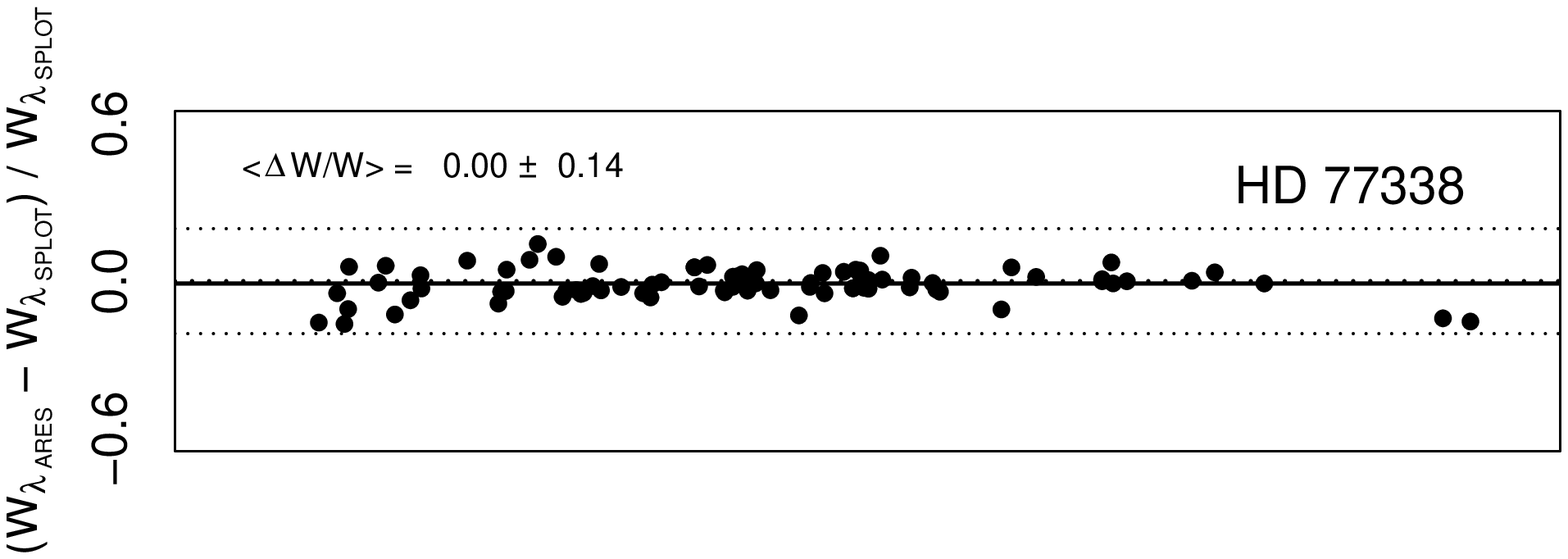}} \\
 \resizebox{0.49\hsize}{!}{\includegraphics{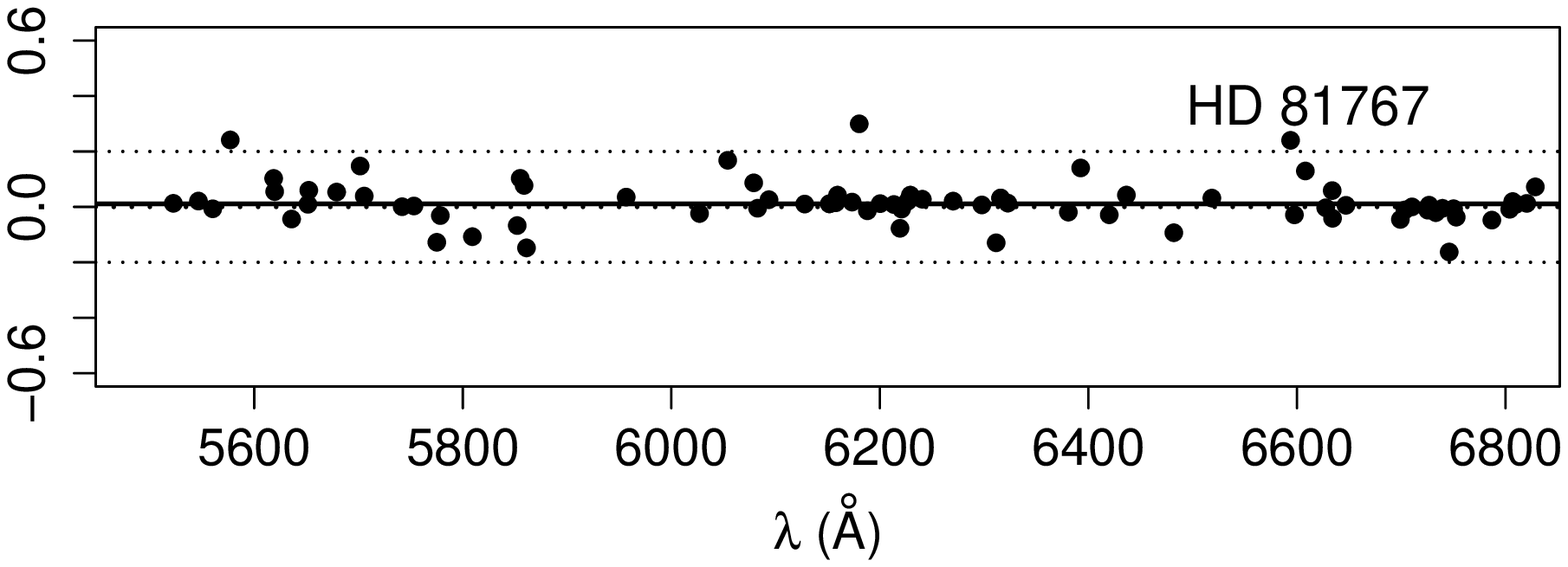}} &
 \resizebox{0.49\hsize}{!}{\includegraphics{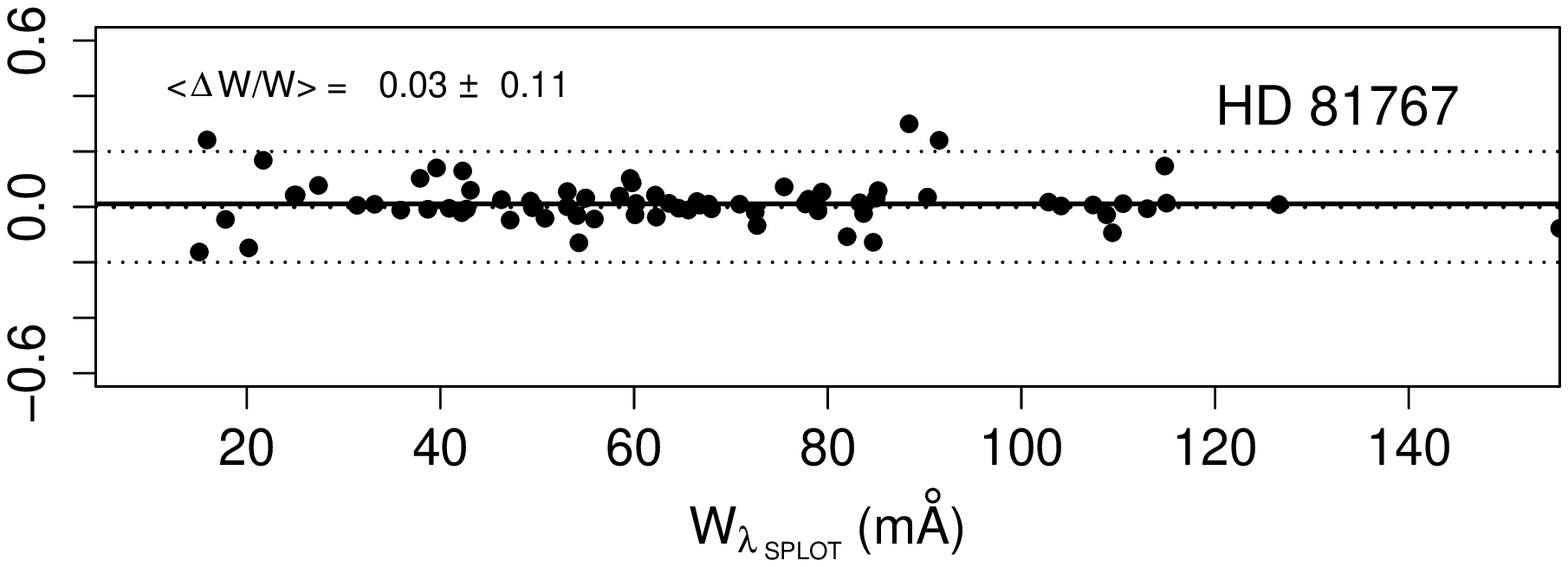}} \\
 \end{tabular}
\caption {Same as Fig. \ref{Fig_ares_iraf_sun} for the stars HD~11608, HD~77338, and HD~81767.}
\label{Fig_ares_iraf}
\end{figure*}

The reliability of the {\tt ARES} measurements was confirmed by comparing equivalent widths
 of iron lines measured with both {\tt ARES} and the task {\tt SPLOT} in the {\tt IRAF} context. 
Figures \ref{Fig_ares_iraf_sun} and \ref{Fig_ares_iraf} present the good agreement
 between $W_{\lambda}$s measurements for the Sun and the stars HD~11608, HD~77338, and HD~81767. 

We employed the equivalent widths of \ion{Fe}{I} lines to derive the microturbulence velocity, $\xi$, 
by requiring independence between abundances and the reduced equivalent width, $\log(W_{\lambda}/\lambda)$. 
Only lines with $20 < W_{\lambda} < 100$~m\AA\ were used, since fainter lines show some scatter  in the comparison
between {\tt ARES} and {\tt SPLOT}, and stronger lines 
are not well fitted by Gaussian profiles. 
No clear $\xi$ velocities were found for our cooler stars ($T_{\rm eff} < 5000$~K).
 We considered that $\xi$ can be defined as a function of temperature and gravity,
 and using stars with $T_{\rm eff} > 5200$~K, we defined $\xi$ = $f(T_{\rm eff}$, $\log g)$
 and then extrapolated this function to $T_{\rm eff} < 5200$~K.  We adopted 0.3~km s$^{-1}$
 from  the extrapolation of our fit. The same problem was found by \citet{Feltzing.Gustafsson:1998}
 for their K dwarfs, and they adopted a constant value of 1~km s$^{-1}$. We found that this value is too
 high for our sample, leading to lower metallicities in this temperature range and, consequently, to a positive gradient [Fe/H] {\it vs.} $T_{\rm eff}$.  A variation of $\Delta \xi = +0.2$~km s$^{-1}$ leads to $\Delta$[\ion{Fe}{I}, II/H] $=-0.05$~dex. The errors in $\xi$ were calculated considering the uncertainty in the slope of abundance {\it vs.} W$_{\lambda}$, and are typically $0.1$ km s$^{-1}$.

To compute the errors in the final metallicity, the following sources of uncertainties were taken into account: $i)$ uncertainties in the stellar parameters [$T_{\rm eff}$, $v_{t}$, $\log g$]; $ii)$ errors in the W$_{\lambda}$s measurements, which were estimated by considering several configurations of the ARES code as described above; and $iii)$ uncertainties in $\log gf$ values, which were computed by considering errors in the solar $W_{\lambda}$s. The total errors are $\sim 0.05$ for [\ion{Fe}{I}/H] and $0.09$~dex for [\ion{Fe}{II}/H].

\subsubsection{Metallicities from two different codes}
\label{Sec_codes}

We proceeded with all the calculations described above using both
 the code by the Uppsala group BSYN/EQWI \citep[][and updates since then]{Edvardsson.etal:1993} and the Meudon code ABON2
\citep[][and updates implemented since then]{Spite:1967}. All steps of the calculation were carefully compared: optical depths of lines, 
continuum opacities  $\kappa_c$, line broadening, and final abundances. 

\begin{figure}
\centering
 \resizebox{\hsize}{!}{\includegraphics{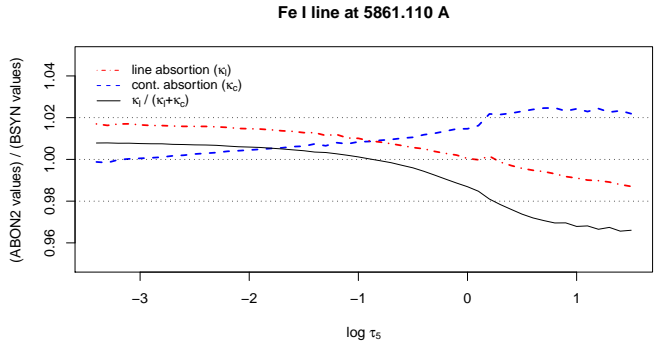}}
\caption {Ratio of opacities computed using the BSYN/EQWI \citep{Edvardsson.etal:1993} and ABON2 \citep{Spite:1967} codes
 for an \ion{Fe}{I} line at $\lambda = 5861.11$ {\rm \AA} for the solar model.}
\label{Fig_opacities}
\end{figure}

The dominating opacity source is the H$^-$ bound-free absorption. The two codes consider different calculations for the H$^-$ photo-detachment cross section $\sigma_{\lambda}$. The ABON2 code \citep{Spite:1967} adopts \citet{Geltman:1962} calculations, represented by the polynomial expressions from \citet{Gingerich:1964}, while cross sections from \citet{Wishart:1979} are adopted in the BSYN/EQWI code \citep{Edvardsson.etal:1993}. Using these sources and considering a \ion{Fe}{I} line at 5861\AA\ and the solar atmosphere model, we found that differences in the continuum absorption are up to 4\% in the upper atmospheric layers ($\tau_5 < -1$), and less than 1\% in the bottom of the photosphere ($\tau_5 > 0$). We updated the ABON2 code \citep{Spite:1967} using new cross sections calculations from \citet{John:1988}, which improve the agreement of $\kappa_c$ between these two codes to $< 1$\% at the upper layers with $\tau_5 < -1$ and $\sim$ 2\% at $\tau_5 > 0$ layers (Fig. \ref{Fig_opacities}).

The line opacity ($\kappa_l$) calculated for an \ion{Fe}{I} line at 5861 \AA\ is shown in Fig. \ref{Fig_opacities} with a solar model. We found that the ratio between line opacities, $\kappa_{l, \rm BSYN} / \kappa_{l, \rm ABON2}$, decreases with optical depth. An agreement between their values in the two codes is found at $\log \tau_5 \approx 0$. The small discrepancy is mainly due to differences in the van der Waals broadening determinations. The ratio $\kappa_{l} / (\kappa_{l} + \kappa_{c})$ is within 1\% at layers with $\tau_5 < -1$.

The same analysis was carried out for other lines at different wavelengths and atmosphere models. Despite the small trends with $\lambda$ and $T_{\rm eff}$, Table \ref{Tab_codes} shows that the differences found between metallicities obtained with the two codes are within $0.02$~dex.

\begin{table}
\centering
\caption{Abundances obtained with two different codes}
 \begin{tabular}{ccccc}
\hline\hline
  Star & $T_{\rm eff}$ (K) & $\log g$ & $\Delta$[\ion{Fe}{I}/H]$^a$ & $\Delta$[\ion{Fe}{II}/H]$^a$ \\
\hline

 Sun      & 5777 & 4.44 & 0.00 & 0.00 \\
HD 15133  & 5223 & 4.47 & 0.01 & 0.02 \\
HD 77338  & 5346 & 4.55 & 0.01 & 0.01 \\
HD 90054  & 6047 & 4.18 & 0.01 & 0.00 \\
HD 177374 & 5044 & 4.38 & 0.01 & 0.02 \\ 
\hline
 \end{tabular}
\tablefoot{
\tablefoottext{a}{$\Delta$[Fe/H] $={\rm [Fe/H]}_{\rm BSYN} - {\rm [Fe/H]}_{\rm ABON2}$.}
}
\label{Tab_codes}
\end{table}

\subsection{Checks on excitation and ionization equilibria}
\label{Sec_equilibria}

\begin{figure}[ht]
\centering
\begin{tabular}{c}
\vspace{-0.8cm}
\includegraphics[width = 8.8cm, height = 3.5cm]{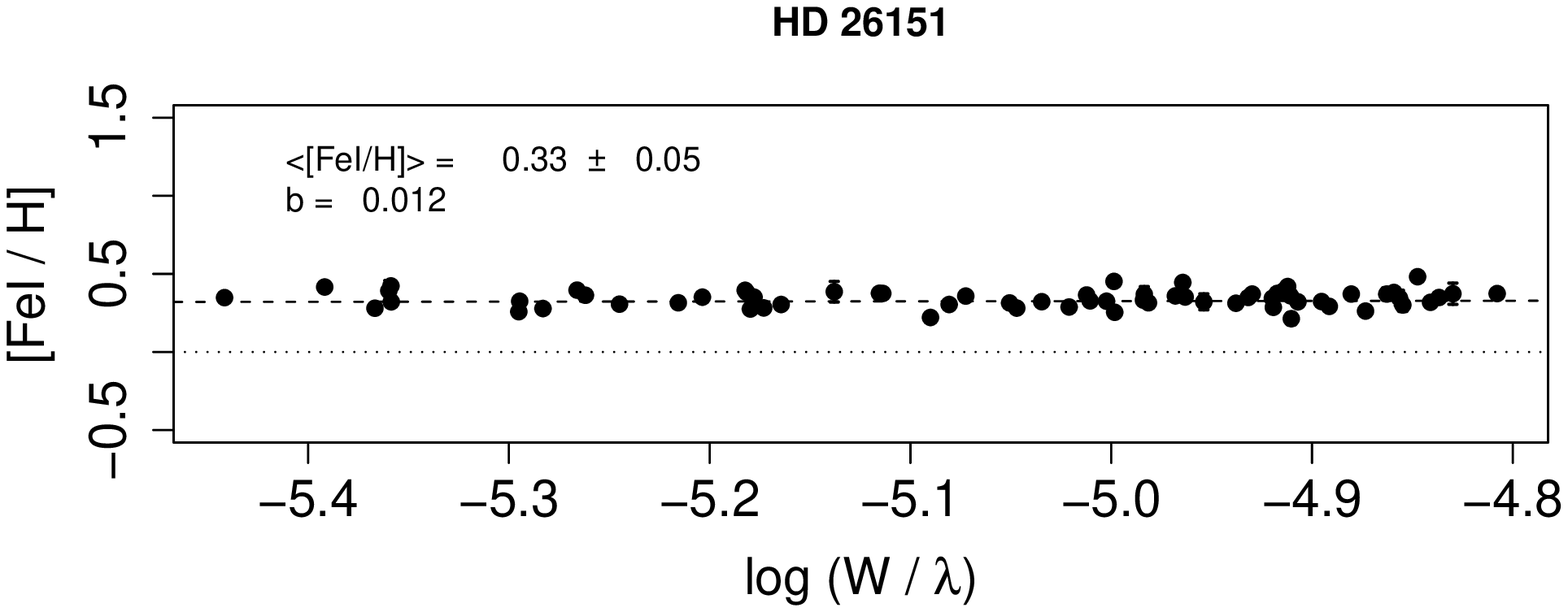}   \\  
\includegraphics[width = 8.8cm, height = 3.5cm]{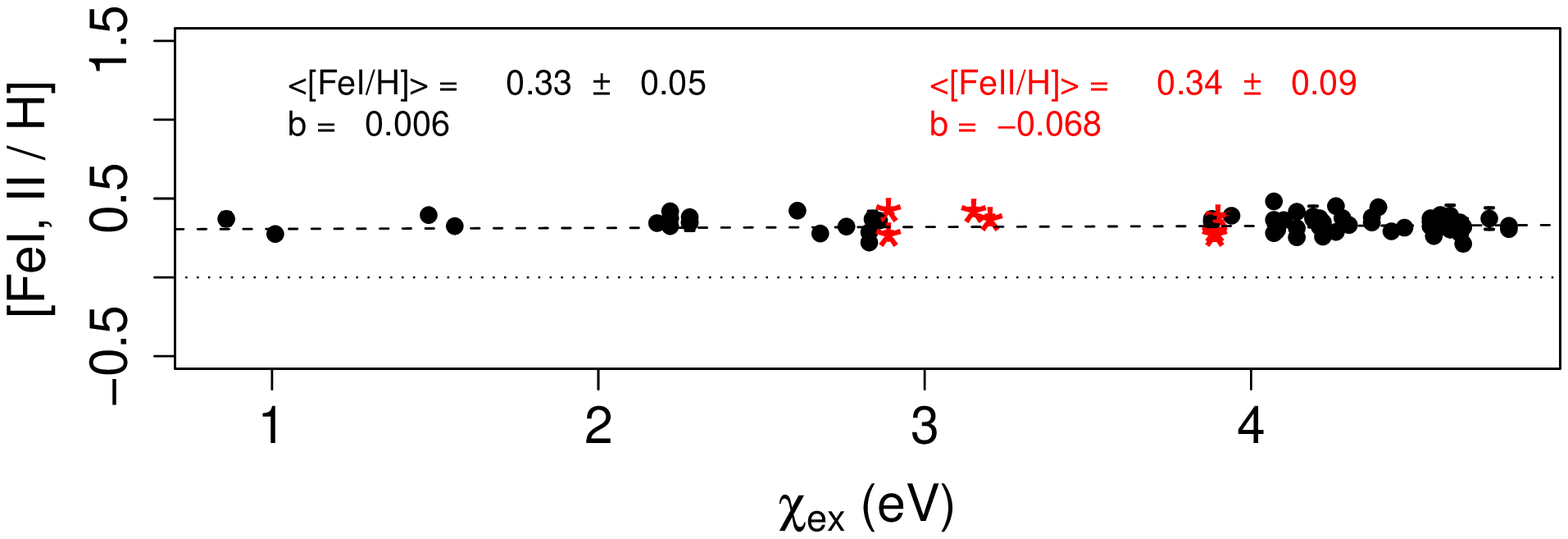} \\ 
\end{tabular}
\caption {Abundances versus normalized equivalent widths, $\log(W/\lambda)$ ({\it top}), and excitation/ionization equilibrium ({\it bottom}) shown for HD~26151, as an example. Red stars represent \ion{Fe}{II} lines. The \ion{Fe}{I} and \ion{Fe}{II} abundances and linear coefficients $b$ are given in the plots.}
\label{Fig_equilibrium_std}
\end{figure}

\begin{figure}
\centering
\begin{tabular}{c}
\vspace{-1cm}
 \resizebox{\hsize}{!}{\includegraphics{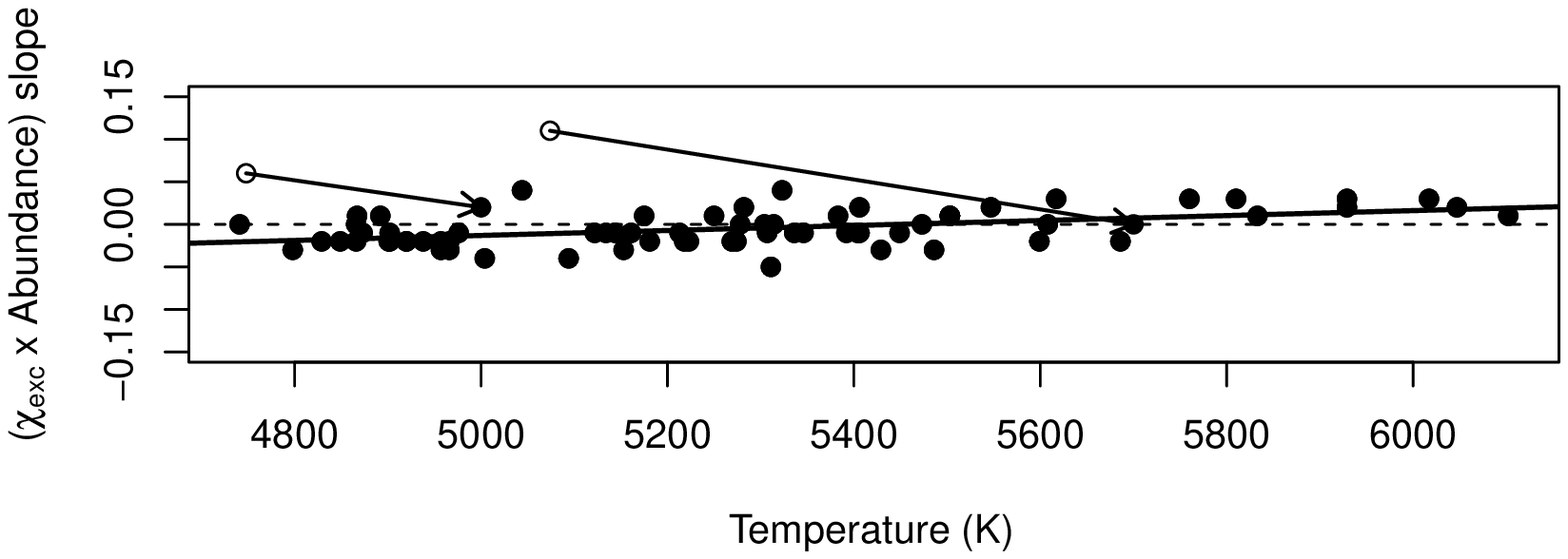}} \\
\vspace{-1cm}
 \resizebox{\hsize}{!}{\includegraphics{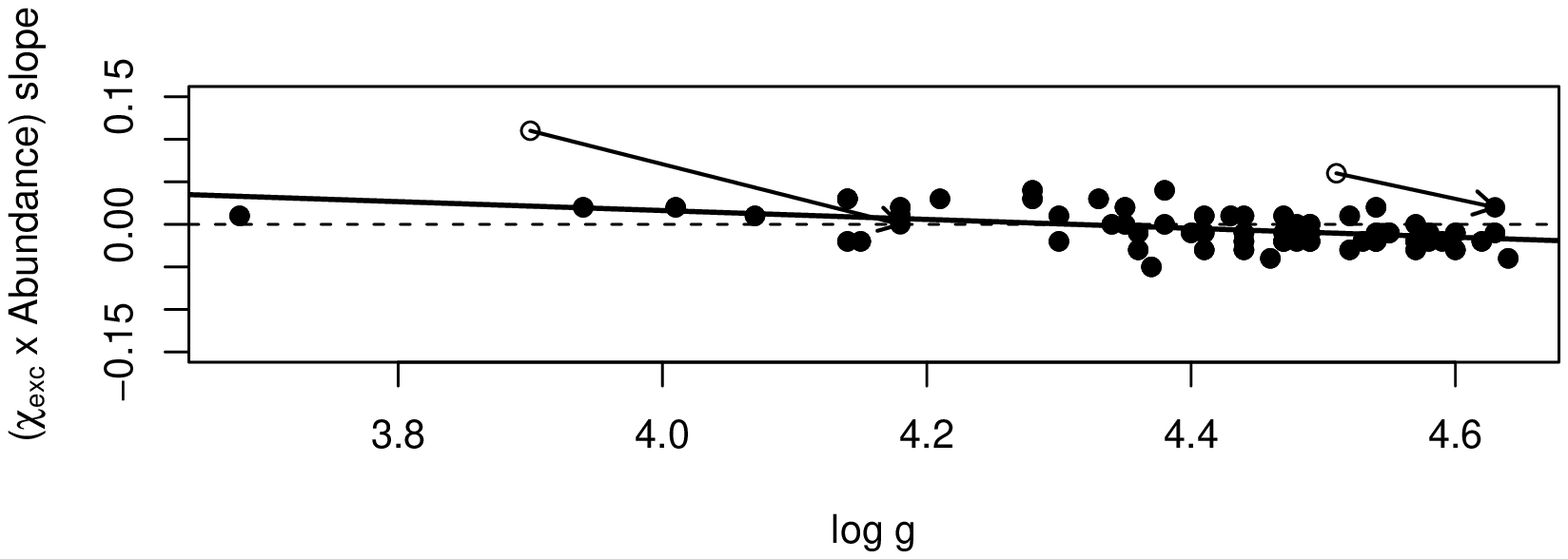}} \\
 \resizebox{\hsize}{!}{\includegraphics{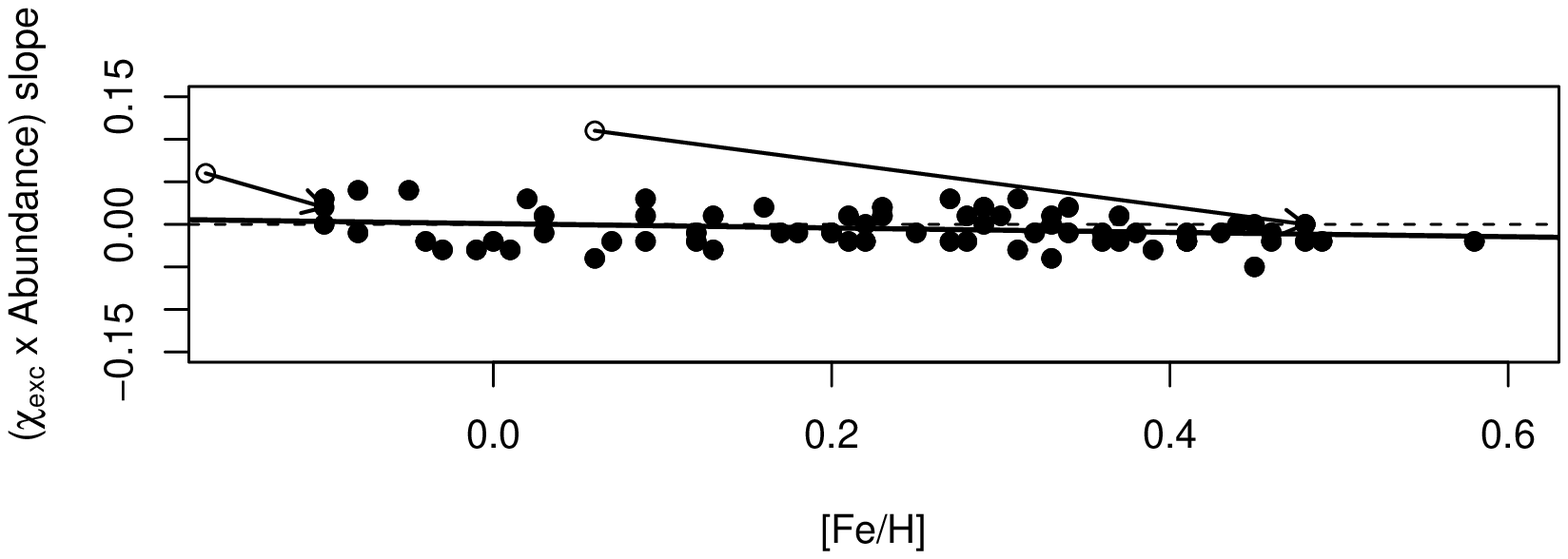}} \\
 \end{tabular}
\caption {Excitation equilibrium as a function of temperature ({\it top}), gravity ({\it middle}), and metallicity ({\it bottom}). The dashed line indicates the perfect excitation equilibrium. The arrows indicate changes in temperature and gravity of HD~94374 and HD~182572 to achieve excitation equilibrium (see text). Open circles indicate
 photometric temperatures obtained as described in Sect. \ref{Sec_Photo_Temp}; filled circles correspond to the adopted, excitation-equilibrium values.}
\label{Fig_boltzmann}
\end{figure}

Since we are using temperatures from photometric scales and trigonometric gravities, it is interesting to check whether excitation and ionization equilibria are reached. For each star, we plotted iron  abundances {\it vs.}
the excitation potential $\chi_{exc}$(eV) and performed a linear fit between these two quantities (Fig. \ref{Fig_equilibrium_std} shows HD~26151 as an example). The excitation equilibria is indicated by the independence between  abundances and $\chi_{exc}$; i. e., the slope $b$ must be zero. Figure \ref{Fig_boltzmann} presents
the slope $b$ as a function of temperature, gravity, and metallicity for all the sample stars. The slope exceeded $2\sigma$  from the mean value for two stars: HD~94374 and HD~182572. HD~182572 is a known variable, with an amplitude variation smaller than 0.2~mag. The photometric calibrations give $T_{\rm eff} \approx 5070$~K, and the excitation temperature ($5700$~K) was adopted for this star. For G~161-029 no parallax measurement is available. Therefore spectroscopic $T_{\rm eff}$ and $\log g$ were adopted for these three stars.

\begin{figure}
\centering
\begin{tabular}{c}
 \resizebox{\hsize}{!}{\includegraphics{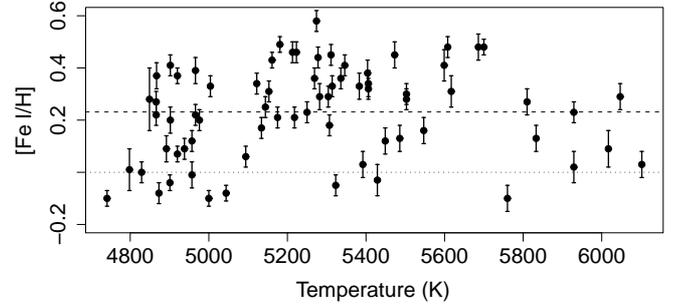}} \\
 \end{tabular}
\caption {[\ion{Fe}{I}/H] {\it vs.} temperatures. Dashed and dotted lines indicate the mean iron abundance and [Fe/H] $= 0$, respectively.}
\label{Fig_abo_teff}
\end{figure}

\begin{figure}
\centering
\begin{tabular}{c}
 \resizebox{\hsize}{!}{\includegraphics{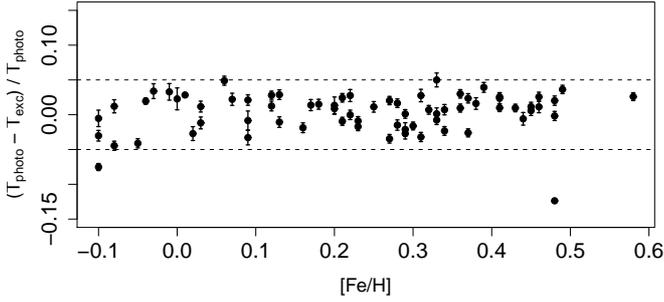}} \\
 \end{tabular}
\caption {Differences between photometric and excitation temperatures {\it vs.} metallicities. The dashed lines indicate $\pm$5\%. }
\label{Fig_tphoto_exc}
\end{figure}

In addition to these outlier stars in Fig. \ref{Fig_boltzmann}, we found small trends between the slope $b$ and temperatures/gravities. This was also found by Feltzing \& Gustafsson (1998) in their analysis of 47 metal-rich stars. On the other hand, the trend between $b$ and [Fe/H] is very small; thus, the effect of choosing photometric temperatures over excitation ones on the final metallicities should be negligible. We quantified this effect by obtaining the excitation temperatures by requiring zero slope from the excitation energy balance diagram, and we compared the excitation and  photometric temperatures, as shown in Fig. \ref{Fig_tphoto_exc}.
 The excitation temperatures agree well with the photometric ones. Differences between temperatures are all below $\pm$~5\%, except for the stars HD~94371 and HD~182572, for which the excitation temperatures were adopted.  The mean difference is only 0.7\%, with a standard deviation of 2.5\%. For [Fe/H] $\gtrsim 0.4$, the photometric temperatures are systematically higher than excitation temperatures by an amount of 2\% in average. If we consider 2\% lower temperatures in these cases, 
the resulting change in abundance is $\sim -0.05$~dex. Finally, in Fig. \ref{Fig_abo_teff},
 we show that the iron abundances derived from \ion{Fe}{I}  show no obvious trend with $T_{\rm eff}$. 

\begin{figure}
\centering
\begin{tabular}{c}
\vspace{-1cm}
 \resizebox{\hsize}{!}{\includegraphics{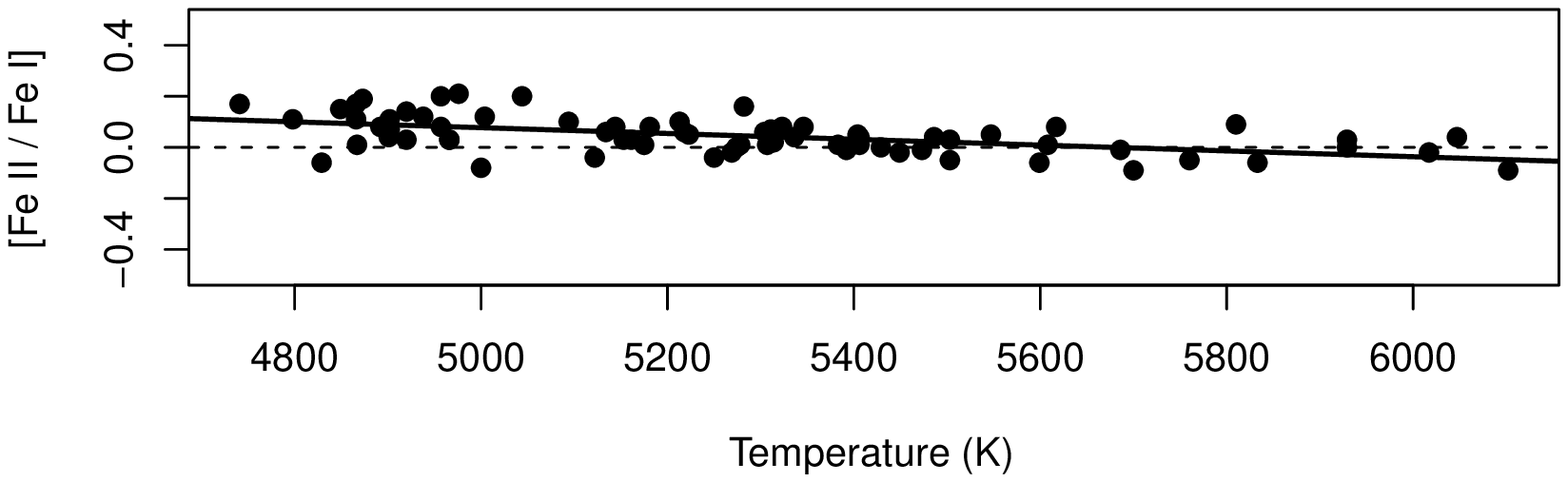}} \\
\vspace{-1cm}
 \resizebox{\hsize}{!}{\includegraphics{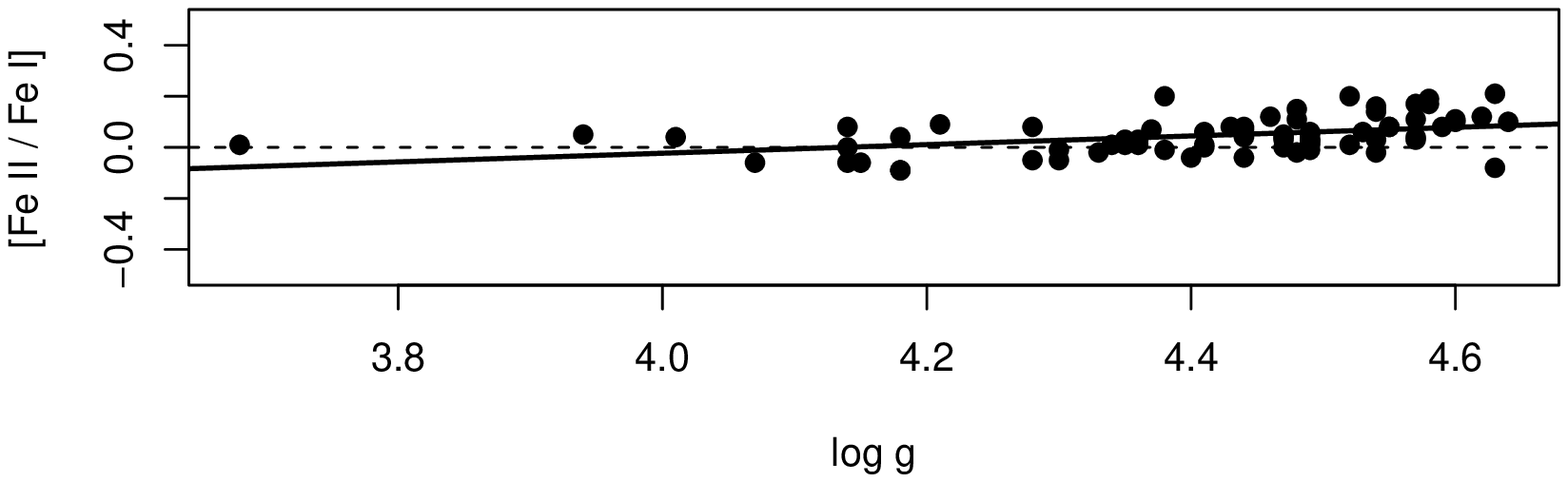}} \\
 \resizebox{\hsize}{!}{\includegraphics{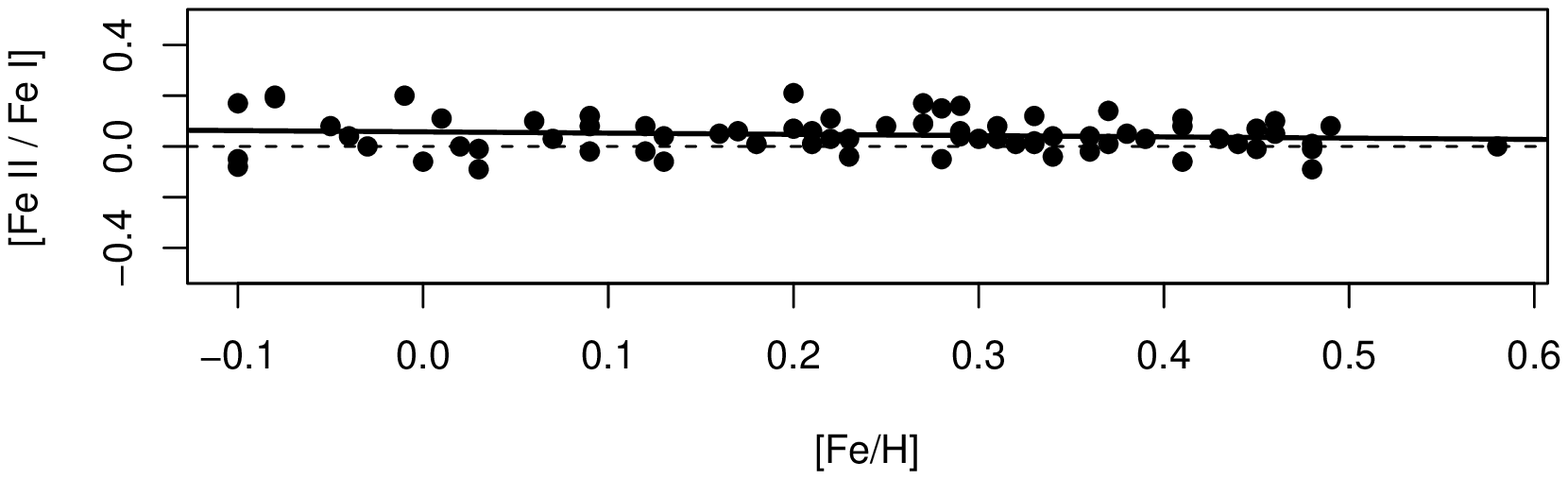}} \\
 \end{tabular}
\caption {Ionization equilibrium  as a function of temperature ({\it top}), gravity ({\it middle}), and metallicity ({\it bottom}). [\ion{Fe}{I}/\ion{Fe}{II}] is the difference between abundances from neutral and single ionized iron lines, [\ion{Fe}{I}/H] $-$ [\ion{Fe}{II}/H].}
\label{Fig_saha}
\end{figure}

Ionization equilibria, indicated by the difference between \ion{Fe}{I} and \ion{Fe}{II} abundances, is shown in Fig. \ref{Fig_saha} for all the sample stars. We found an apparent overionization as compared to expectations from LTE calculations for the cooler stars ($T_{\rm eff} \leq 5200$~K) in the sample, with an upper limit of [\ion{Fe}{II}$-$/\ion{Fe}{I}] $< 0.2$~dex. Even though lines of ionized iron atoms are less susceptible to non-LTE effects than \ion{Fe}{I}
\citep{Thevenin.Idiart:1999}, we considered the abundances derived from \ion{Fe}{I} lines to be the final metallicities of our stars, since \ion{Fe}{I} lines are more numerous.

%

\subsection{Stellar masses and ages}

\label{Sec_masses_ages}

Isochrone fitting techniques can provide estimates of stellar masses and ages. \citet{AllendePrieto.Lambert:1999} compared masses derived from interpolation of isochrones and the direct estimates from observations in eclipsing spectroscopic binaries, and they concluded that masses can be estimated with uncertainties
 below 8\%. More recently, Mel\'endez et al. (2011, in prep.) show that masses and ages can be estimated with even higher accuracy, provided the isochrones are well calibrated to reproduce the solar ages and masses.

Therefore, we used a grid of Yonsei-Yale isochrones to determine the masses and ages for the sample stars. The isochrone points were characterized by the effective temperature ($T_{\rm eff}$), the absolute magnitude ($\mathcal{M}_{\rm V}$), and the metallicity ([Fe/H]). Using $T_{\rm eff}$, parallaxes, apparent magnitudes, and [Fe/H] as input values, we recovered the possible solutions for $\log g$, masses, and ages, which are within the errors in $T_{\rm eff}$, $\mathcal{M}_{\rm V}$ and [Fe/H], and computed the mean values. This procedure was repeated 200 times, and each time, the input values  were varied following a normal distribution with mean $X$ 
and standard deviation of $\sigma_X$, where $X$ (with X =  $T_{\rm eff}$, $\mathcal{M}_{\rm V}$ and [Fe/H])
 is the parameter value and $\sigma_X$ is the error associated with it. The internal errors on masses and ages were then computed as the standard deviation of the output values of these 200 realizations; errors of about 3\% were found.

To check the accuracy of our mass determination method, we derived the masses of stars listed in \citet{Torres.etal:2010}. These authors have produced the most recent compilation of high-accuracy mass determinations in binaries ($<$ 3\%).
 We selected stars with temperatures and distances similar to those of
 our sample stars: parallaxes in the range 11 to 70 mas and temperatures from 4500 to 6500 K. We found 14 stars 
satisfying these criteria. Using the temperatures from these authors and solar metallicity, we derived the masses using the same method as was applied to our stars. We found that masses obtained with $Y^2$ isochrones are 4\% lower than masses from Torres et al., with a standard deviation of 5\%.  The good accuracy found in the present work comes from the narrower range of parameters considered here. 
Despite having found a systematic difference between masses from isochrones and those from dynamical considerations, 
no corrections were applied to the masses of our sample stars, since this difference would lead to
 lower $\log g$ only by an amount of $< 0.02$~dex, and the effect in the resulting abundances 
would be negligible. We added these uncertainties to the internal errors of masses and $\log g$.

We also derived the stellar masses from the Padova \citep{Girardi.etal:2000} isochrones with the tool PARAM \citep{daSilva.etal:2006} (see also Sect. \ref{Sec_grav}).  We found that masses estimated from 
Padova isochrones are $\sim 5\%$ lower than $Y^2$ masses, and this difference leads to lower gravities by  $\sim 0.02$~dex. 

Using the same procedure for mass determinations, we obtained the ages for all sample stars. The ages of 36 stars could be determined with errors smaller than 30\%, and for 22 of them, the errors are within 20\%. The stellar masses,
 ages, and their uncertainties for all  sample stars are reported in Table \ref{Tab_final}.

\subsection{Final parameters and comparison with other studies}
\label{Sec_Final_pars}

The final adopted temperatures, gravities, and metallicities were compared with data available in the literature. 
For comparison purposes, the parameters ($T_{\rm eff}$, $\log g$, [Fe/H]) were retrieved from the PASTEL catalogue \citep{Soubiran.etal:2010}, which compiles stellar atmospheric parameters obtained from the analysis of highresolution, high signal-to-noise spectra.
 We only took into account analyses more recent than the year 1997 into account, keeping only from previous year
the reference paper of \citet{McWilliam:1990}. The parameters of 38 of our sample stars are available in this catalogue.
 Table~\ref{Tab_pastel_average} presents the mean values and standard deviations (when more than one value is available) of ($T_{\rm eff}$, $\log g$, [Fe/H]). The differences are also reported. The selected list of stellar parameters given
 in this catalogue is reported in Table \ref{Tab_pastel}.

\begin{figure}
\centering
\begin{tabular}{c}
\vspace{-1.8cm}
 \resizebox{\hsize}{!}{\includegraphics{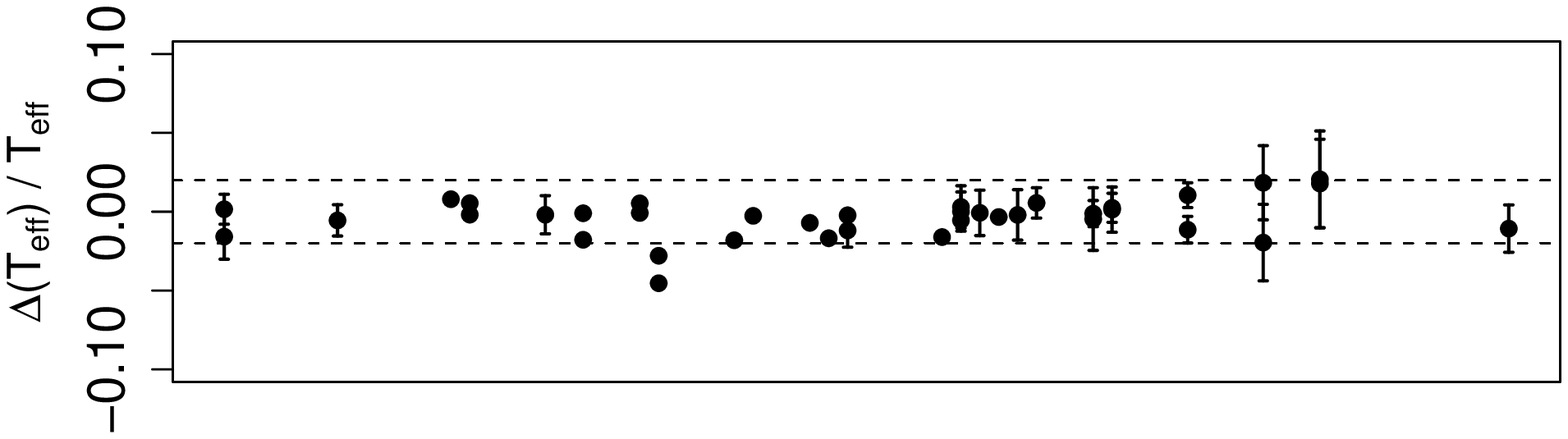}} \\
\vspace{-1.8cm}
 \resizebox{\hsize}{!}{\includegraphics{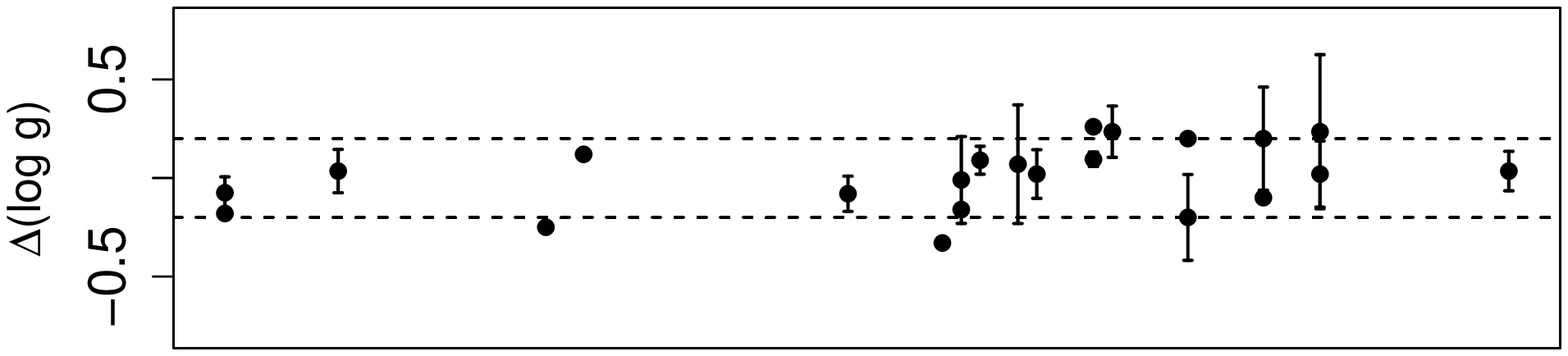}} \\
 \resizebox{\hsize}{!}{\includegraphics{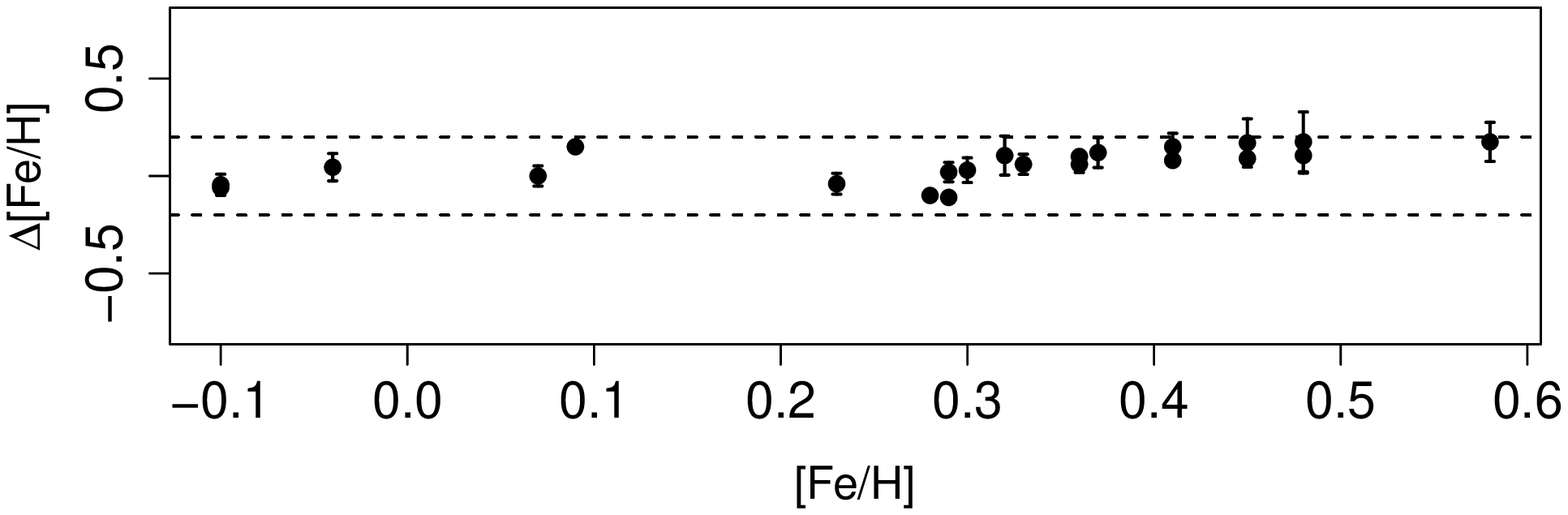}} \\ 
 \end{tabular}
\caption {Relative differences between temperatures ({\it top}), gravities ({\it middle}), and metallicities ({\it bottom}) derived
in the present work and from the PASTEL catalogue. Differences are {\it this work - PASTEL}. Dashed lines indicate $\pm 2$\% in the $T_{\rm eff}$ panel
and $\pm 0.2$~dex in the $\log g$ and metallicity panels. }
\label{Fig_pars_literature}
\end{figure}

The comparisons are presented in Fig. \ref{Fig_pars_literature}. Differences between temperatures considered in the present work and those from the PASTEL catalogue do not exceed $2$\%, except for two stars (HD~31827 and HD~35854). We also found good agreement between gravities, with differences within $\sim 0.2$~dex. 

The metallicities derived in this work are in good agreement with the values reported in the literature for
 [Fe/H]$ \lesssim 0.3$. At higher metallicities ([Fe/H] $ \gtrsim 0.3$), the abundances determined in this work are systematically higher than the values reported in the literature  by $\sim 0.1$~dex on average. On the other hand, there are no systematic differences between temperatures and gravities for [Fe/H] $\gtrsim 0.3$; thus, it is unlikely that differences in stellar parameters are the source of our higher metallicities.   

We also compared the present final metallicities and the photometric metallicities from the GCS survey. 
Our sample contains 17 stars in common with GCS, and the mean difference between the metallicities of these stars is [Fe/H]$_{\rm present} - $ [Fe/H]$_{\rm GCS} = 0.08 \pm 0.12$. Improved new calibrations of the GCS data from \citet{Casagrande.etal:2011} brings the GCS metallicity scale into agreement with ours: using the temperatures from CRM10 calibrations, Casagrande et al. also found higher [Fe/H] by an amount of $0.1$~dex.

Finally, the metallicities derived from the Geneva photometry, presented in Table \ref{Tab_final},
show differences of spectroscopic iron abundances derived in the 
present work being $-0.12 \pm 0.16$~dex lower than the Geneva photometric metallicities.

\begin{table*}
\centering
\caption{Data from the PASTEL catalogue}
\begin{tabular}{clclrlr}
\hline\hline

Star    & \multicolumn{1}{c}{$T_{\rm eff}$ (K)} & \multicolumn{1}{c}{$\Delta T_{\rm eff}$ (K)} & \multicolumn{1}{c}{$\log g$} & \multicolumn{1}{c}{$\Delta \log g$} & \multicolumn{1}{c}{[Fe/H]} & \multicolumn{1}{c}{$\Delta$[Fe/H]} \\  
\hline
HD     8389 & 5330 $\pm$  67 & -56 &	 4.44  $\pm$   0.09  &   0.03 &    0.41 $\pm$	 0.09  &   0.17  \\
HD     9424 & 5420 $\pm$ ... &  29 & \multicolumn{1}{c}{...} &    ... &\multicolumn{1}{c}{...} &    ...  \\
HD    10576 & 5882 $\pm$ ... &  47 & \multicolumn{1}{c}{...} &    ... &\multicolumn{1}{c}{...} &    ...  \\
HD    13386 & 5294 $\pm$  95 & -24 &	 4.28  $\pm$   ...   &   0.26 &    0.26 $\pm$	 ...   &   0.10  \\
HD    15555 & 4855 $\pm$  20 &  12 & \multicolumn{1}{c}{...} &    ... &\multicolumn{1}{c}{...} &    ...  \\
HD    25061 & 5321 $\pm$ ... & -14 & \multicolumn{1}{c}{...} &    ... &\multicolumn{1}{c}{...} &    ...  \\
HD    26151 & 5353 $\pm$  22 &  30 &	 4.39  $\pm$   0.11  &   0.02 &    0.27 $\pm$	 0.01  &   0.06  \\
HD    26794 & 4930 $\pm$  29 & -10 &	 4.74  $\pm$   ...   &  -0.25 &    0.07 $\pm$	 0.04  &   0.00  \\
HD    27894 & 4914 $\pm$  54 &   6 &	 4.30  $\pm$   0.12  &   0.24 &    0.25 $\pm$	 0.07  &   0.12  \\
HD    30295 & 5417 $\pm$  74 & -11 &	 4.29  $\pm$   0.30  &   0.07 &    0.22 $\pm$	 0.09  &   0.10  \\
HD    31452 & 5262 $\pm$ ... & -12 & \multicolumn{1}{c}{...} &    ... &\multicolumn{1}{c}{...} &    ...  \\
HD    31827 & 5508 $\pm$ 150 & 100 &	 4.12  $\pm$   0.39  &   0.23 &    0.30 $\pm$	 0.15  &   0.17  \\
HD    35854 & 4928 $\pm$  32 & -27 &	 4.54  $\pm$   0.11  &   0.04 &   -0.09 $\pm$	 0.06  &   0.05  \\
HD    37986 & 5507 $\pm$  66 &  -4 &	 4.38  $\pm$   0.06  &   0.09 &    0.27 $\pm$	 0.05  &   0.03  \\
HD    39213 & 5372 $\pm$ 120 & 100 &	 4.18  $\pm$   0.25  &   0.20 &    0.28 $\pm$	 0.11  &   0.17  \\
HD    39715 & 4816 $\pm$  25 & -75 &	 4.75  $\pm$   ...   &  -0.18 &   -0.04 $\pm$	 ...   &  -0.06  \\
HD    77338 & 5290 $\pm$   0 &  56 &	 4.75  $\pm$   0.21  &  -0.20 &    0.26 $\pm$	 0.06  &   0.15  \\
HD    82943 & 6000 $\pm$  44 & -70 &	 4.43  $\pm$   0.08  &  -0.08 &    0.27 $\pm$	 0.04  &  -0.04  \\
HD    86065 & 5026 $\pm$ ... & -88 &	 4.50  $\pm$   ...   &   0.12 &   -0.06 $\pm$	 ...   &   0.15  \\
HD    86249 & 4961 $\pm$ ... &  -4 & \multicolumn{1}{c}{...} &    ... &\multicolumn{1}{c}{...} &    ...  \\
HD    87007 & 5282 $\pm$  29 &   0 &	 4.55  $\pm$   0.21  &  -0.01 &    0.27 $\pm$	 0.00  &   0.02  \\
HD    90054 & 6080 $\pm$ ... & -33 & \multicolumn{1}{c}{...} &    ... &\multicolumn{1}{c}{...} &    ...  \\
HD   104212 & 5996 $\pm$ ... &-163 & \multicolumn{1}{c}{...} &    ... &\multicolumn{1}{c}{...} &    ...  \\
HD   107509 & 6069 $\pm$ ... &  33 & \multicolumn{1}{c}{...} &    ... &\multicolumn{1}{c}{...} &    ...  \\
HD   120329 & 5636 $\pm$ ... & -19 & \multicolumn{1}{c}{...} &    ... &\multicolumn{1}{c}{...} &    ...  \\
HD   148530 & 5402 $\pm$ ... & -10 & \multicolumn{1}{c}{...} &    ... &\multicolumn{1}{c}{...} &    ...  \\
HD   149933 & 5735 $\pm$ ... &-249 & \multicolumn{1}{c}{...} &    ... &\multicolumn{1}{c}{...} &    ...  \\
HD   165920 & 5342 $\pm$   5 &  -6 &	 4.38  $\pm$   0.02  &   0.09 &    0.30 $\pm$	 0.01  &   0.06  \\
HD   171999 & 5288 $\pm$  55 &  16 &	 4.65  $\pm$   ...   &  -0.16 &    0.40 $\pm$	 ...   &  -0.11  \\
HD   180865 & 5255 $\pm$ ... & -37 & \multicolumn{1}{c}{...} &    ... &\multicolumn{1}{c}{...} &    ...  \\
HD   181234 & 5415 $\pm$ 121 &-104 &	 4.47  $\pm$   ...   &  -0.10 &    0.36 $\pm$	 ...   &   0.09  \\
HD   181433 & 4958 $\pm$   6 & -56 &	 4.37  $\pm$   ...   &   0.20 &    0.33 $\pm$	 ...   &   0.08  \\
HD   182572 & 5583 $\pm$ 172 & 117 &	 4.16  $\pm$   0.17  &   0.02 &    0.38 $\pm$	 0.09  &   0.10  \\
HD   197921 & 4948 $\pm$ ... & -82 & \multicolumn{1}{c}{...} &    ... &\multicolumn{1}{c}{...} &    ...  \\
HD   211706 & 6023 $\pm$ ... &  -6 & \multicolumn{1}{c}{...} &    ... &\multicolumn{1}{c}{...} &    ...  \\
HD   218566 & 4927 $\pm$ ... & -78 &	 4.81  $\pm$   ...   &  -0.33 &    0.38 $\pm$	 ...   &  -0.10  \\
HD   218750 & 5227 $\pm$ ... & -93 & \multicolumn{1}{c}{...} &    ... &\multicolumn{1}{c}{...} &    ...  \\
HD   224383 & 5751 $\pm$  14 &   9 &	 4.36  $\pm$   0.06  &  -0.08 &   -0.06 $\pm$	 0.02  &  -0.04  \\
\hline
\end{tabular}
\label{Tab_pastel_average}
\end{table*}

\section{Abundance determination}
\label{Sec_abonds}

\subsection{Carbon, oxygen, magnesium, and calcium}

To derive the abundance of C, O, Mg, and Ca, we performed spectral synthesis, and the abundances were obtained by minimizing the $\chi^2$ between the observed and synthetic spectra. The synthetic spectra were obtained using the PFANT code described in \citet{Cayrel.etal:1991}, \citet{Barbuy.etal:2003}, and \citet{Coelho.etal:2005} which includes molecular lines in the ABON2 code (Sect. \ref{Sec_Met}). Again, the same MARCS model atmospheres are employed. 
 
\subsubsection*{Carbon}
The carbon abundances were derived from the \ion{C}{I} $5380$~{\rm \AA} line, adopting the line list given in
 \citet{Spite.etal:1989}. To check the atomic parameters, we derived the solar carbon abundance by $\chi^2$ minimization between observed and synthetic solar spectra, obtaining (C/H)$_{\odot} = 8.53$. This value is in good  agreement with abundances from \citet{Grevesse.etal:1996}, (C/H)$_{\odot} = 8.55$, and \citet{Grevesse.Sauval:1998}, (C/H)$_{\odot} = 8.52$.

\subsubsection*{Oxygen}

The oxygen abundances were determined using the forbidden line at $6300$~\AA. The blend with nickel was taken into account using the atomic data from \citet{Bensby.etal:2004} (Table \ref{Tab_lines6300}). Since previous studies suggest that the [Ni/Fe] ratio increases at higher metallicities \citep{Bensby.etal:2005}, the contribution of the Ni blend at 6300~\AA\ may be important for our sample stars. For this reason, the abundances of Ni were previously derived,
 in order to consider the correct ratio [Ni/Fe] for each star. 

To check the atomic parameters, we derived the solar oxygen abundance, obtaining (O/H)$_{\odot} = 8.63$. This value is in good agreement with (O/H)$_{\odot} = 8.66$ from \citet{Asplund.etal:2005}.

\begin{table}
\centering
\caption{Nickel blend at 6300 \AA.}
 \begin{tabular}{cccc}
\hline\hline
Species & $\lambda$ & $\chi_{exc}$  & $\log gf$ \\
        &  (\AA)    &   (eV)        &           \\ 
\hline
\ion{Ni}{I}       & 6300.335  & 4.27  &  -2.275 \\ 
\ion{Ni}{I}       & 6300.355  & 4.27  &  -2.695 \\ 
$[$\ion{O}{I}$]$  & 6300.340  & 0.00  &  -9.820 \\
\hline
 \end{tabular}
\label{Tab_lines6300}
\end{table}

\subsubsection*{Magnesium}
\begin{figure}
\centering
 \resizebox{\hsize}{!}{\includegraphics{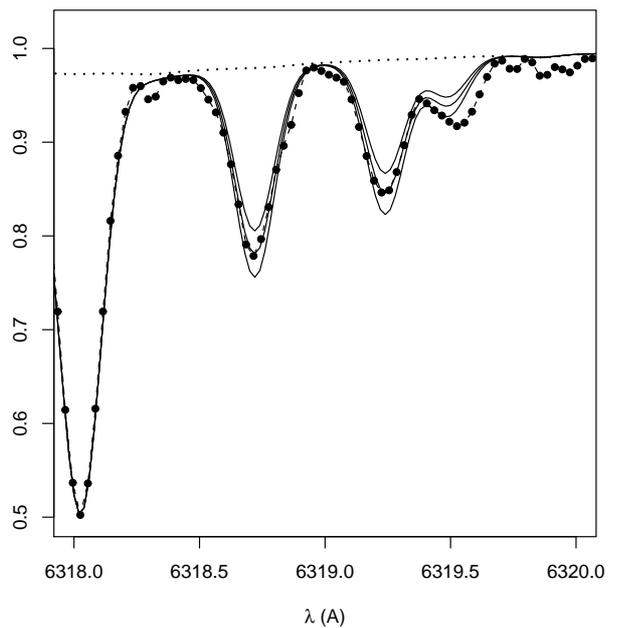}}
\caption {Solar spectra in the region of the $6319$~\AA\ Mg triplet. Solid lines indicate the synthetic spectra for (Mg/H) $ = 7.50,\ 7.60,\ 7.70$; the dots indicate the observed spectrum. The dotted line shows the contribution of the \ion{Ca}{I} autoionization line.}
\label{Fig_solar_MGlines}
\end{figure}

\begin{figure}
\centering
 \resizebox{\hsize}{!}{\includegraphics{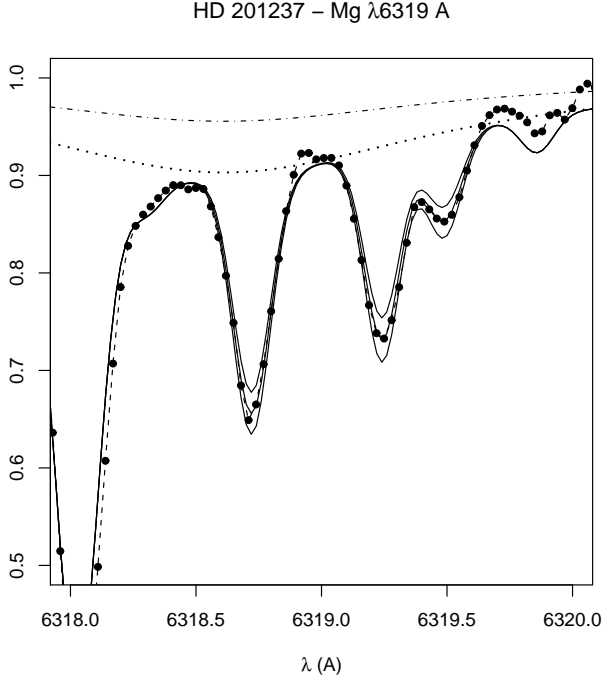}}
\caption {Spectrum of HD~201237 in the region of the $6319$~\AA\ Mg triplet. Solid lines indicate the synthetic spectra for (Mg/H) $ = 7.53,\ 7.63,{\rm\ and\ } 7.73$; the dots indicate the observed spectrum. We considered the calcium abundance of HD~201237 to calculate the synthetic spectrum, and the contribution of the \ion{Ca}{I} autoionization line is represented by the dotted curve. The dot-dashed line indicates the \ion{Ca}{I} dip when we consider [Ca/Fe] $ = 0.00$.}
\label{Fig_hd201237_MGlines}
\end{figure}

The magnesium abundances were obtained using the triplet \ion{Mg}{I} lines at $6319$~\AA. A \ion{Ca}{I} line at  $6318.3$~\AA\ showing autoionization effects, producing a $\sim$5\AA\ broad line, can affect the determination of the continuum placement \citep[e.g.][]{Lecureur.etal:2007}. The \ion{Ca}{I} autoionization line was treated by increasing its radiative broadening to reflect the much reduced lifetime of the level suffering autoionization compared with
 the radiative lifetime of this level. The radiative broadening had to be increased by 16 000 of its standard value ($\propto 1/ \lambda^2$, based on the radiative lifetimes alone) to reproduce the \ion{Ca}{I} dip in the solar spectrum (Fig. \ref{Fig_solar_MGlines}). In addition, the abundances
 of Ca of each star were derived before the calculations of the synthetic spectra at the 6319 \AA\ region,
 in order to take the correct [Ca/Fe] ratio into account in the computation of the Ca line. 
Even if the majority of the stars in the sample are not affected,
 since their abundance ratios are close to solar ([Ca/Fe]~$\sim$~$0.00$), 
for some of the sample stars
 the effect can be important. Figure \ref{Fig_hd201237_MGlines} presents the spectrum of HD~201237 at the  6319 \AA\ region. The contribution of the \ion{Ca}{I} autoionization line is shown considering both [Ca/Fe] $ = 0.00$ and [Ca/Fe] $ = 0.37$. The latter is the abundance ratio of HD~201237 before the correction of the trend with temperature (Sect. \ref{Sec_final_abo}). It is clear that the
Ca abundance of this star must be taken into account to reproduce the \ion{Ca}{I} dip. The resulting differences in the Mg abundance considering solar and non-solar [Ca/Fe] ratio are $\sim 0.07$ dex in average.

To check the atomic parameters of the lines at the \ion{Mg}{I} triplet region (Table \ref{Tab_linesMg}), we derived the solar Mg abundance. We obtained (Mg/H)$_{\odot} = 7.60$, in good agreement with (Mg/H)$_{\odot} = 7.58$ from \citet{Grevesse.Sauval:1998}. 

\begin{table*}
\centering
\caption{Lines at the $6319$~\AA\ region.}
\begin{tabular}{ccccccc}
\hline\hline  
Species &  $\lambda$  & $\chi_{ex}$  &  $\log gf$ & $\log gf$ & $\log gf$ & $\log gf$\\
        &   (\AA)     &   (eV)       &    (Sun)   & (NIST)    & (VALD)    & (BZO+09)  \\
\hline
\ion{Fe}{I} & 6318.03  & 2.45  &  -1.80 & -1.80 & -2.26 &  ...  \\
\ion{Ti}{I} & 6318.03  & 1.43  &  -0.94 &  ...  &  ...  &  ...  \\
\ion{Ca}{I} & 6318.35  & 4.43  &   0.06 &  ...  &  0.06 &  ...  \\
\ion{Mg}{I} & 6318.72  & 5.11  &  -1.98 & -2.10 & -1.73 & -2.10 \\
\ion{Mg}{I} & 6319.24  & 5.11  &  -2.23 & -2.32 & -1.95 & -2.36 \\
\ion{Mg}{I} & 6319.49  & 5.11  &  -2.80 & -2.80 & -2.43 & -2.80 \\
\hline
\label{Tab_linesMg}
\end{tabular}
\tablefoot{
BZO+09: \citet{Barbuy.etal:2009}
}
\end{table*}

\subsubsection*{Calcium}

The \ion{Ca}{I} lines were selected from \citet{Bensby.etal:2004}, \citet{Spite.etal:1987}, and \citet{Barbuy.etal:2009}, and they
are listed in Table \ref{Tab_lines_Si_Ca_Ti}. The $\log gf$ values were fitted to the solar line profiles, using (Ca/H)$_{\odot} = 6.36$ \citep{Grevesse.Sauval:1998}. The lines that give unreliable abundances were identified using the same procedure
as applied to Fe  lines, and the differences $A_{\lambda i} - \left\langle A \right\rangle_i$ for all the sample stars are plotted in Fig. \ref{Fig_CaSiTi_lines}. 

\subsection{Silicon, titanium and nickel}

Silicon, titanium, and nickel abundances were determined by recovering the measured equivalent widths through LTE analysis with the ABON2 code \citep{Spite:1967}. Again, the equivalent widths were measured with the {\tt ARES} code, and errors in $W_{\lambda}$ were estimated by carrying out the same procedure as described in Sect. \ref{Sec_Met}. Ti, Si, and Ni spectral lines were selected from \citet{Bensby.etal:2004}, \citet{Cohen.etal:2009}, and \citet{Pompeia.etal:2007}. Through a similar procedure used in the iron abundance determination, the $\log gf$ values were fitted to the solar equivalent widths, adopting (Ti/H)$_{\odot} = 5.02$, (Si/H)$_{\odot} = 7.55$ and (Ni/H)$_{\odot} = 6.25$ \citep{Grevesse.Sauval:1998}. The final line list and the astrophysical $\log gf$ values are presented in Table \ref{Tab_lines_Si_Ca_Ti}.

\begin{figure}[ht]
\centering
 \resizebox{\hsize}{!}{\includegraphics{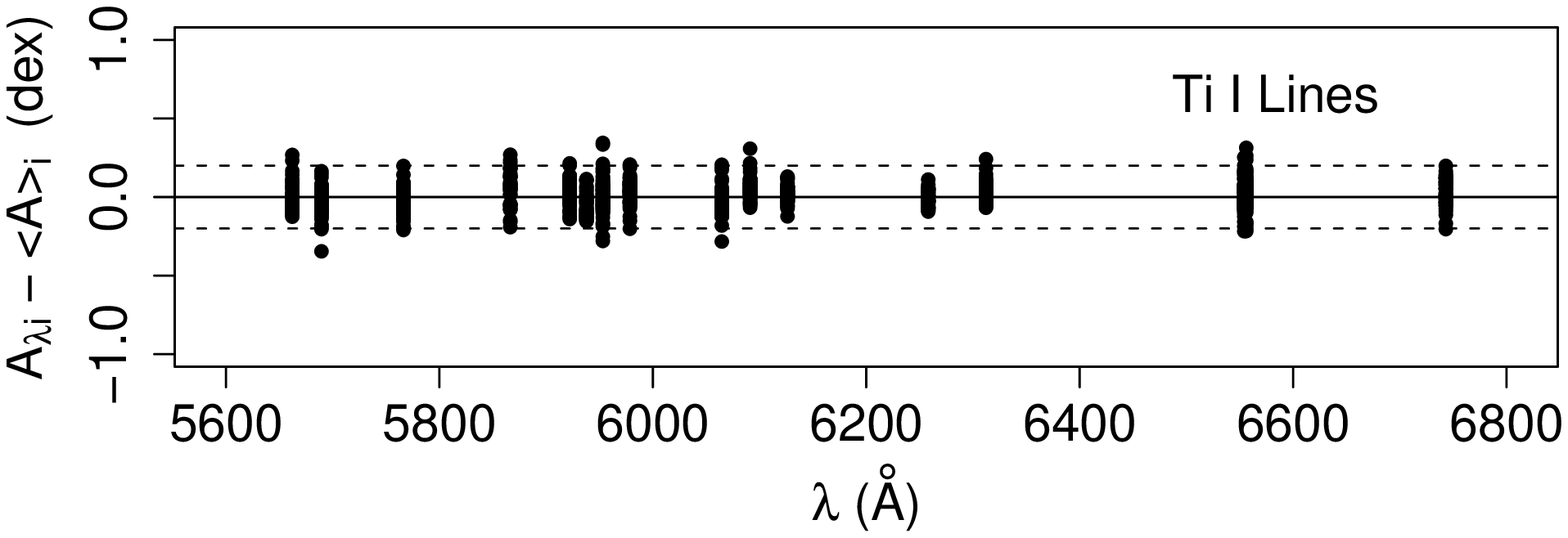}}
 \resizebox{\hsize}{!}{\includegraphics{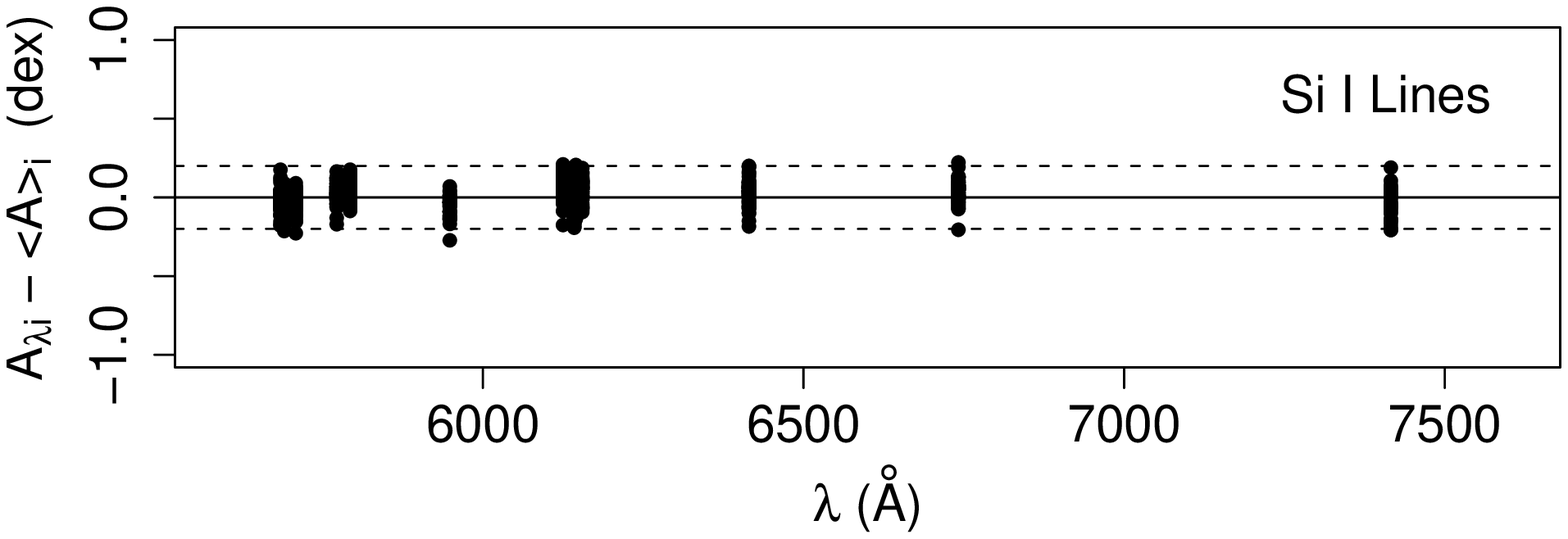}}
 \resizebox{\hsize}{!}{\includegraphics{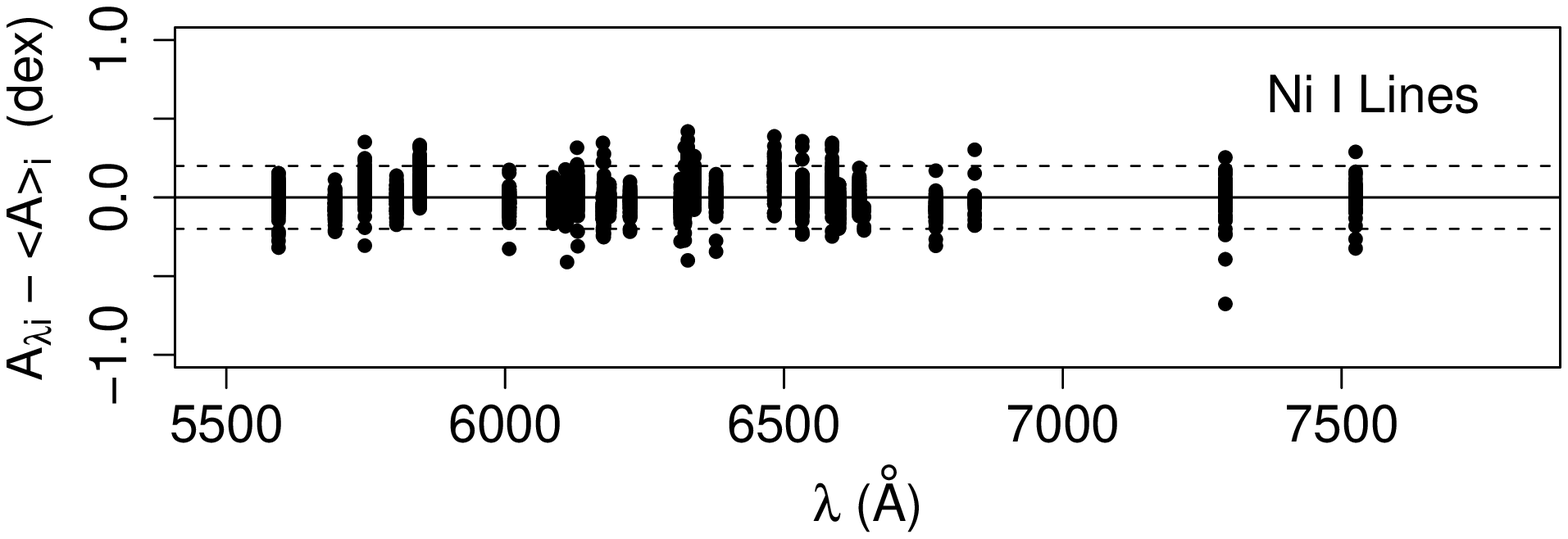}}
 \resizebox{\hsize}{!}{\includegraphics{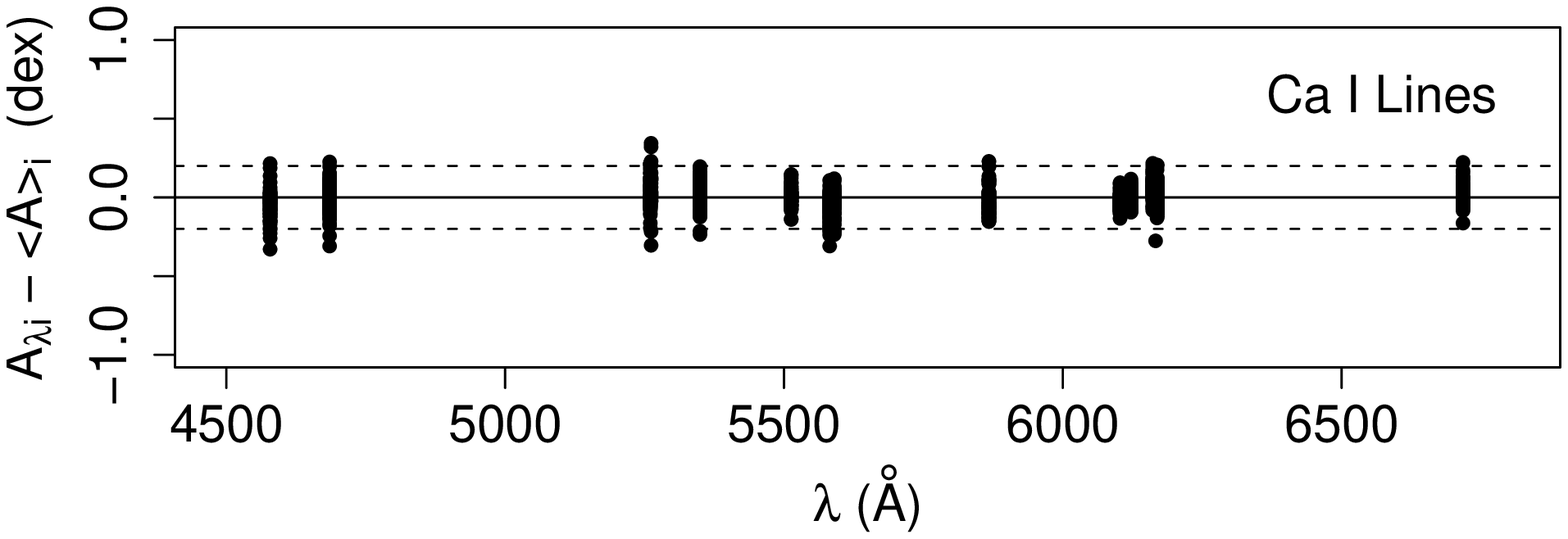}}
\caption {Process of selection of stable lines of \ion{Ti}{I}, \ion{Si}{I}, and \ion{Ca}{I}. The dashed lines indicate $\pm 0.2$ dex.}
\label{Fig_CaSiTi_lines}
\end{figure}

Lines that give unreliable abundances were identified with the same procedure as used for iron lines. We computed the differences between the abundance of each star, $\left\langle A \right\rangle_i$, and the abundance derived from an individual line, $A_{\lambda i}$. The differences $A_{\lambda i} - \left\langle A \right\rangle_i$ for all sample stars are plotted in Fig. \ref{Fig_CaSiTi_lines}.

\subsection{Spurious abundance trends and errors}
\label{Sec_final_abo}

Trends with effective temperature were found in previous studies. In an analysis of 1040 F, G, and K dwarf stars, \citet{Valenti.Fischer:2005} find that the abundances of Na, Si, Ti,  Ni, and Fe present trends with the temperature of the star. \citet{Neves.etal:2009} derived the chemical abundances of 12 elements for a sample of 451 stars of the HARPS GTO planet search programme, and similar trends with temperatures were found. 

\begin{figure*}
\centering
\begin{tabular}{ccc}
\vspace{-0.8cm}
 \resizebox{0.32\hsize}{!}{\includegraphics{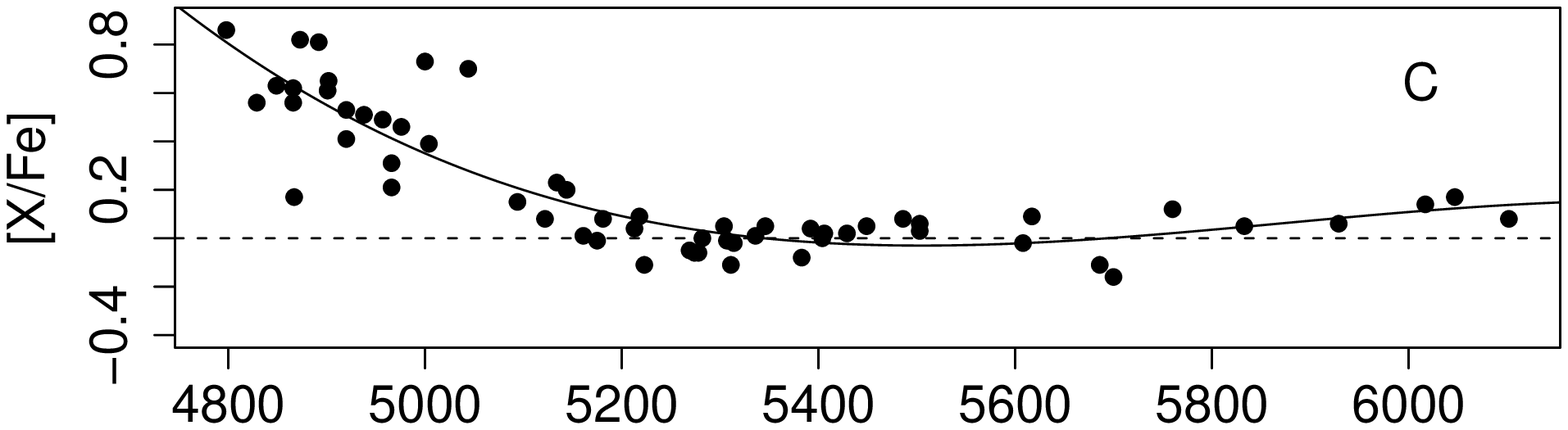}} & & \\
\vspace{-0.8cm}
 \resizebox{0.32\hsize}{!}{\includegraphics{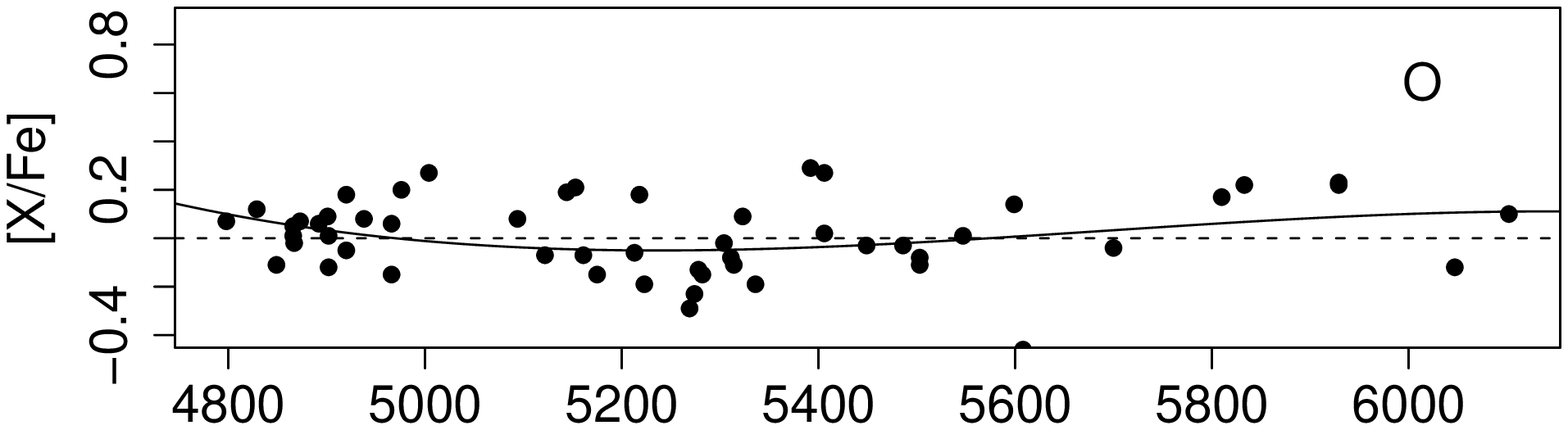}} & 
 \resizebox{0.32\hsize}{!}{\includegraphics{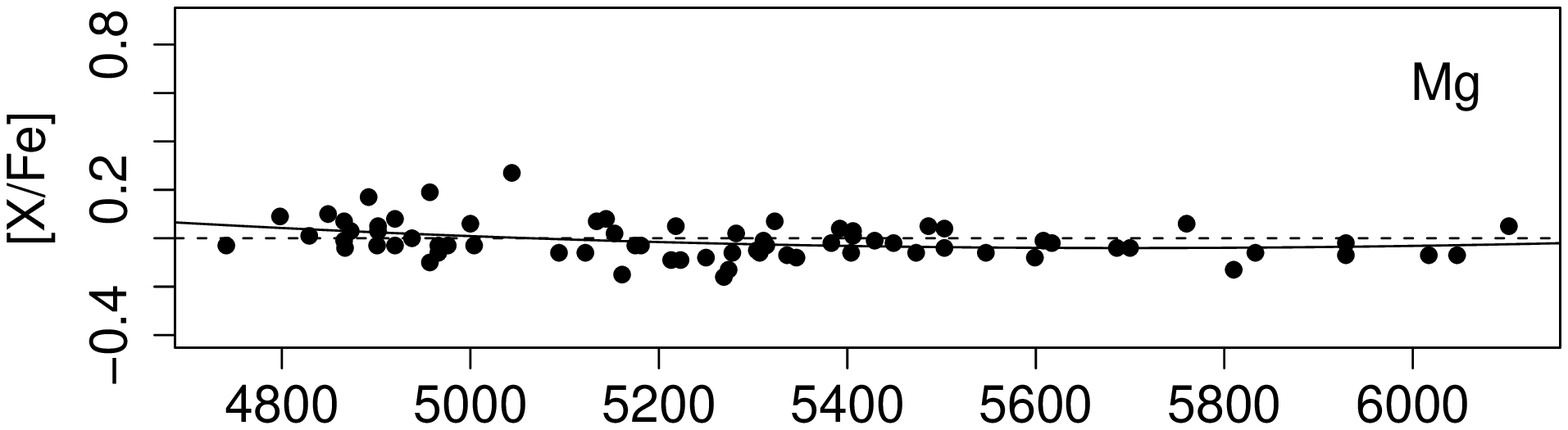}} &
 \resizebox{0.32\hsize}{!}{\includegraphics{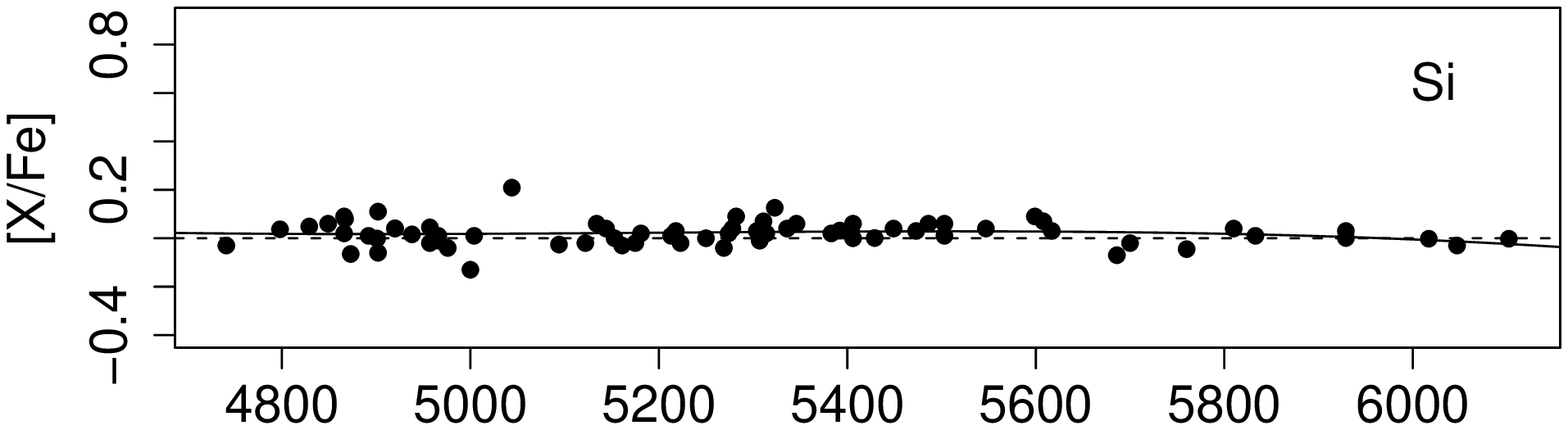}} \\
 \resizebox{0.32\hsize}{!}{\includegraphics{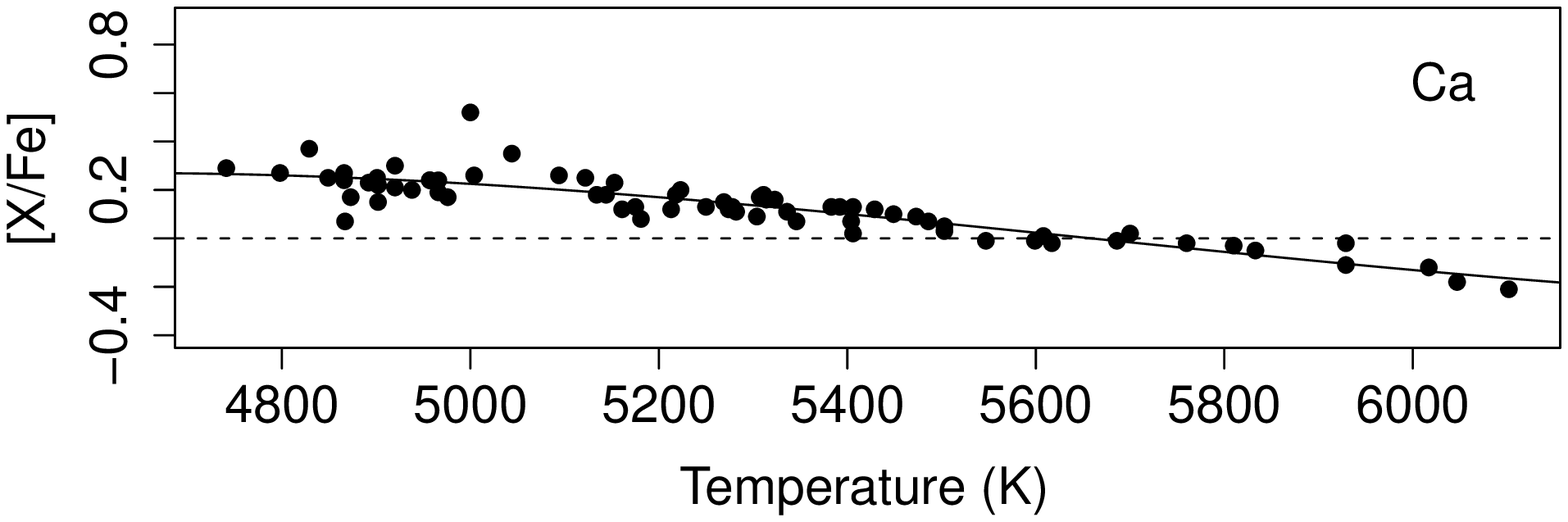}} &
 \resizebox{0.32\hsize}{!}{\includegraphics{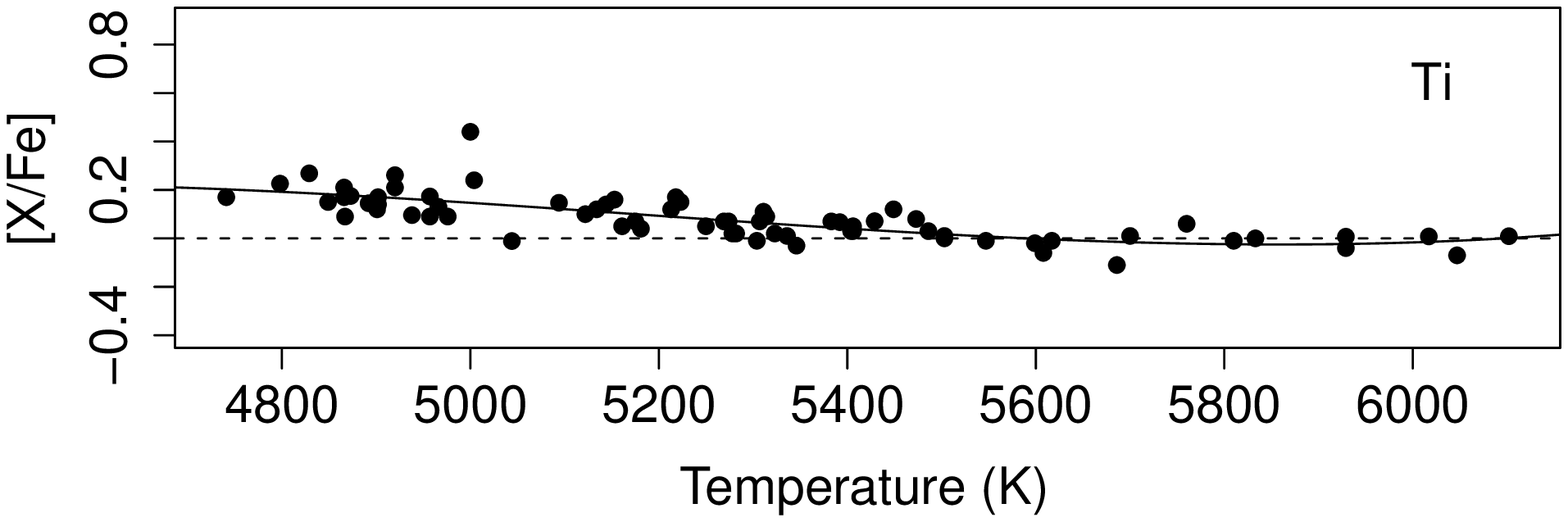}} &
 \resizebox{0.32\hsize}{!}{\includegraphics{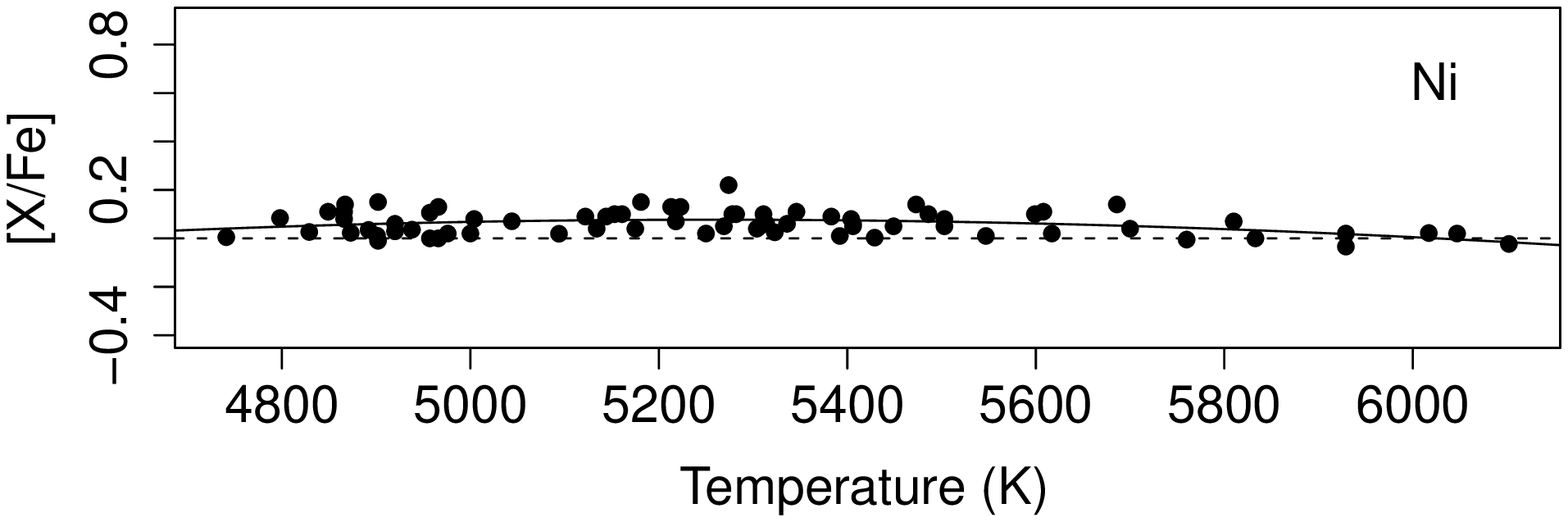}} \\

 \end{tabular}
\caption {Abundances of C, O, Mg, Si, Ca, Ti, and Ni {\it vs.} temperature. A pronounced trend is observed for C, Ca and Ti.}
\label{Fig_teff_abonds}
\end{figure*}

\begin{figure*}
\centering
\begin{tabular}{ccc}
 \resizebox{0.32\hsize}{!}{\includegraphics{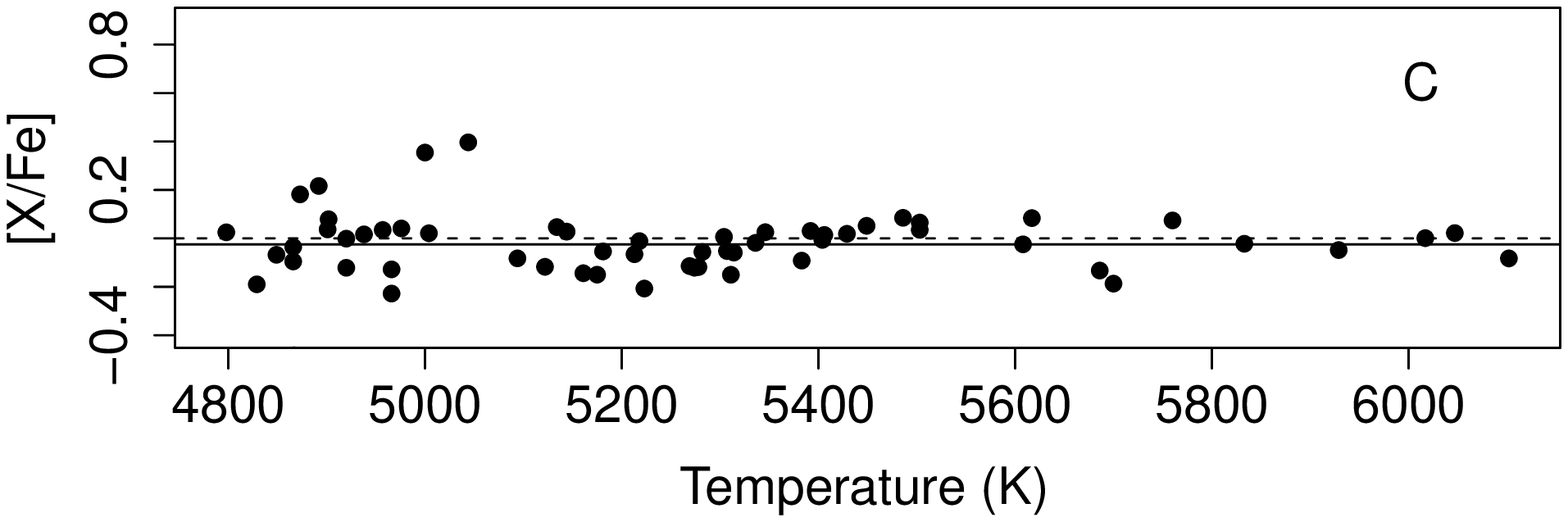}} & 
 \resizebox{0.32\hsize}{!}{\includegraphics{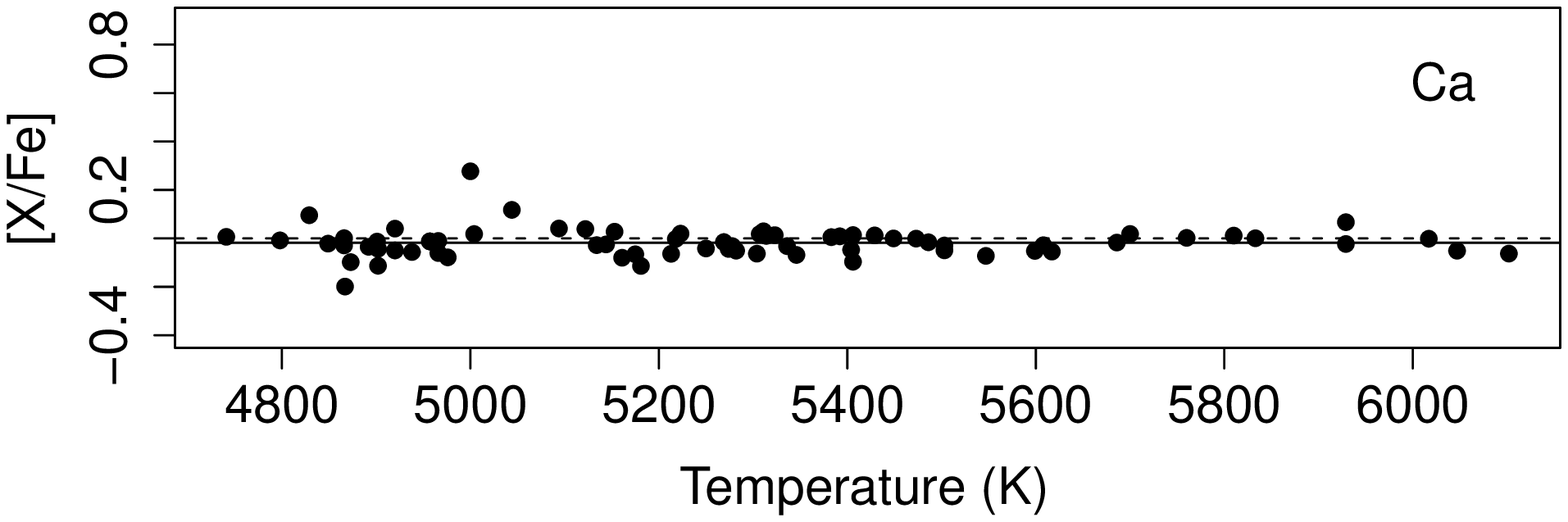}} &
 \resizebox{0.32\hsize}{!}{\includegraphics{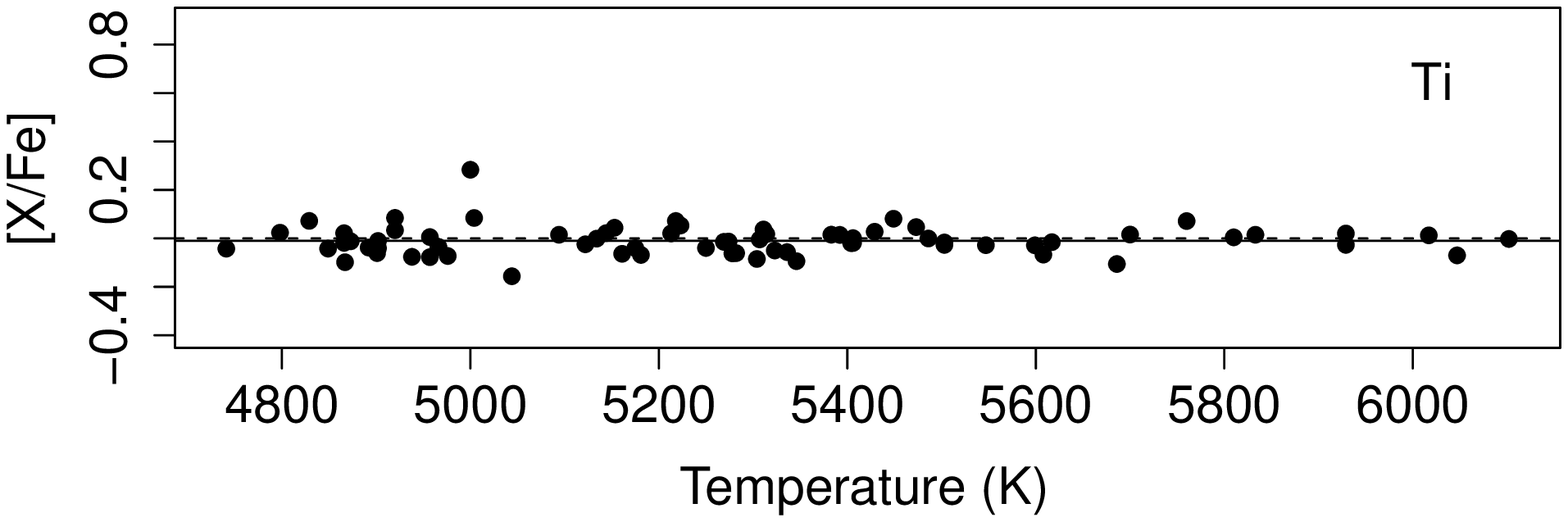}} \\
 \end{tabular}
\caption{Abundances of C, Ca, and Ti {\it vs.} temperature after the correction. We considered the mean abundance in the range  $5680 < T_{\rm eff} < 5880$ K as the zero point of the correction, which is $-0.03$, $-0.02$, and $-0.01$ for C, Ca, and Ti, respectively.}
\label{Fig_teff_abonds_corr}
\end{figure*}

To check such trends in our results, we plotted our final abundances against temperatures. Figure \ref{Fig_teff_abonds} shows the abundances {\it vs.} $T_{\rm eff}$, and a significant trend is observed for C, Ca, and Ti abundances. Following a
 procedure similar to the one adopted by  \citet{Valenti.Fischer:2005}
and \citet{Melendez.Cohen:2009}, 
 we corrected this trend by fitting a second-order polynomial and applied the correction for C, Ca, and Ti (Fig. \ref{Fig_teff_abonds_corr}). We assumed that abundance trend corrections are zero at $T_{\rm eff} = 5777$~K. This assumption eliminates
 the possibility that the Sun itself may have peculiar abundances  \citep[e.g.][]{AllendePrieto.etal:2004}.

The errors on abundances were estimated as follows. The errors on abundances of C, O, and Mg were derived by taking the uncertainties on the stellar parameters into account. Temperatures, gravities, and metallicities were varied individually according to their errors, and the resulting variations on the abundance were added quadratically. This procedure were performed for ten stars in the sample, and the mean error was assumed to be the characteristic error for the sample stars ($0.12$, $0.17$, and $0.08$ dex for C, O, and Mg, respectively). More than one spectral line was used to derive the abundances of  Ni, Si, Ca, and Ti, and for these elements, the errors are assumed to be the standard deviation of the abundances derived from individual lines. The errors on these four abundances are presented in Tables \ref{Tab_abonds1} and \ref{Tab_abonds2}.

%

%

\subsection{Final abundances and comparison with other studies}
\label{Sec_Final_abo}

\begin{figure}[ht]
\centering
 \resizebox{\hsize}{!}{\includegraphics{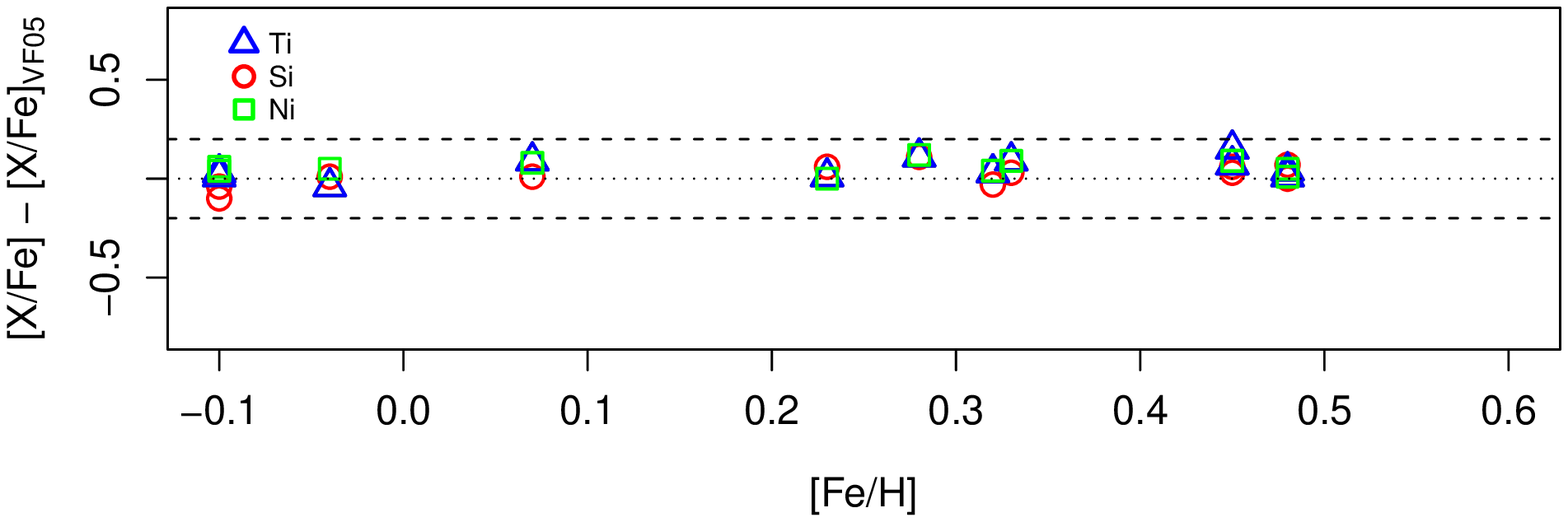}}
 \resizebox{\hsize}{!}{\includegraphics{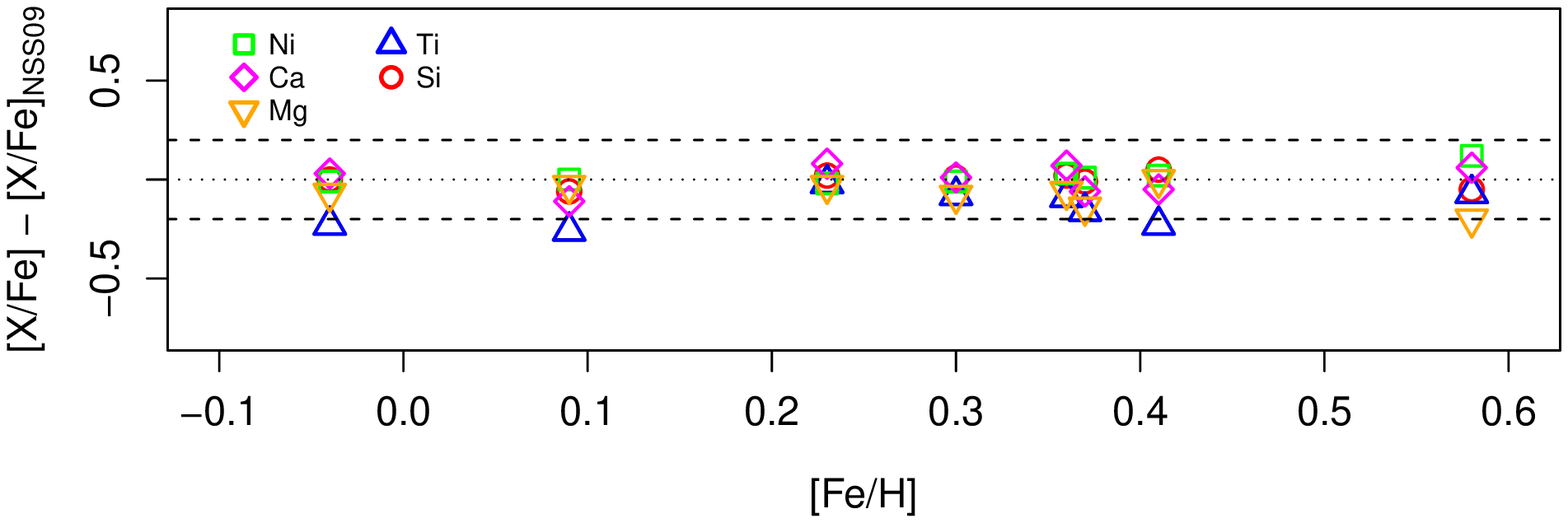}}
\caption {Comparison of abundances with other studies. The upper panel shows differences between our abundances of Si, Ti, and Ni and those from \citet{Valenti.Fischer:2005}. The bottom panel presents the comparisons of Si, Ti, Ca, Mg, and Ni abundances from \citet{Neves.etal:2009}.}
\label{Fig_abonds_deltas}
\end{figure}

To test the reliability of our results, we compared the abundances derived in the present work and derivations from previous studies of \citet{Valenti.Fischer:2005} and \citet{Neves.etal:2009}. Our sample contains 12 stars in common with the sample studied by Valenti \& Fischer and eight in common with Neves et al. In Table \ref{Tab_abonds_deltas}, we give the mean difference between the abundance ratios [X/Fe], and in Fig. \ref{Fig_abonds_deltas}, these differences are shown as a function of [Fe/H]. Our results are in good agreement with the abundances derived by these authors. Differences are all within $\sim 0.2$~dex, and there is no evident trend with metallicity. 

\begin{figure*}
\centering
\begin{tabular}{cc}
\vspace{-1cm}
 \resizebox{0.48\hsize}{!}{\includegraphics{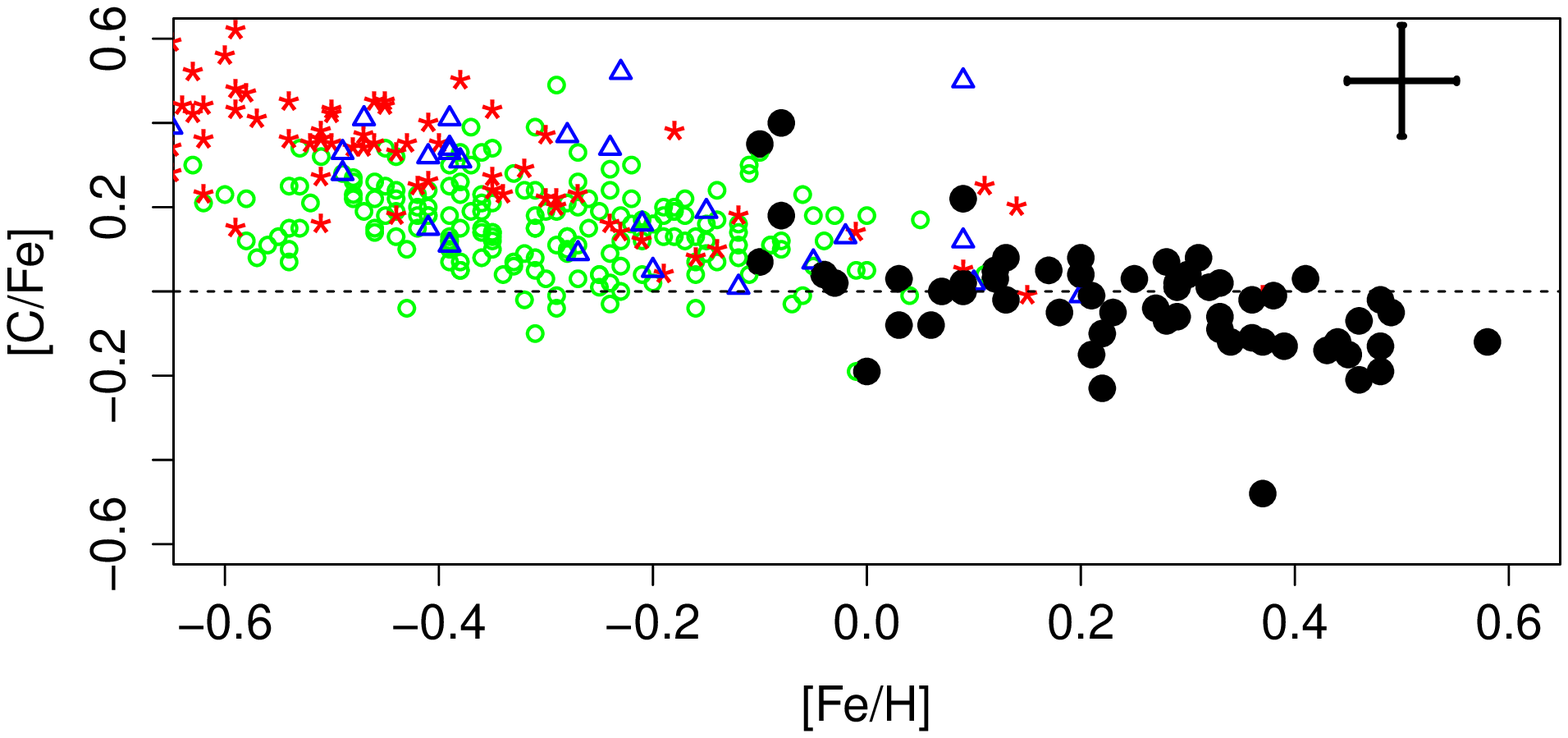}} & \\
\vspace{-1cm}
 \resizebox{0.48\hsize}{!}{\includegraphics{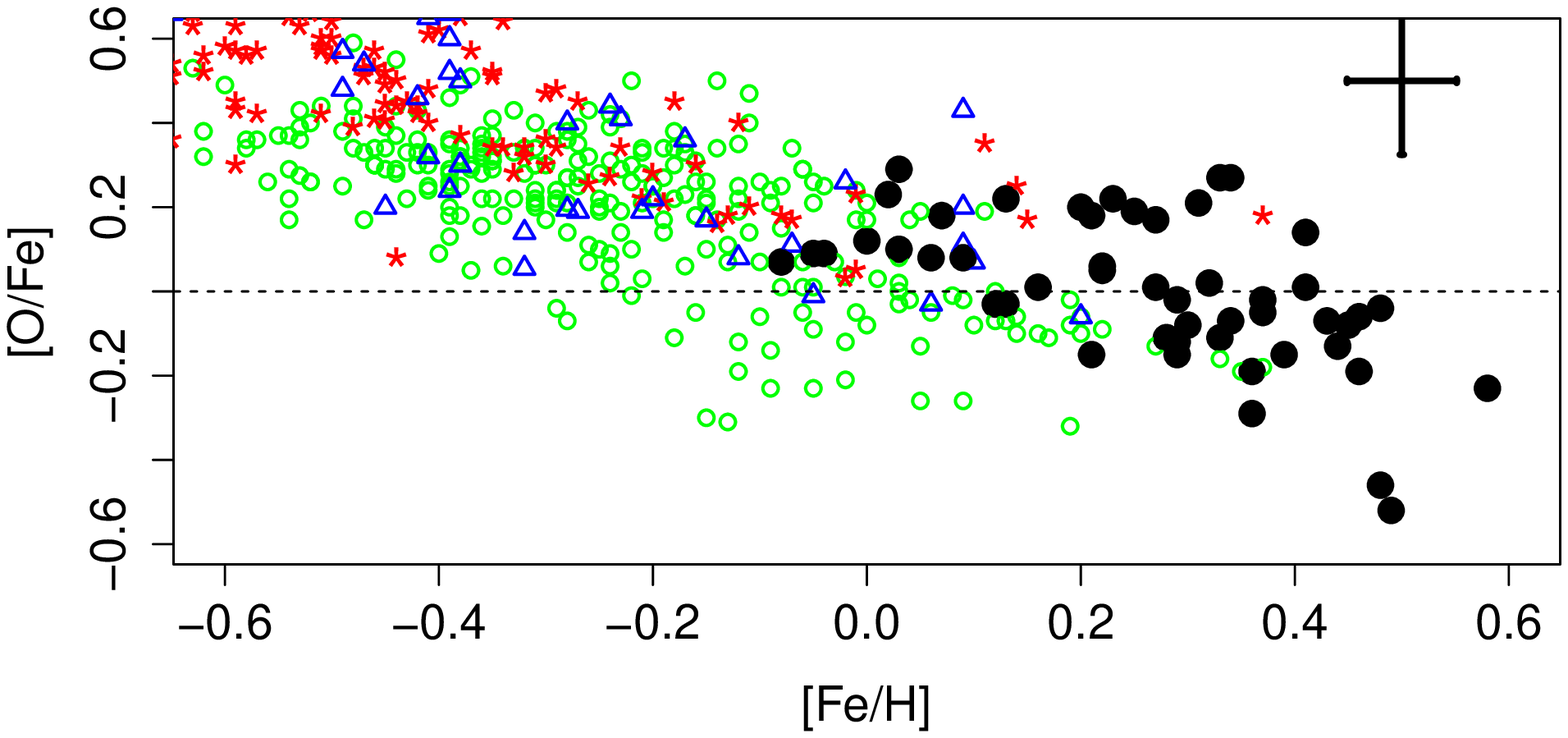}} & 
 \resizebox{0.48\hsize}{!}{\includegraphics{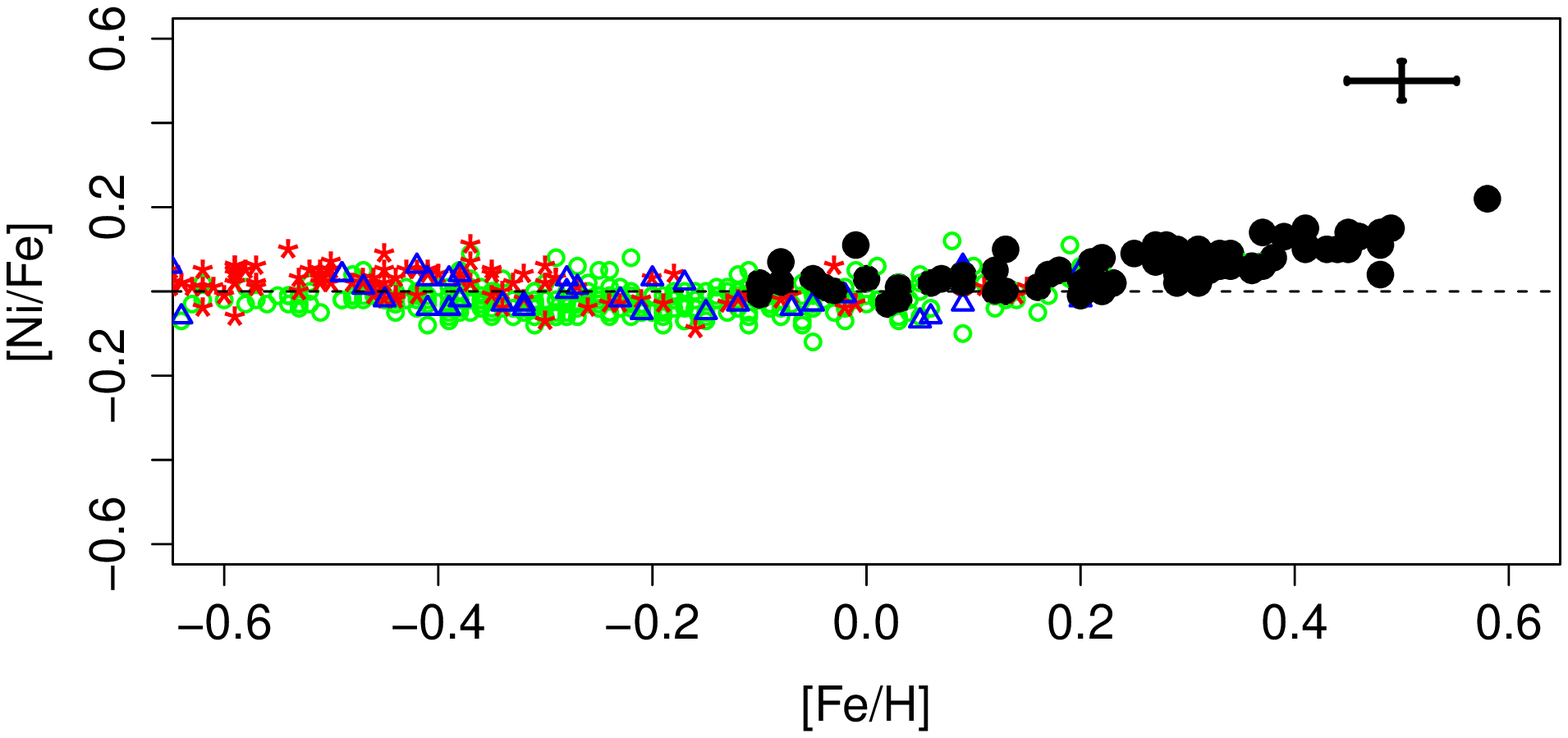}} \\
\vspace{-1cm}
 \resizebox{0.48\hsize}{!}{\includegraphics{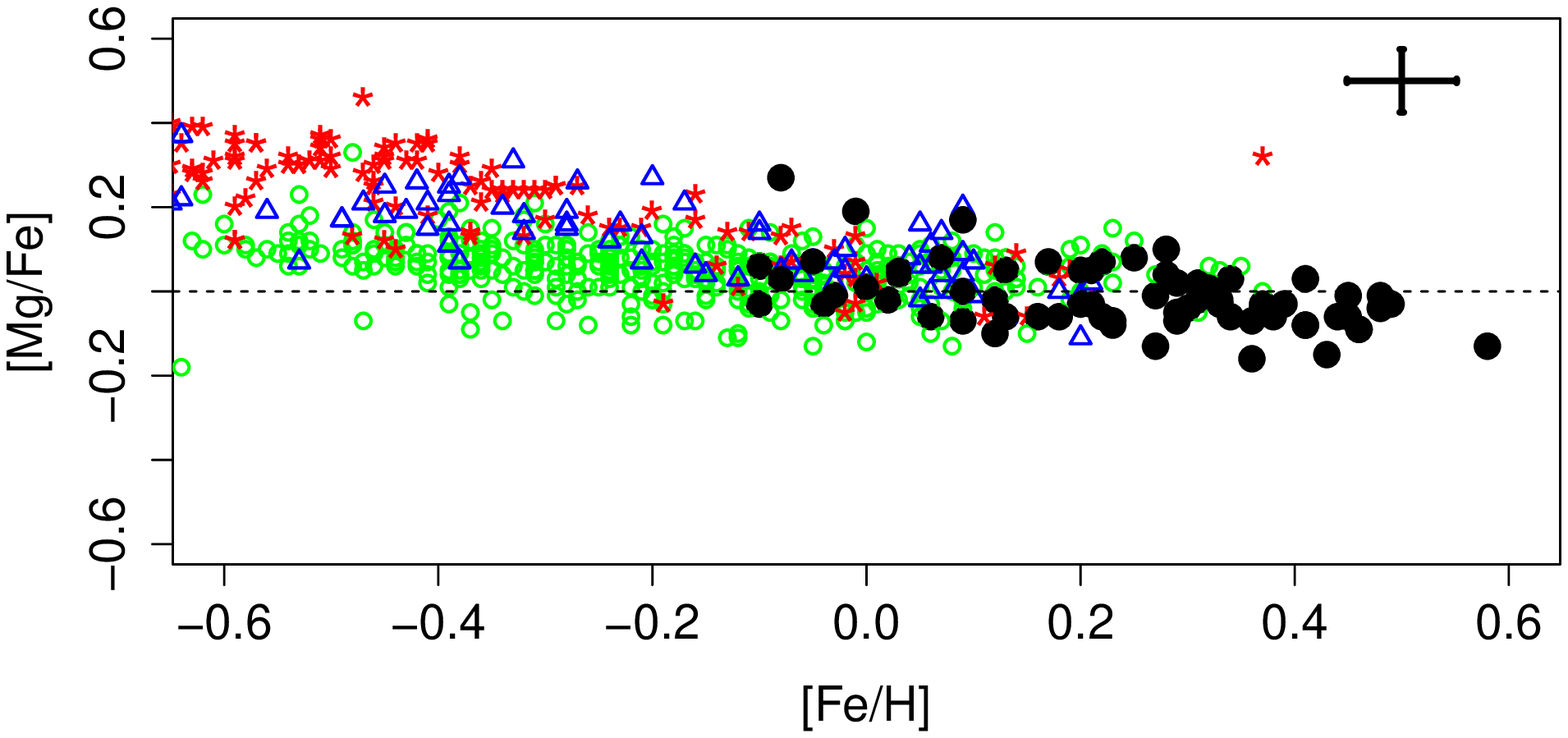}} &
 \resizebox{0.48\hsize}{!}{\includegraphics{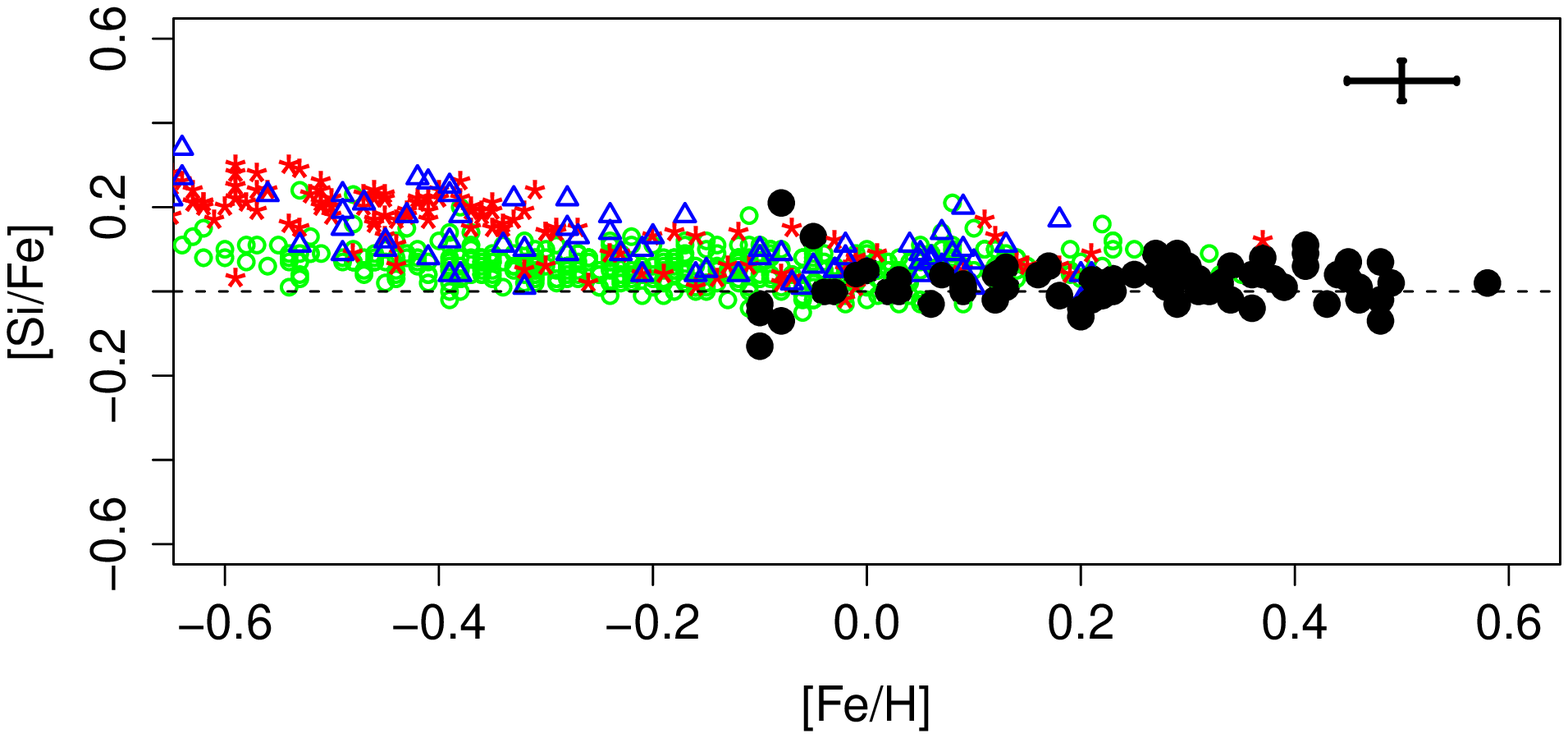}} \\
 \resizebox{0.48\hsize}{!}{\includegraphics{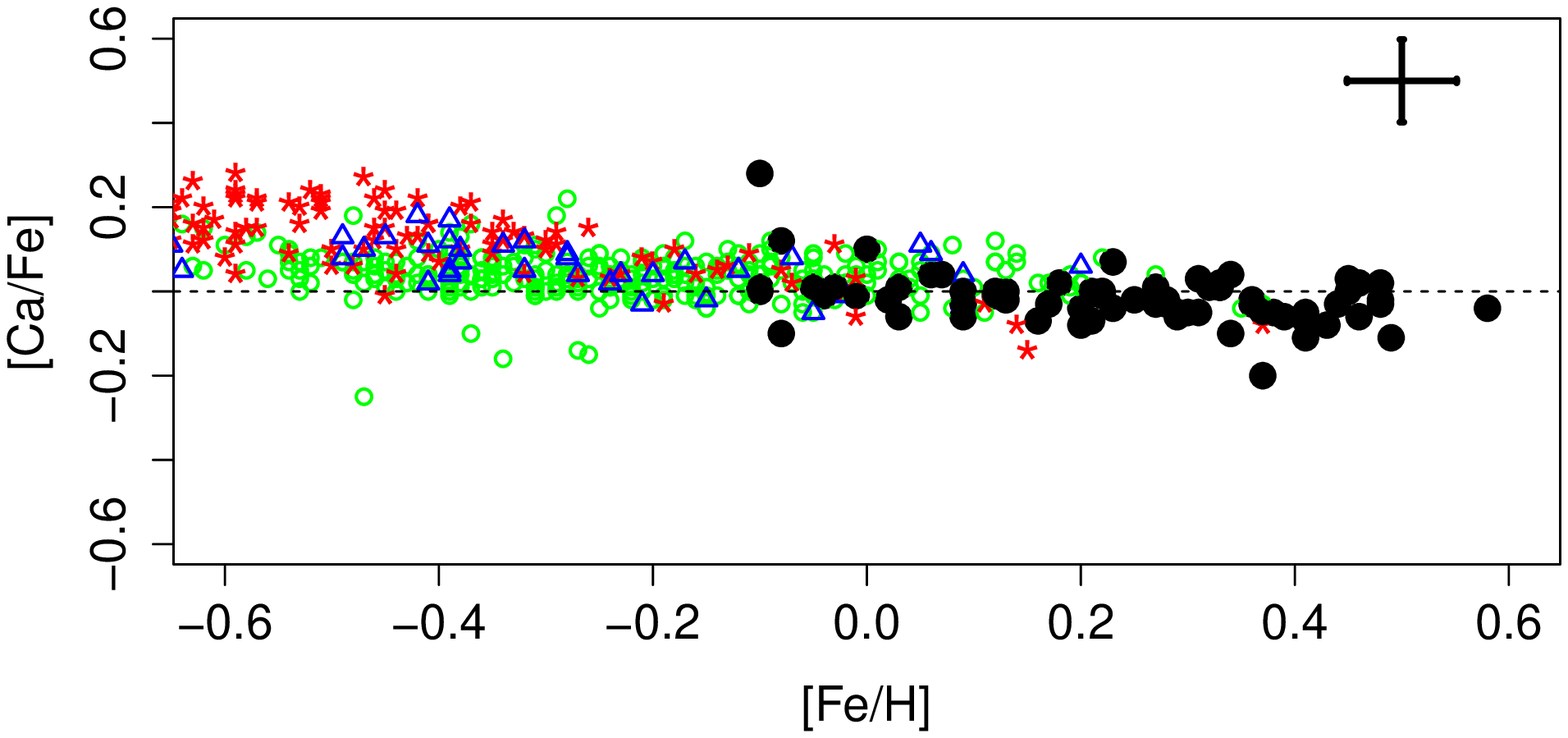}} &
 \resizebox{0.48\hsize}{!}{\includegraphics{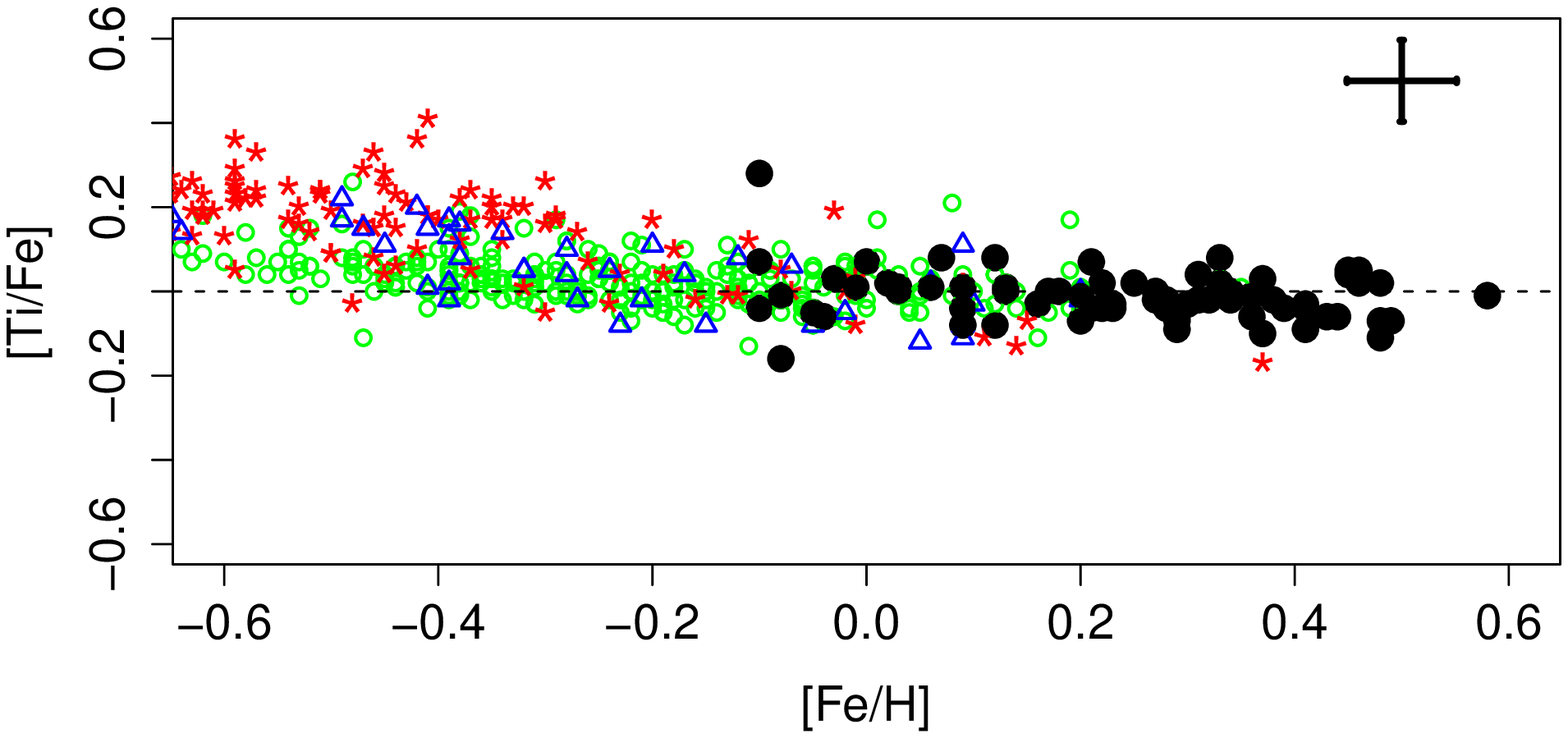}} \\
 \end{tabular}
\caption {Abundances of C, O, Ni, Mg, Ca, Si,, and Ti {\it vs.} metallicities. The abundances derived in this work (black circles) are compared with the thin (green circles), thick (red stars) disk stars, and the intermediate population (blue triangles) from \citet{Bensby.etal:2003, Bensby.etal:2004},  \citet{Mishenina.etal:2004, Mishenina.etal:2008}, and \citet{Reddy.etal:2003, Reddy.etal:2006}.}
\label{Fig_abonds}
\end{figure*}

The final abundances relative to iron, [X/Fe], {\it vs.} [Fe/H] are presented in Fig. \ref{Fig_abonds}. We plotted the abundances from \citet{Bensby.etal:2003}, \citet{Bensby.etal:2004} and \citet{Mishenina.etal:2004, Mishenina.etal:2008}, and \citet{Reddy.etal:2003, Reddy.etal:2006} for
a comparison.  We applied the same procedure as described in Sect. \ref{Sec_disk_member} to separate the samples of these
 studies into thin disk, thick disk, and intermediate populations\footnote{To perform a consistent comparison, we re-classified the samples from these authors, using the same procedure as for our sample. Hereafter, the thin disk, the thick disk, and intermediate populations from \citet{Bensby.etal:2003, Bensby.etal:2004}, \citet{Mishenina.etal:2004, Mishenina.etal:2008}, and \citet{Reddy.etal:2003, Reddy.etal:2006} refer to the classification described in Sect.~\ref{Sec_disk_member}.}.  We stress that the comparison must be considered carefully: different approaches and methods can lead to systematic differences between abundances from different authors. These systematic differences could be determined by direct comparison of abundances for stars in common with other samples. However, our sample contains only two stars in common with the sample of  \citet{Bensby.etal:2004} and one in common with \citet{Mishenina.etal:2004, Mishenina.etal:2008}, so that the systematic differences could not be quantified accurately. By comparing the abundances of these stars
(even if they are so few), we found that our results are 0.04~dex lower than [$\alpha$/Fe] ratios from these studies.
 Differences at this level do not affect the main conclusions of this work.

We found the following trends of [X/Fe] {\it vs.} [Fe/H] for each element:

\begin{description}
 \item {\it Carbon:} [C/Fe] is a decreasing function of metallicity. The median value of [C/Fe] at lower metallicities ([Fe/H] $< 0.2$) is $0.03$, while we found [C/Fe] $-0.07$ for the more metal-rich stars ([Fe/H] $> 0.2$). 

 \item {\it Oxygen: } [O/Fe] decreases with increasing metallicity. Eight stars
with metallicity 0.2 $<$ [Fe/H] $<$ 0.4 have [O/Fe] $\approx$ 0.2. Among these, five are assigned to the thick disk,
two to the intermediate population, and one to the thin disk. This overabundance is within the errors; therefore 
higher S/N spectra would be interesting to verify their oxygen abundances with higher precision, since these could be a distinct category of
stars, but  there is no  evidence for such a conclusion with the present data.

 \item {\it Nickel: } The nickel-to-iron ratio is constant up to [Fe/H] $\sim 0.2$, and increases at higher metallicities. This trend has already been suggested by data from \citet{Bensby.etal:2004}, and is confirmed here.

 \item {\it Magnesium, silicon, calcium, and titanium: } The abundance of these elements present a low scatter and follows the general trend of thin disk stars.
\end{description}

\begin{table*}
\centering
\caption{Comparison of abundances with other studies}
 \begin{tabular}{cccccccc}
\hline\hline
 Reference                   & $\Delta$[Fe/H]  & $\Delta$[Ca/Fe]  & $\Delta$[Mg/Fe]  & $\Delta$[Si/Fe]  & $\Delta$[Ti/Fe]  & $\Delta$[Ni/Fe] & \# \\
\hline
\citet{Valenti.Fischer:2005} & $0.01 \pm 0.07$ &    ...          &      ...        & $0.02 \pm 0.06$ & $0.05 \pm 0.05$ & $0.06 \pm 0.04$ & 12  \\
\citet{Neves.etal:2009}      & $0.10 \pm 0.08$ & $0.12 \pm 0.11$ & $0.04 \pm 0.07$ & $0.00 \pm 0.04$ &$-0.14 \pm 0.09$ & $0.02 \pm 0.04$ & 8 \\
\hline
 \end{tabular}
\tablefoot{
$\Delta$[X/Fe] $= \langle {\rm [X/Fe]}_{\rm Our} - {\rm [X/Fe]}_{\rm Other} \rangle$
}
\label{Tab_abonds_deltas}
\end{table*}
\normalsize

%

\section{Discussion}
\label{Sec_discussion}

 In the present work we concentrate efforts
on studying  metal-rich stars with [Fe/H] $>0.0$, from a sample of high proper-motion, NLTT-selected stars
as described in Sect. \ref{Sec_Sample}. 
In \citet{Pompeia.etal:2003}, we studied the behaviour of [$\alpha$/Fe] as a function of metallicity in the range $-0.8 <$ [Fe/H] $<$ +0.4,
for stars with similar kinematics to the present sample. It was found that the enhancement of 
$\alpha$-elements relative to Fe drops with increasing metallicity, reaching solar ratios
at around [Fe/H] $\approx -0.4$ for Si, Ca, and Ti, and at  [Fe/H] $\approx -0.2$ for Mg and O.
This behaviour is compatible with the thick disk characteristics. \citet{Bensby.etal:2003} shows
a drop in [$\alpha$/Fe] at [Fe/H]~$\approx -0.4$, reaching the solar ratio at [Fe/H]~$\approx 0.0$.
It was shown in Sect. \ref{Sec_abonds} that  $\alpha$-element abundances are not enhanced
in the metal-rich sample stars, a result compatible with the behaviour previously shown
by \citet{Pompeia.etal:2003}.

In terms of kinematical properties,
we analysed the U, V, W velocities of the sample stars to identify members of the
thick disk, thin disk and intermediate ones, according to definitions by \citet{Soubiran.etal:2003}.
The membership with thin or thick disk components discussed in Sect. \ref{Sec_disk_member} leads to
 42 ($59\%$) of the sample stars to be identified with the thick disk.

\subsection{Comparison with thin and thick disk stars}

We compared the characteristics of the sample stars with thin disk, thick disk, and intermediate populations from
\citet{Bensby.etal:2003, Bensby.etal:2004}, \citet{Mishenina.etal:2004, Mishenina.etal:2008}, and \citet{Reddy.etal:2006}.  

\begin{figure*}
\centering
\begin{tabular}{cc} 
 \resizebox{0.49\hsize}{!}{\includegraphics{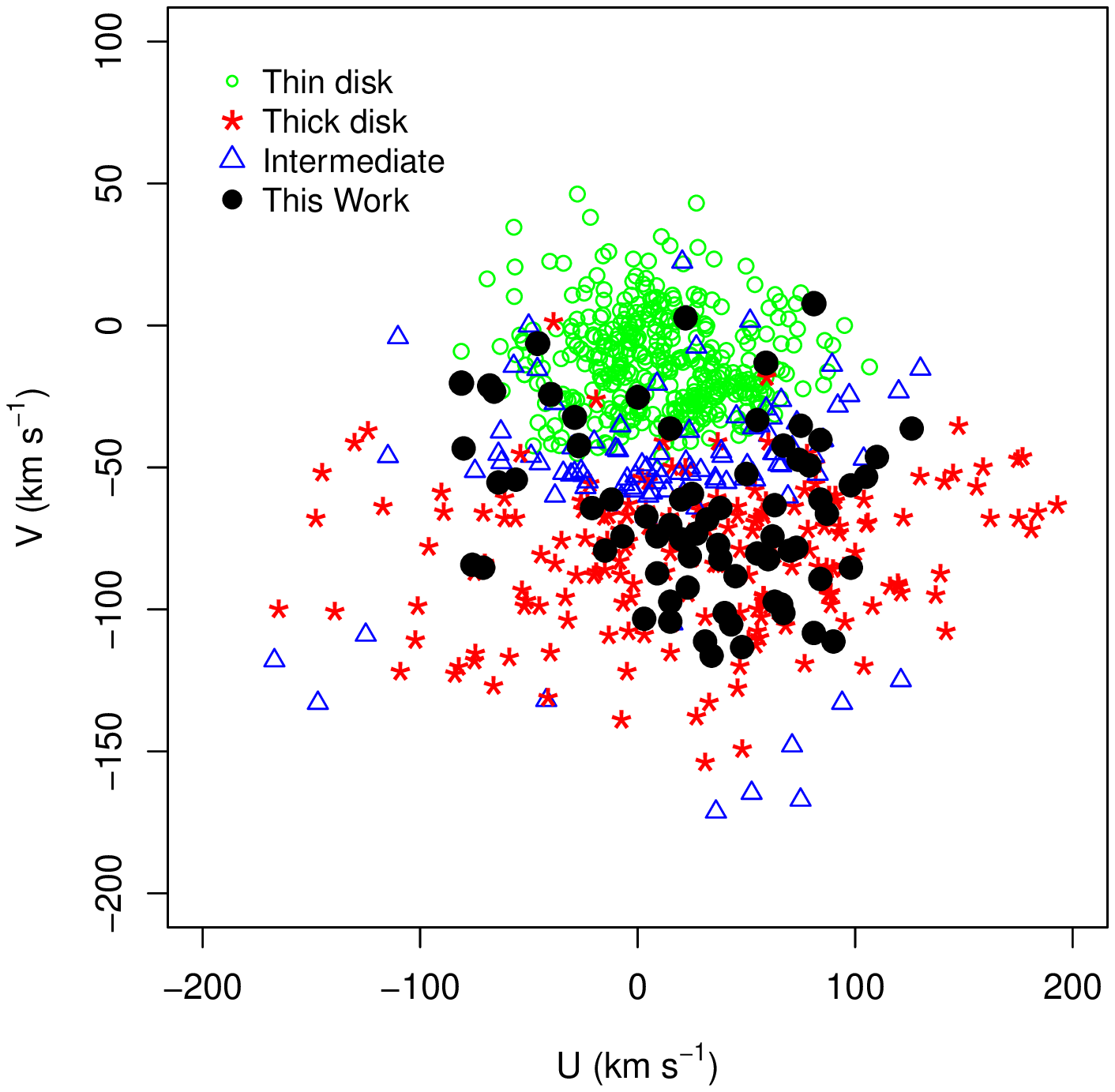}} &
 \resizebox{0.49\hsize}{!}{\includegraphics{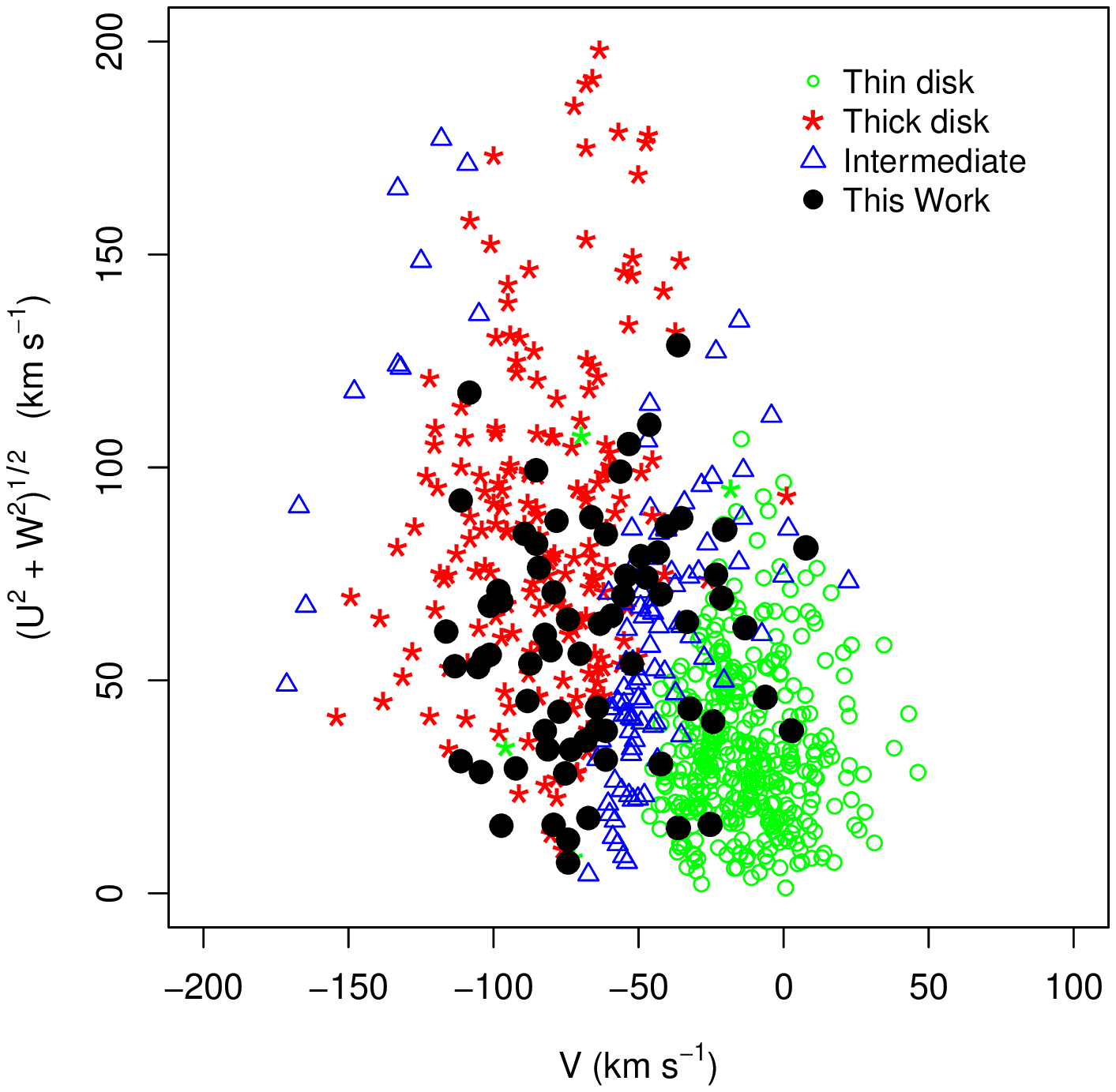}} 
 \end{tabular}
\caption {{\it Left}: UV plane. {\it Right}: Toomre diagram. In both panels, the present sample is indicated by black circles. The blue open circles and red stars are the thin and thick disks, respectively. The stars shown are the samples from \citet{Reddy.etal:2003, Reddy.etal:2006}, \citet{Mishenina.etal:2004, Mishenina.etal:2008}, and \citet{Bensby.etal:2004}. The velocities and population fractions used (Eqs. \ref{Eq_membership1} and \ref{Eq_membership2}) were taken from \citet{Soubiran.etal:2003}.}
\label{Fig_uvplane2}
\end{figure*}

The sample is dominated by stars having the metallicities indicative of thin disk population, as seen by the behaviour of
 [$\alpha$-elements/Fe] vs. [Fe/H] presented in Fig. \ref{Fig_abonds}. On the other hand, the kinematics of the sample stars would 
suggest membership with the thick disk, as shown in the UV plane and Toomre diagram (Fig. \ref{Fig_uvplane2}).
 If confirmed as members of the thick disk, these metal-rich stars
provide an interesting sample for testing models of thick disk formation. They show  the kinematics of a thick
disk, together with the metallicities and abundance ratios of thin disk stars. 

\begin{figure}
\centering
\begin{tabular}{c} 
 \resizebox{\hsize}{!}{\includegraphics{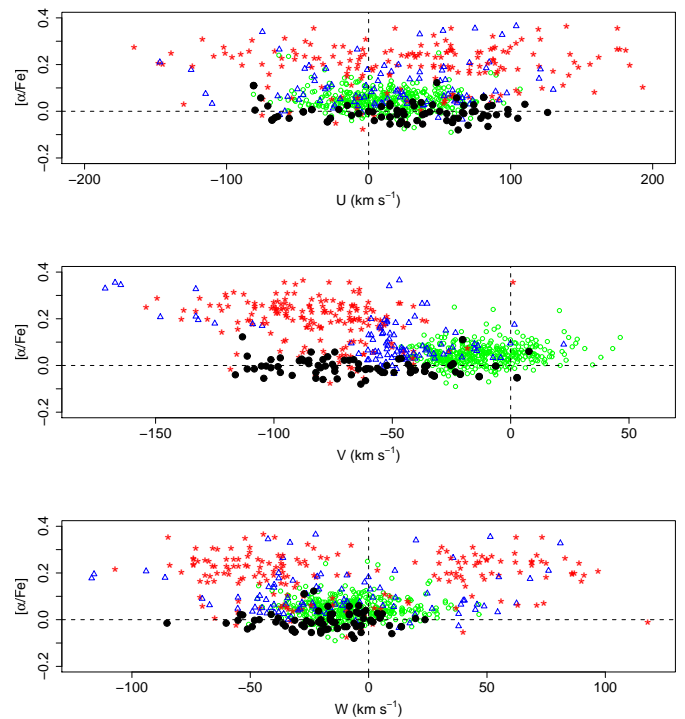}} 
 \end{tabular}
\caption {UVW {\it vs.} [$\alpha$/Fe]. The black circles show the present sample. Data from \citet{Reddy.etal:2003, Reddy.etal:2006, Mishenina.etal:2004, Mishenina.etal:2008, Bensby.etal:2004} are shown as red stars (thick disk), open green circles (thin disk), and blue triangles (intermediate population). }
\label{Fig_uvw_alpha}
\end{figure}

In Fig. \ref{Fig_uvw_alpha}, we show the space velocities U, V, and W against [$\alpha$/Fe]. 
The sample stars show a lower
rotational velocity V than the thin disk stars. The $\vert$W$\vert$ velocity is somewhat higher
 than the thin disk, showing essentially only negative values, which may be further
 investigated in terms of migration effects in the Galaxy.

Grenon (1987) proposed that the average radius of 
the orbit, $R_{\rm m} = (R_{\rm max} + R_{\rm min}) / 2$, is kept close to the initial 
galactocentric radius of the stellar birthplace. Therefore, we are able to use $R_{\rm m}$ to derive radial constraints,
 such as the abundance gradients for each population. 
Figure \ref{Fig_rm} shows how the metallicity and [$\alpha$/Fe] vary with respect to $R_{\rm m}$
 for thin, thick, and intermediate stars. The thin and thick disk stars appear to have $R_{\rm m} \sim 8$ 
and $6$~kpc, whereas the intermediate population has $R_{\rm m} \sim 7$~kpc. These distances agree
 with $R_{\rm m}$ distances of our thin disk, thick disk, and intermediate subsamples.
 Therefore, we investigate whether $R_{\rm m}$ for each component remains the same when considering 
the more complete sample from GCS. This is shown in Fig. \ref{Fig_rm_gcs}, and it is clear that 
the GCS data also show that thin, thick, and intermediate stars have these typical $R_{\rm m}$ values.

\begin{figure}
\centering
\begin{tabular}{c} 
 \resizebox{\hsize}{!}{\includegraphics{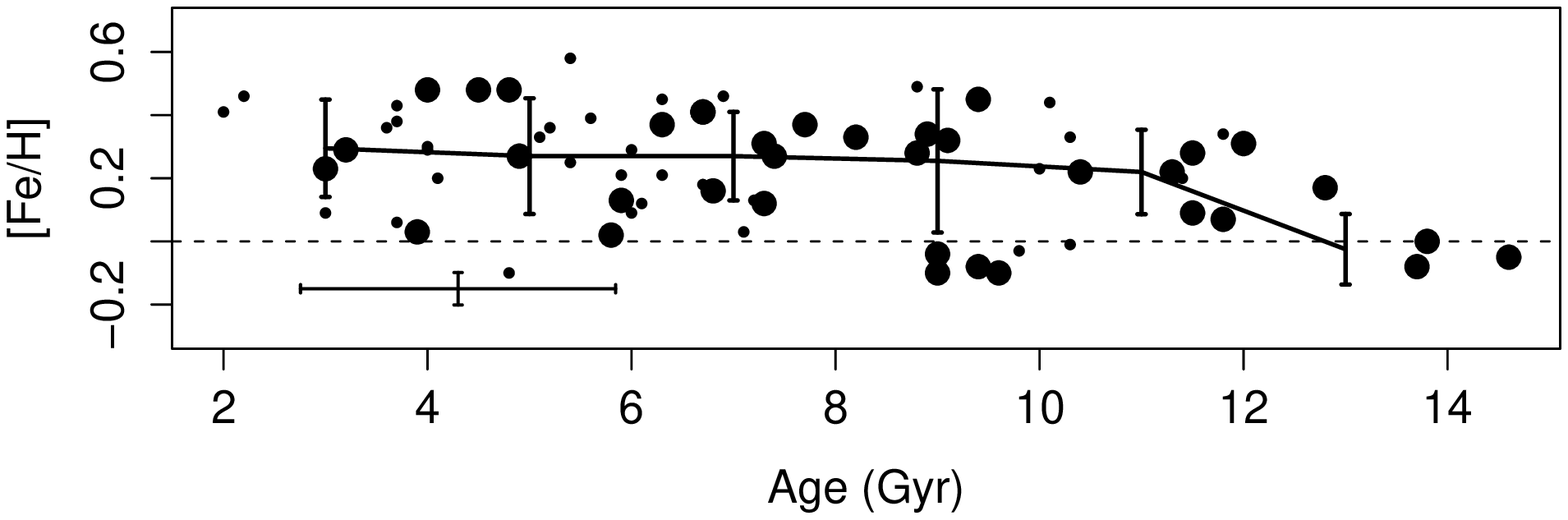}} \\
 \resizebox{\hsize}{!}{\includegraphics{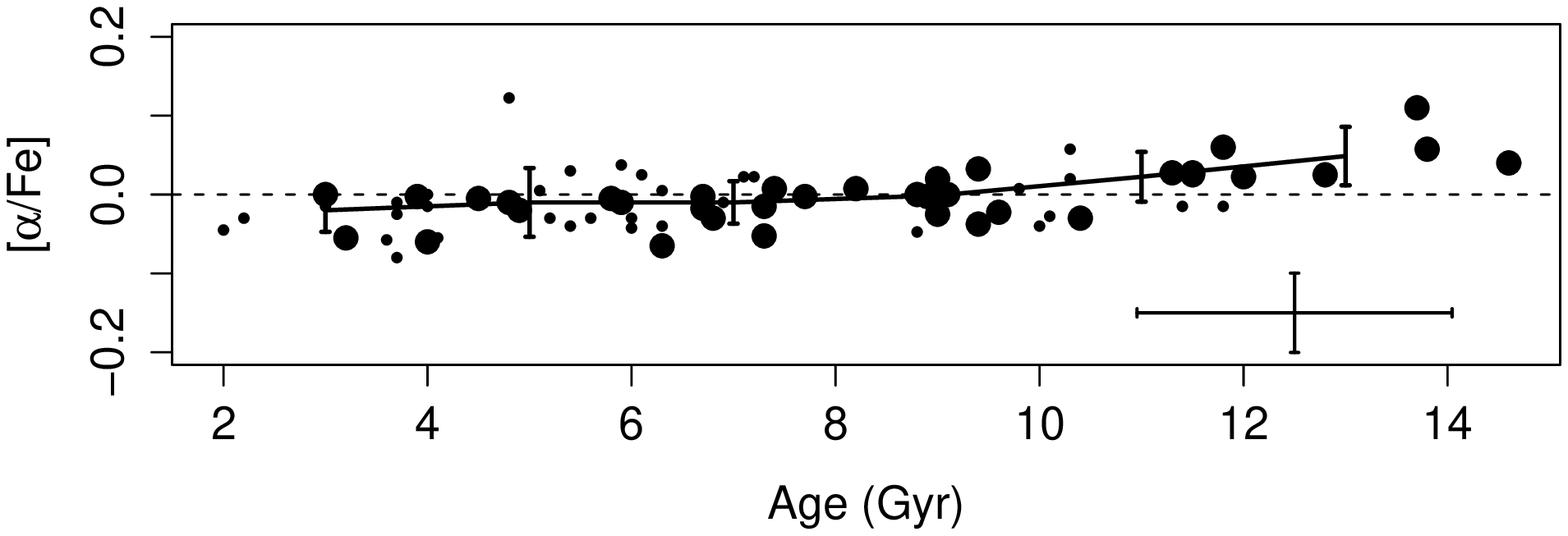}} 
 \end{tabular}
\caption {Ages {\it vs.} [Fe/H] ({\it top}) and [$\alpha$/Fe] {\it bottom}. The solid lines indicate the median value of [Fe/H] and [$\alpha$/Fe] in bins of 2~Gyr. Small symbols represent stars with age errors greater than 30\%.}
\label{Fig_ages}
\end{figure}

\begin{figure}
\centering
\begin{tabular}{c} 
 \resizebox{\hsize}{!}{\includegraphics{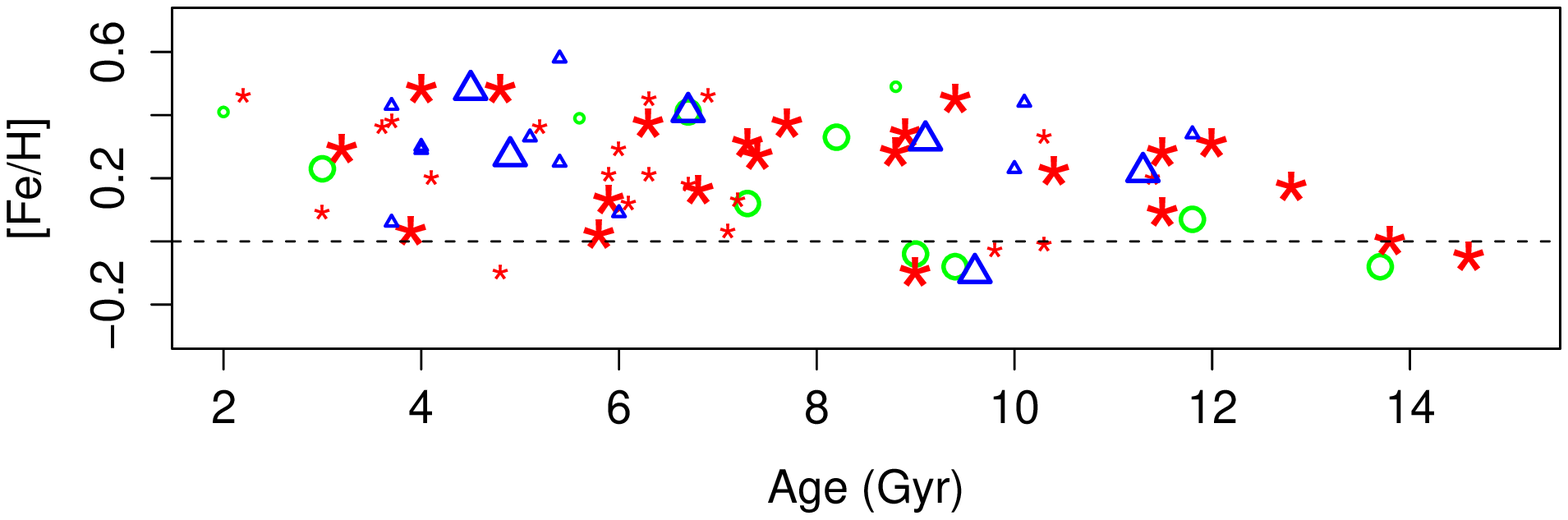}} \\
 \resizebox{\hsize}{!}{\includegraphics{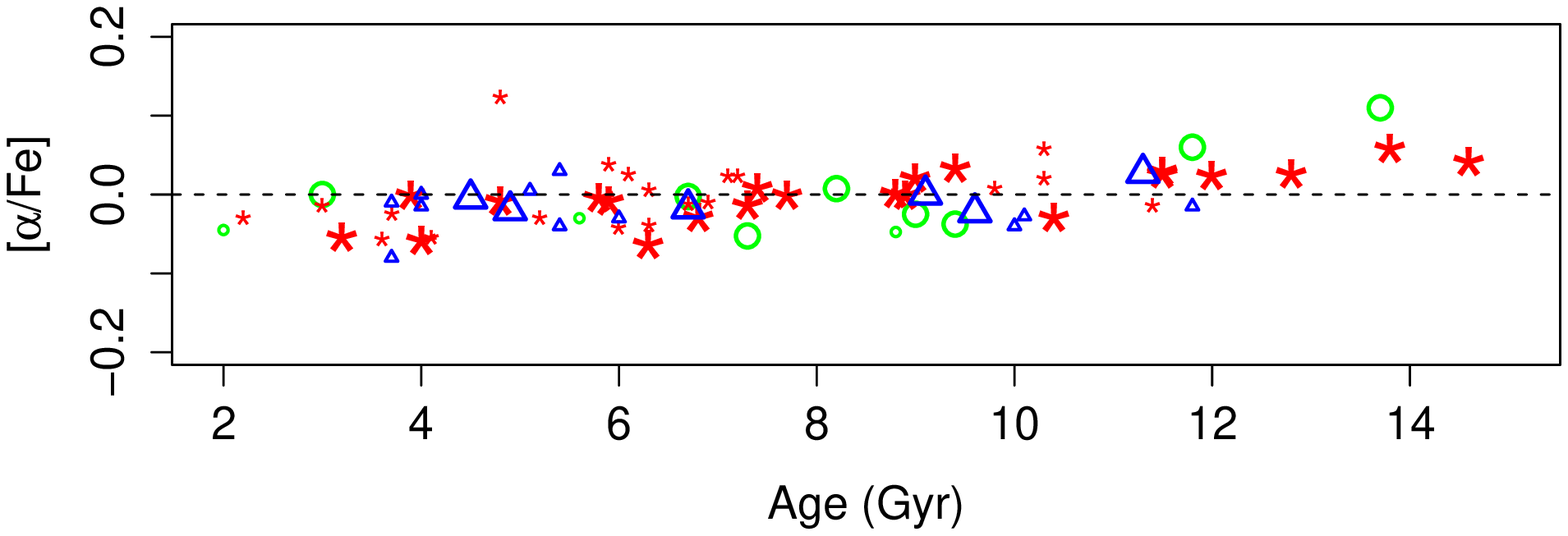}} 
 \end{tabular}
\caption {Ages {\it vs.} [Fe/H] ({\it top}) and [$\alpha$/Fe] {\it bottom}. Stars belonging to thin disk, thick disk, and intermediate populations are indicated as green circles, red stars, and blue triangles, respectively. Small symbols represent stars with age errors greater than 30\%.}
\label{Fig_ages2}
\end{figure}

Figure \ref{Fig_ages} shows [Fe/H] and [$\alpha$/Fe] {\it vs.} the stellar ages.
 The ages of the sample stars span from $\sim 2$ to $\sim 14$~Gyr, 
with mean age of 7 to 8  Gyr. In this plot, the large symbols represent stars for which
 ages could be determined with uncertainties lower than $30\%$, and the remaining stars 
are shown as small symbols. The older stars in the sample present lower metallicities
 and higher $\alpha$-element enhancement. 
To verify if different ages, [Fe/H] and [$\alpha$/Fe] correspond to different populations, 
in Figure \ref{Fig_ages2} we show the same as Fig. \ref{Fig_ages} for each subsample. 
Thin disk, thick disk, and intermediate populations are indicated by different symbols. 
It seems that the three subsamples span the same range of ages, and the trends of
 decreasing metallicity with increasing age and increasing [$\alpha$/Fe] with increasing age
 are observed for both thin and thick disk stars. Therefore, the evolution of abundances appear
 to be very similar for the three populations.

The maximum height from the Galactic plane Z$_{\rm max}$ of our selected 42 thick disk stars, of
 Z$_{\rm max}$ = 380 pc (Table \ref{Tab_populations}),
could be considered to be lower than a mean thick-disk height 
\citep[e.g.][]{Ivezic.etal:2008}.

\subsection{Comparison with bulge stars}

Very metal-rich stars can be found in the Galactic bulge. 
In the most extensive high-resolution spectroscopic survey available so far,
 \citet{Zoccali.etal:2008} studied stars in three different fields along the 
Galactic minor axis and find that in the most central region of their bulge
sample (b $ = -4^{\circ}$),
 $\sim 30$\% of the stars have [Fe/H] $> 0.2$, and more than $50\%$ have [Fe/H] $> 0.0$.
 The fraction of very metal-rich stars decreases with increasing Galactic latitudes. 
We investigate the similarity between the stars studied in this work and bulge stars. 

\citet{Gonzalez.etal:2011} determined the abundances of the $\alpha$ elements
Mg, Si, Ca, and Ti, and obtained a mean [$\alpha$/Fe] ratio of bulge stars. 
 At solar metallicities, the bulge stars are Mg-Si-Ca-Ti-enhanced by $\sim 0.1$~dex
at the solar metallicity, when compared with our sample stars.
However, the bulge stars are giants, and the present sample consists
of dwarfs, therefore systematic effects  of model atmospheres and other differences are expected on the abundance analysis.

\begin{figure*}
\centering
\begin{tabular}{cc} 
 \resizebox{0.48\hsize}{!}{\includegraphics{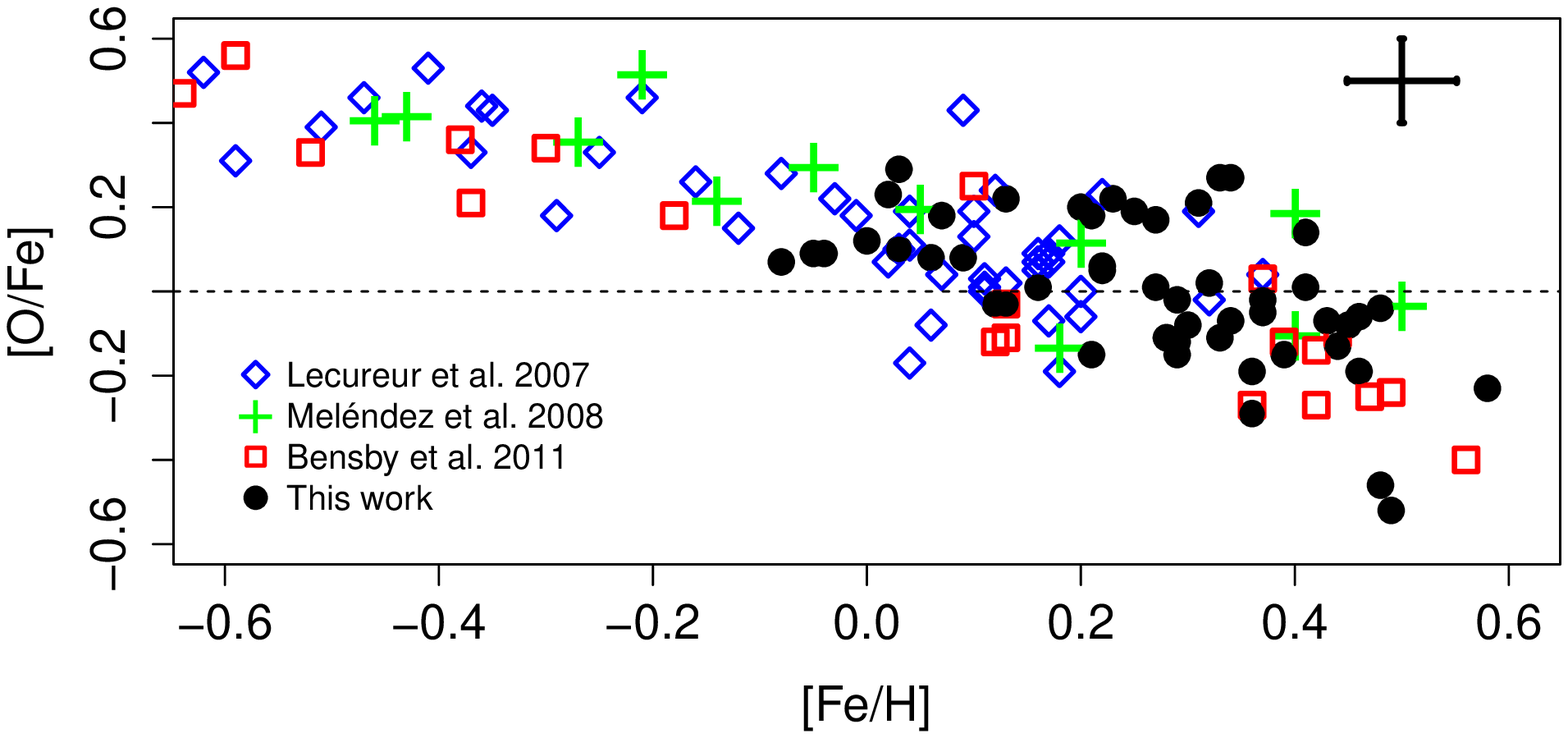}} &
 \resizebox{0.48\hsize}{!}{\includegraphics{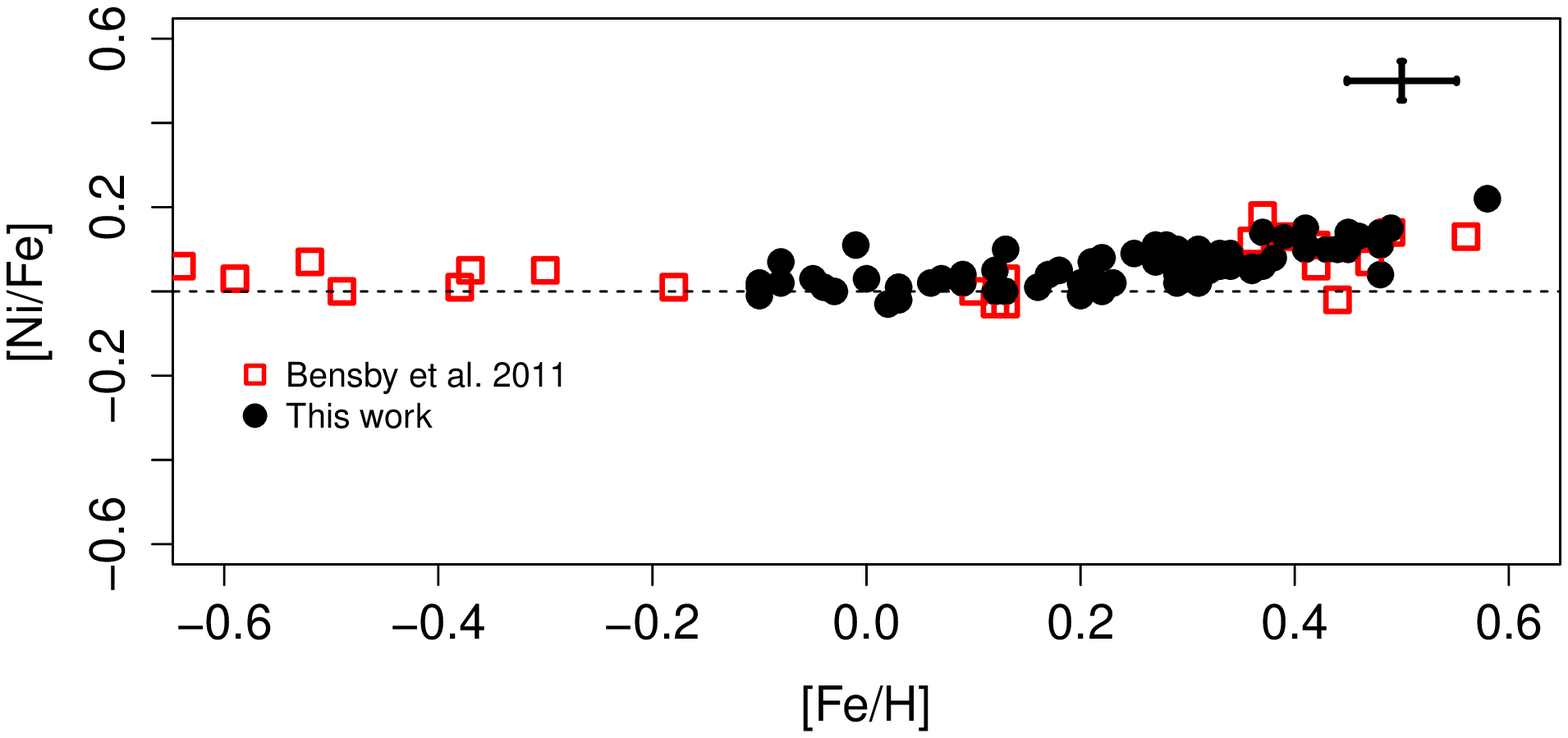}} \\ 
  \resizebox{0.48\hsize}{!}{\includegraphics{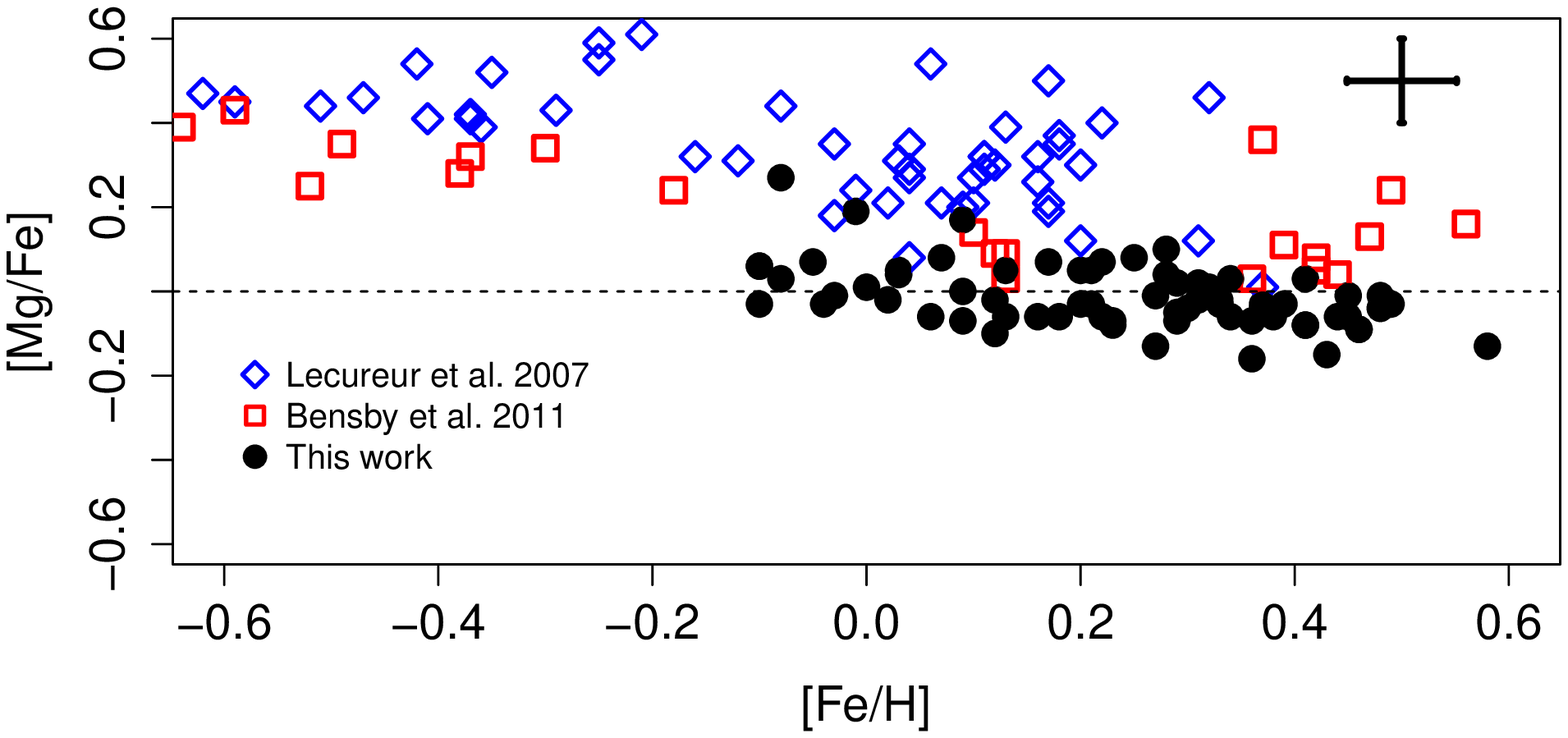}} & 
  \resizebox{0.48\hsize}{!}{\includegraphics{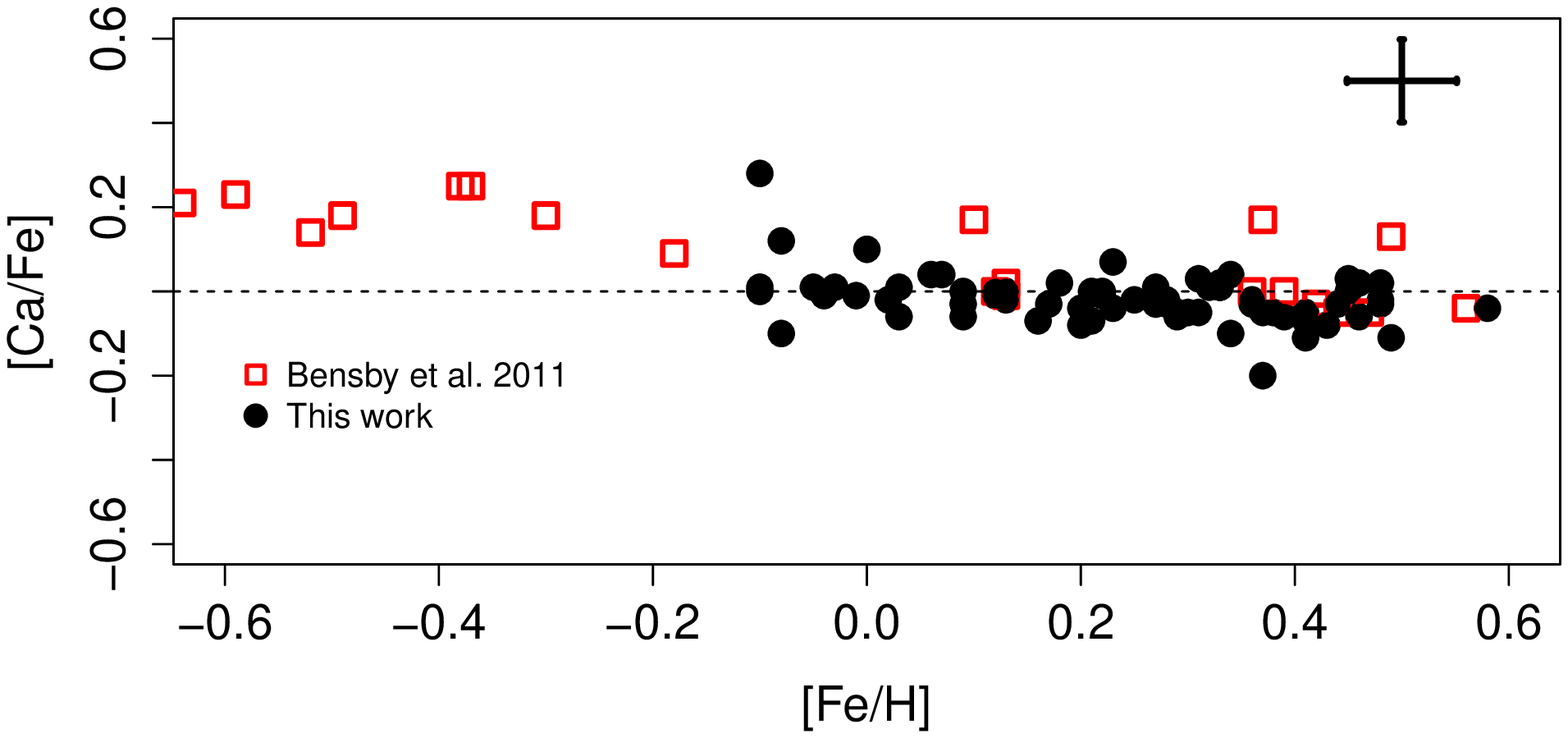}} \\
  \resizebox{0.48\hsize}{!}{\includegraphics{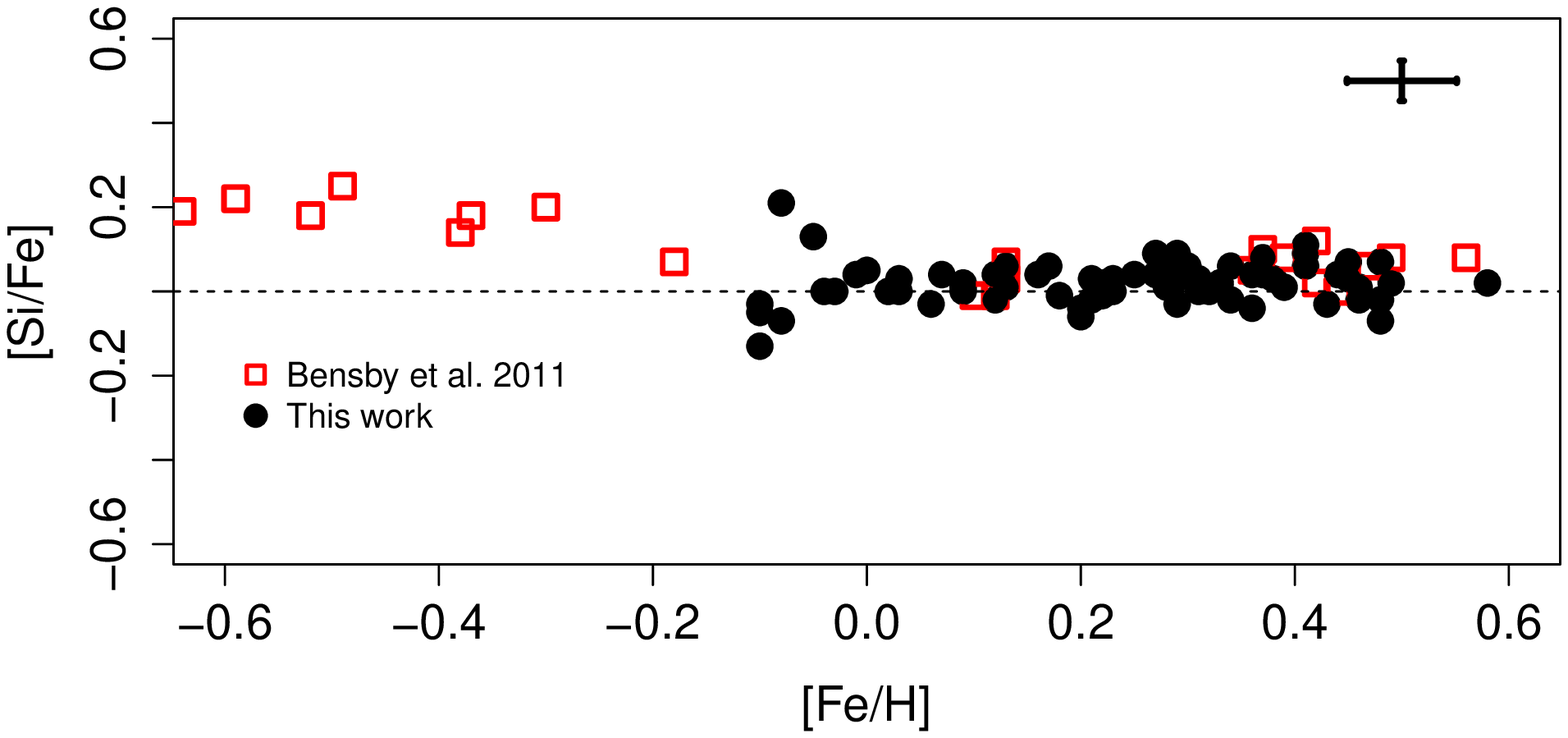}} & 
  \resizebox{0.48\hsize}{!}{\includegraphics{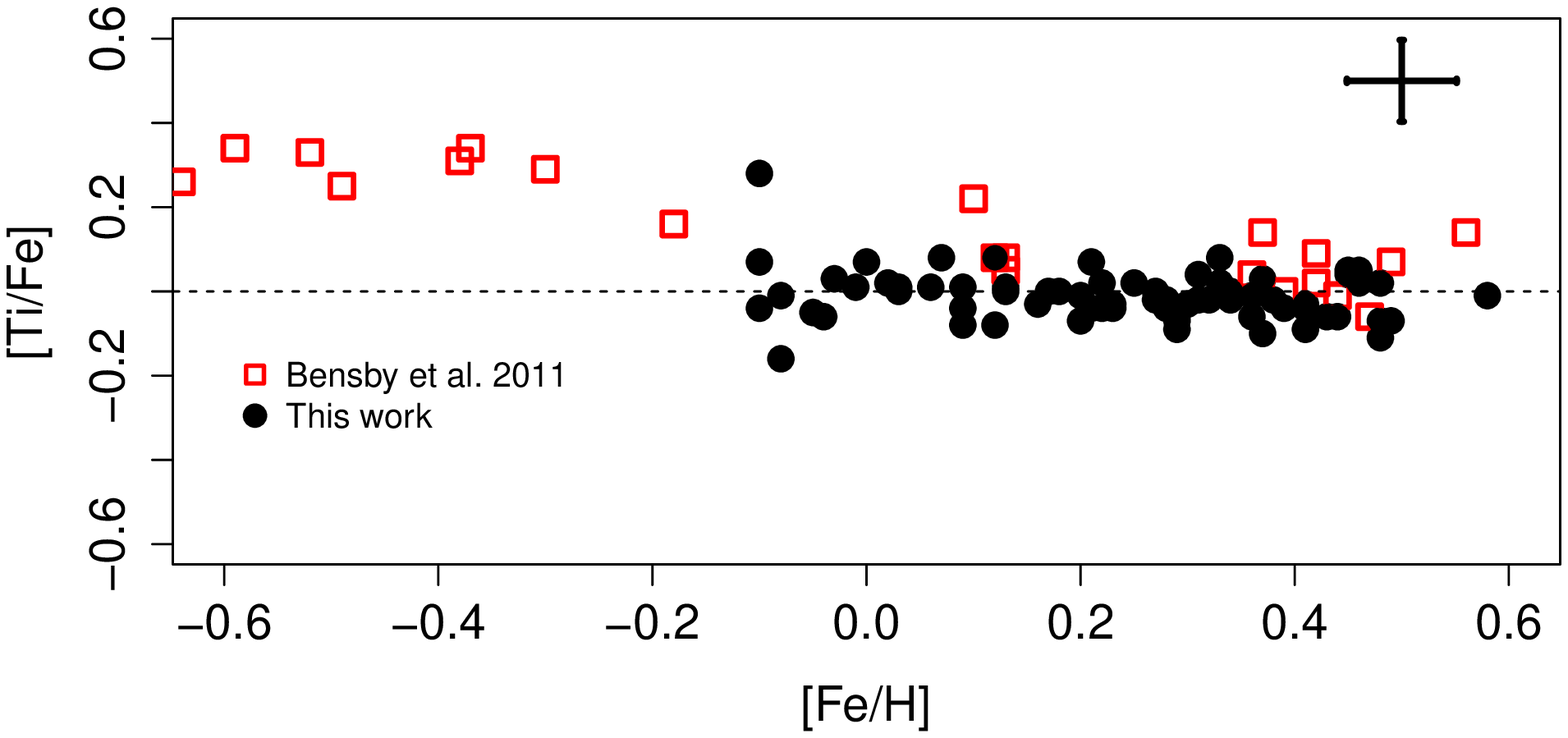}} \\
 \end{tabular}
\caption {Abundances of O, Ni, Mg, Ca, Si, and Ti {\it vs.} [Fe/H]. 
Sample stars from thick disk are indicated as black dots, blue diamonds, green crosses, and red squares are 
bulge stars from \citet{Lecureur.etal:2007}, \citet{Melendez.etal:2008}, and bulge microlensed 
dwarfs by \citet{Bensby.etal:2011}, respectively.}
\label{Fig_Oxy_bulge}
\end{figure*}

\begin{figure}
\centering
\begin{tabular}{c} 
 \resizebox{\hsize}{!}{\includegraphics{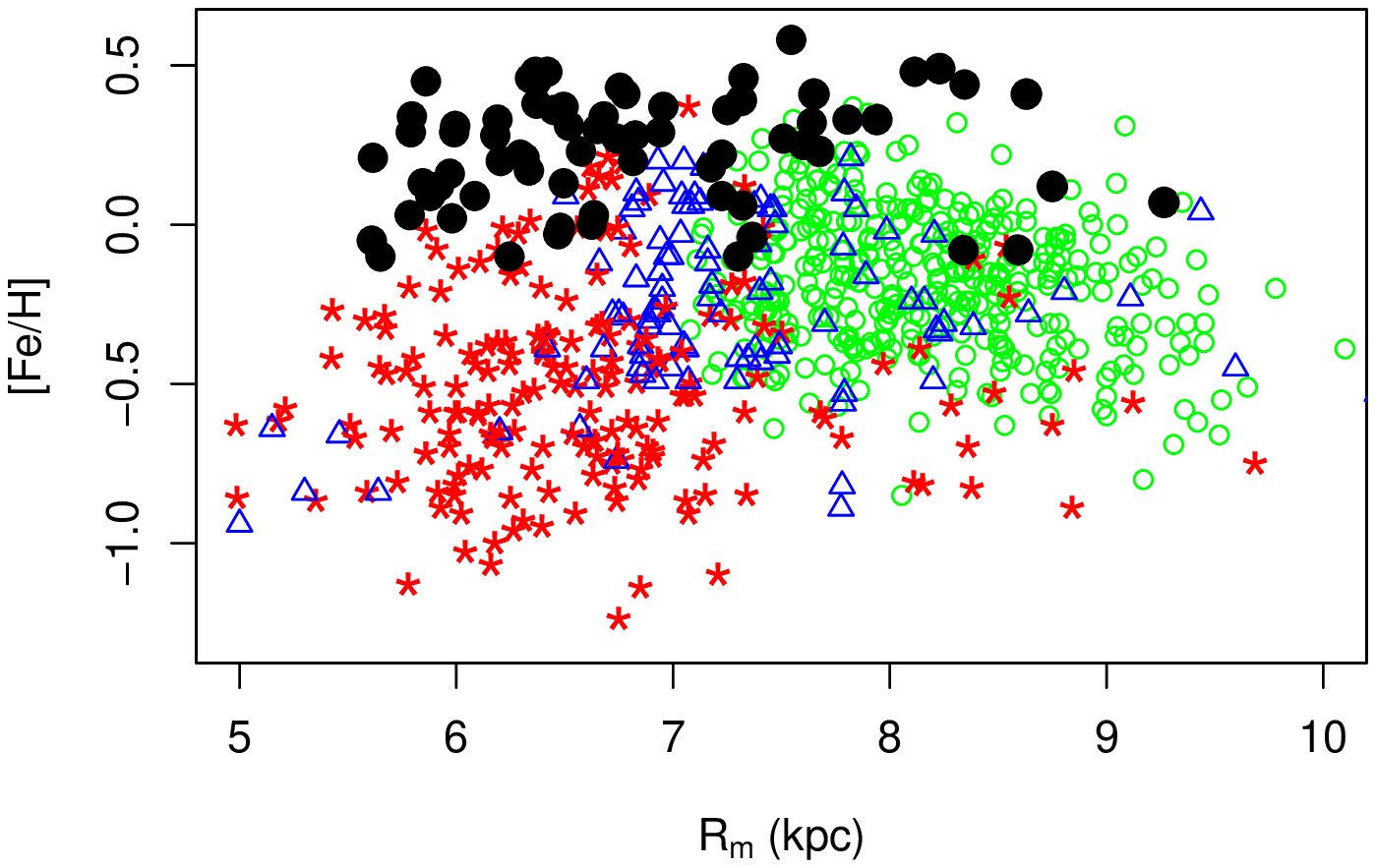}} \\
 \resizebox{\hsize}{!}{\includegraphics{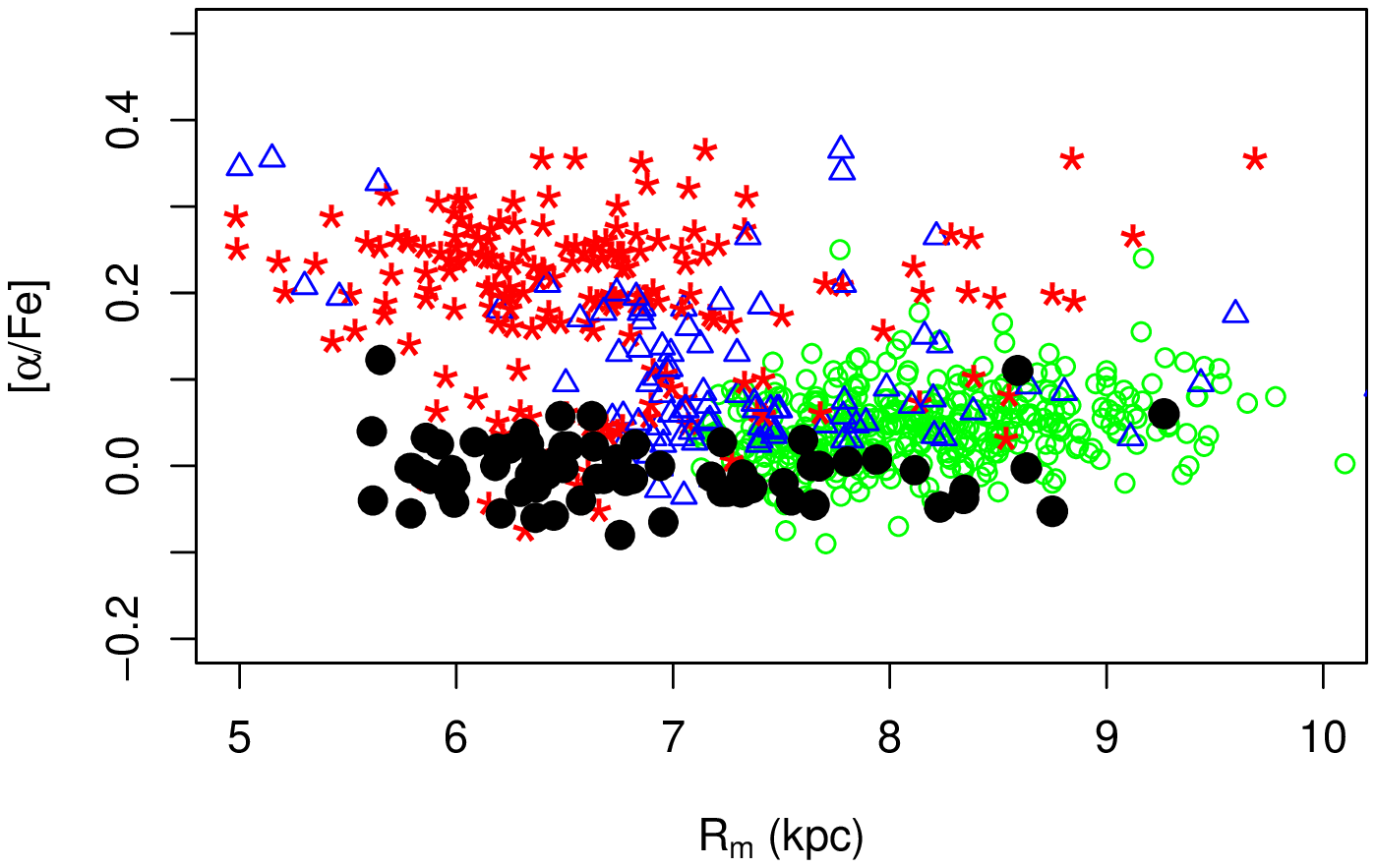}} \\
 \end{tabular}
\caption {Mean Galactocentric distance {\it vs.}  [Fe/H] ({\it top}) [$\alpha$/Fe] ({\it bottom}). Stars belonging to thin disk, thick disk, and intermediate populations are indicated as green circles, red stars, and blue triangles, respectively.}
\label{Fig_rm}
\end{figure}

\begin{figure}
\centering
\begin{tabular}{c} 
 \resizebox{\hsize}{!}{\includegraphics{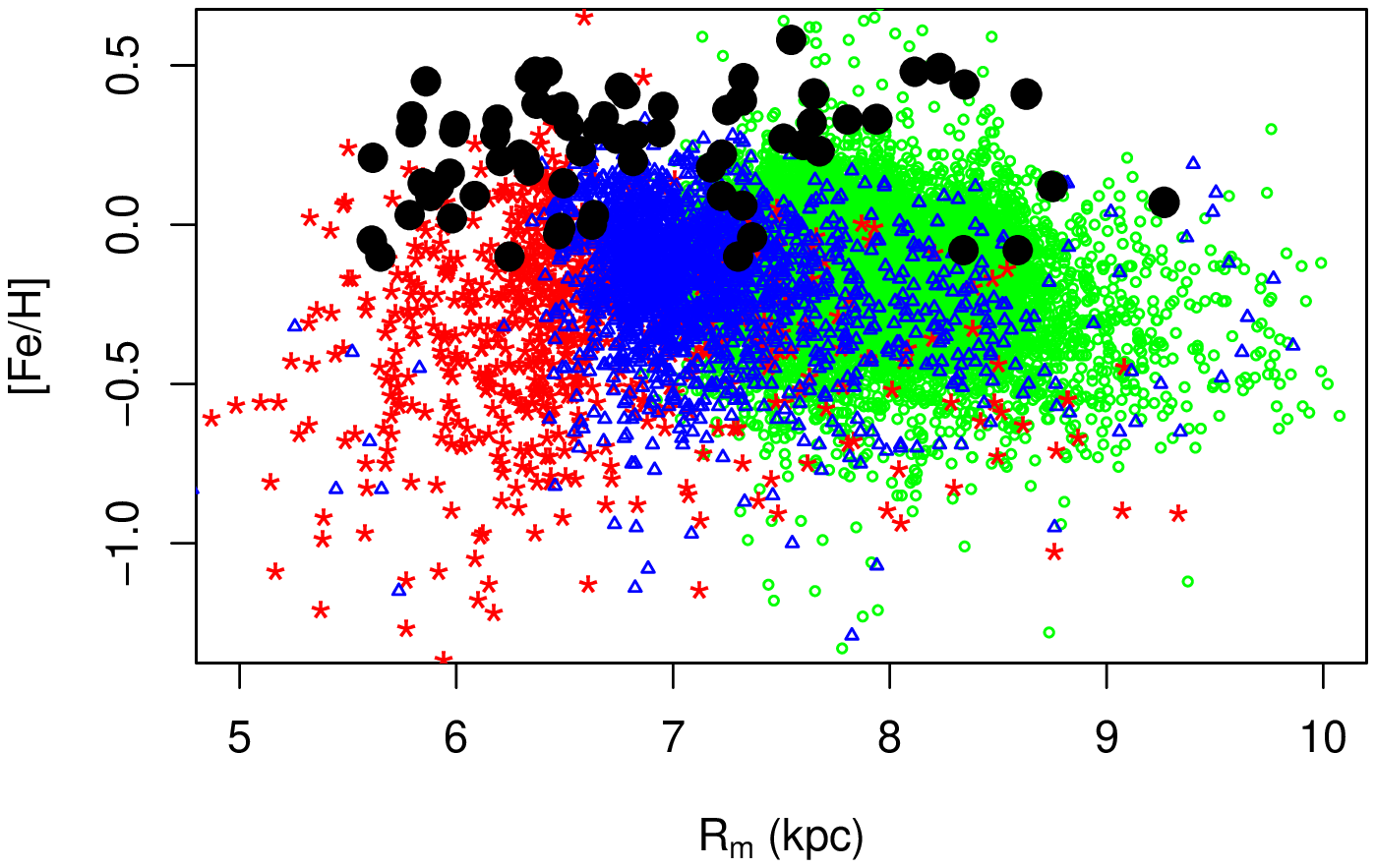}} \\
 \resizebox{\hsize}{!}{\includegraphics{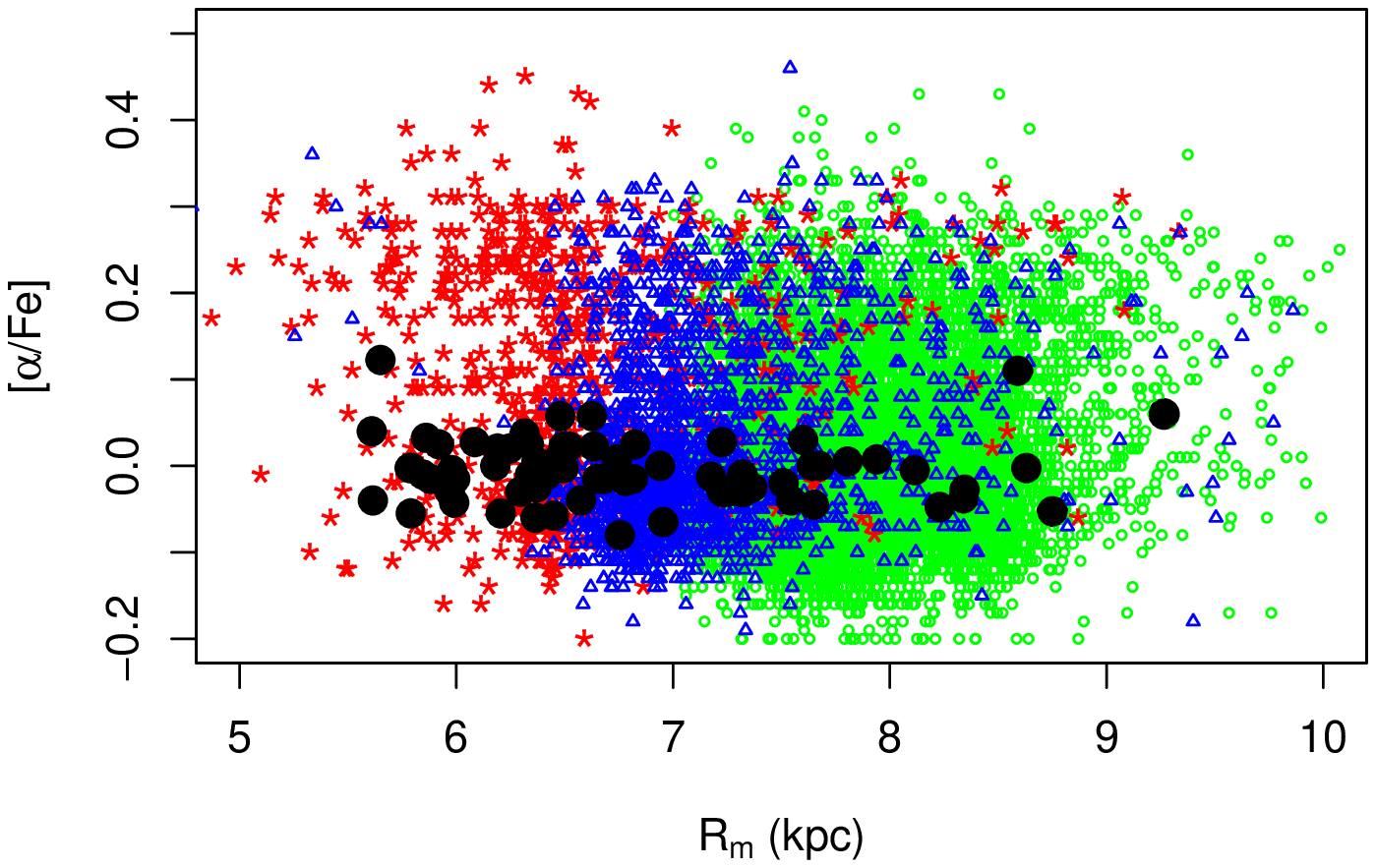}} 
 \end{tabular}
\caption {Mean Galactocentric distance {\it vs.}  [Fe/H] ({\it top}) [$\alpha$/Fe] ({\it bottom}) for the GCS stars. Symbols are the same as Fig. \ref{Fig_rm}.}
\label{Fig_rm_gcs}
\end{figure}

Another interesting piece of information comes from the observation
of microlensed dwarf and subgiant stars. In Fig. \ref{Fig_Oxy_bulge} we show the abundances of
O, Ni, Mg, Ca, Si, and Ti for 26 such stars, presented by \citet{Bensby.etal:2011}, and we plot
the abundances of our sample stars for comparison.
Except for an enhanced Mg in a few of the microlensed dwarfs,
the results for their six metal-rich stars are also compatible with the present results, therefore
our metal-rich thick-disk star subsample could be identified with
a bulge origin as well.

\subsection{Theoretical predictions}

The kinematical and chemical characteristics of the sample stars might be explained by
models of radial migration of stars. \citet{Fux:1997} and \citet{Raboud.etal:1998} identified
``hot'' orbits produced by effects of the bar, moving stars between regions inside the
bar, and outside corrotation.
\citet{Raboud.etal:1998} found that the old disk stars in their large
sample appeared to show a positive mean U motion, with an imbalance between positive and negative U velocities 
 reaching up to 50 km s$^{-1}$. A U anomaly of +29$\pm$2 km s$^{-1}$ with respect to the Sun 
and +19$\pm$9 km s$^{-1}$ with respect to the Galactic centre was identified. 
Raboud et al. suggest that the metal-rich stars within this sample appeared to wander from inside the bar,
reaching the solar neighbourhood. Therefore,
 the kinematical anomaly for the old disk (see Sect. \ref{Sec_Sample})
detected by Raboud et al. could
be a signature of the bar.
More recently, \citet{Sellwood.Binney:2002} have shown that the transient spiral arms
have a dominant effect on radial migration.
 If these mechanisms prove to be the origin of our thick disk
sample stars, it could be that these stars are  bulge or inner thick disk stars
reaching the solar neighbourhood.

 In recent years, radial migration has been the subject
 of several studies, such as
\citet{Haywood:2008}, \citet{Minchev.Famaey:2010}, \citet{Schonrich.Binney:2009a, Schonrich.Binney:2009b}, and
\citet{Brunetti.etal:2010}, among others. For example, 
\citet{Schonrich.Binney:2009b} predict  that there are old very metal-rich stars in the solar neighbourhood, at a 
relatively low rotational velocity. In this case as well, these metal-rich stars would have an origin
in the inner Galaxy. 

 Our subsample of 42 stars with kinematics of thick disk and solar 
$\alpha$-to-Fe ratios seems to be similar to a sample identified by 
\citet{Haywood:2008}: his identified subsample, shown as diamonds
 in his Fig. 12, has kinematics of thick disk,
[$\alpha$/Fe] $<$ +0.1, and they are old with ages in the range 8-12 Gyr, 
ages characteristic of an old thin disk. \citet{Haywood:2008} assigns a status of
transition objects between the two disks, but closer to an old thin disk.
Despite the higher metallicity of our 42 such stars, they seem otherwise to
be identical. It therefore seems that this subsample should be an inner disk,
closer to the Galactic centre than Haywood's subsample, an old thin disk component.

Indeed, radial migration from the inner disk (or bulge?)
 is the most probable origin of these stars. A need for more substantial
radial mixing as first discussed in Wielen et al. (1996), was shown by
\citet{Sellwood.Binney:2002} to be possible through the passage of
 recurrent transient spiral patterns. \citet{Lepine.etal:2003} and \citet{Roskar.etal:2011}
 present calculations demonstrating that resonant scattering
with spiral arms trigger efficient migration of stars from regions at
R$\sim$4-5 kpc into the solar system region. Roskar et al. (2011) conclude
that 50\% of stars in the solar neighbourhood have come from
R $<$ 6kpc.

Radial migration of stars could be caused by spiral and/or bar ressonance. 
\citet{Minchev.Famaey:2010} studied the combined effect of a central bar
and spiral structure on the dynamics of a galactic disk, 
and they find that the 
spiral-bar ressonance overlap induces a nonlinear response leading to a strong redistribution of
angular momentum in the disk. They show that a large population of stars from the bar's corotation resonance
(r $\sim 4.5$ kpc) enters the solar circle (their Fig. 6).

%

%

%

%

%

%

%

\section{Summary and conclusions}
\label{Sec_Summary}

In the present work, we analysed 71 metal-rich dwarf and turn-off stars,
most of them old, selected from the high proper motion NLTT catalogue,
as described in \citet{Raboud.etal:1998} (see Sect. \ref{Sec_Sample}). The aim 
of this work is   to better understand these stellar populations, as well 
as to verify their high metallicities. 

To be confident about the high-metallicity values, we compared the calculations 
carried out with two codes, from the Meudon \citep[ABON2 code,][]{Spite:1967} and the Uppsala \citep[BSYN/EQWI code, ][]{Edvardsson.etal:1993} groups, and the results are similar within
 [Fe/H]$\pm$0.02. The metallicities derived are in the range  -0.10 $<$ [\ion{Fe}{I}/H] $<$ +0.58 
from \ion{Fe}{I}, and  -0.18 $<$ [\ion{Fe}{II}/H] $< +0.56$ from \ion{Fe}{II}.

The present sample was studied by means of their kinematics and abundances. 
Our sample of 71 metal-rich stars can be kinematically subclassified in 
samples of thick disk, thin disk, and intermediate 
stellar populations, with
mean ages of about  $7.8 \pm 3.5$, $7.5 \pm 3.1$, and $6.8 \pm 2.9$~Gyr, respectively.
It seems definitely clear that some of the sample stars are quite old, and still quite metal rich.
 A most interesting feature of the sample stars is that 42 of them can
 be identified as belonging to the
thick disk. In particular, 70\% of the sample stars have space
 velocity V $< -50$~km~s$^{-1}$, 
which is more typical of a thick disk, but show solar $\alpha$-to-iron ratios that are 
more compatible with thin disk 
members. This subsample appears similar to one identified by 
\citet{Haywood:2008}, having kinematics of thick disk, together with
[$\alpha$/Fe] $<$ +0.1, and old  ages in the range 8-12 Gyr; 
\citet{Haywood:2008} interprets these stars as old thin disk, or
transition objects between the two disks, but as closer to an old thin disk.
Our subsample has higher metallicities than Haywood's subsample and could have
an origin closer to the Galactic centre than Haywood's old thin disk/transition component.

The presence of very metal-rich stars in the solar neighbourhood, at a 
relatively low rotational velocity give evidence of radial migration
in the Galaxy, induced by the bar and/or interaction of bar and spiral arms,
such as proposed by \citet{Fux:1997}, \citet{Raboud.etal:1998},
 \citet{Sellwood.Binney:2002}, \citet{Lepine.etal:2003},
\citet{Haywood:2008}, \citet{Minchev.Famaey:2010}, 
 \citep{Schonrich.Binney:2009a, Schonrich.Binney:2009b}, or \citet{Brunetti.etal:2010}.

 Finally, we can conclude that the sample stars, all metal-rich,
 should be old thin stars from the inner disk,
 as suggested by Haywood (2008), including 
the 42 ones identified to have kinematics of the thick disk,
and [$\alpha$/Fe]$<$ +0.1. On the other hand, it is natural
that the very metal-rich stars have low $\alpha$-to-iron ratios,
as discussed in Roskar et al. (2011), i.e. all stars with [Fe/H] $>$ 0
show such low  $\alpha$-to-iron.
In other words, the decreasing trend of [$\alpha$-elements/Fe] with increasing
 metallicity means that the SNIa enrichment in iron
 occurs at the same pace for our
sample, thick disk, and bulge stars. Therefore,
 for identifying bulge stars
and thick disk as a same population, as suggested by \citet{Bensby.etal:2011},
this cannot be inferred from the present results.

\section*{Acknowledgments}
The observations were carried out within Brazilian time in an ESO-ON agreement and within an IAG-ON agreement funded by FAPESP project n$^{\circ}$ 1998/10138-8. We thank the referee for the very insightful comments that led to a number of significant improvements in our manuscript. We are grateful to Giovanni Carraro for sharing his {\tt GRINTON} integrator code  used in Galactic orbits calculations. BB acknowledges partial financial support from CNPq and Fapesp. MT acknowledges an FAPESP fellowship no. 2008/50198-3.

\clearpage



\onecolumn
\scriptsize
\begin{longtable}{ccccrrrrrcc}

\caption{\label{Tab_kinematics} Kinematical data} \\

\hline\hline

{\rm star} &      \multicolumn{1}{c}{U}   &  \multicolumn{1}{c}{V}   &   \multicolumn{1}{c}{W}  & \multicolumn{1}{c}{v$_{\rm Helio}$}  & \multicolumn{1}{c}{$R_{\rm min}$} & \multicolumn{1}{c}{$R_{\rm max}$}  & \multicolumn{1}{c}{$Z_{\rm max}$}  & \multicolumn{1}{c}{$e$} & $p_{\rm thin}$ & $p_{\rm thick}$ \\
           &  \multicolumn{3}{c}{(km s$^{-1}$)}  & (km s$^{-1}$) & \multicolumn{1}{c}{(kpc)}     & \multicolumn{1}{c}{(kpc)}        & \multicolumn{1}{c}{(kpc)}  &   & (\%) & (\%)  \\

\hline
\endfirsthead
\caption{continued.}\\
\hline\hline

{\rm star} &      \multicolumn{1}{c}{U}   &  \multicolumn{1}{c}{V}   &   \multicolumn{1}{c}{W}  & \multicolumn{1}{c}{v$_{\rm Helio}$}  & \multicolumn{1}{c}{$R_{\rm min}$} & \multicolumn{1}{c}{$R_{\rm max}$}  & \multicolumn{1}{c}{$Z_{\rm max}$}  & \multicolumn{1}{c}{$e$} & $p_{\rm thin}$ & $p_{\rm thick}$ \\
           &  \multicolumn{3}{c}{(km s$^{-1}$)}  & (km s$^{-1}$) & \multicolumn{1}{c}{(kpc)}     & \multicolumn{1}{c}{(kpc)}        & \multicolumn{1}{c}{(kpc)}  &   & (\%) & (\%)  \\
\hline
\endhead
\hline
 \\
\endfoot
 
\\
\multicolumn{11}{c}{Thin disk} \\
HD    11608  & -37 &  -37 &   21 &  -25.34 $\pm$ 0.16 & 5.58 $\pm$ 0.32 &  9.05 $\pm$ 0.52 & 0.17 $\pm$ 0.03 & 0.24 $\pm$ 0.01  & 84 & 16\\
HD    26151  & -50 &  -19 &   12 &   -6.87 $\pm$ 0.11 & 6.20 $\pm$ 0.36 &  9.68 $\pm$ 0.58 & 0.07 $\pm$ 0.02 & 0.22 $\pm$ 0.01  & 95 &  5\\
HD    26794  &  71 &   13 &    3 &   56.49 $\pm$ 0.17 & 7.39 $\pm$ 0.41 & 11.14 $\pm$ 0.67 & 0.06 $\pm$ 0.03 & 0.20 $\pm$ 0.01  & 95 &  5\\
HD    35854  &   5 &  -31 &    4 &   23.05 $\pm$ 0.23 & 6.18 $\pm$ 0.38 &  8.55 $\pm$ 0.51 & 0.04 $\pm$ 0.02 & 0.16 $\pm$ 0.01  & 92 &  8\\
HD    77338  & -39 &  -27 &  -25 &    8.47 $\pm$ 0.11 & 6.05 $\pm$ 0.37 &  9.25 $\pm$ 0.56 & 0.46 $\pm$ 0.06 & 0.21 $\pm$ 0.01  & 84 & 16\\
HD    82943  & -10 &  -20 &   -9 &    8.25 $\pm$ 0.13 & 6.71 $\pm$ 0.41 &  8.64 $\pm$ 0.50 & 0.19 $\pm$ 0.04 & 0.13 $\pm$ 0.01  & 95 &  5\\
HD    86249  &  12 &	8 &  -24 &   -7.84 $\pm$ 0.09 & 8.51 $\pm$ 0.49 &  8.99 $\pm$ 0.58 & 0.43 $\pm$ 0.06 & 0.03 $\pm$ 0.01  & 96 &  4\\
HD    93800  &  49 &   -8 &  -13 &    3.81 $\pm$ 0.11 & 7.13 $\pm$ 0.42 &  9.33 $\pm$ 0.55 & 0.26 $\pm$ 0.04 & 0.13 $\pm$ 0.01  & 94 &  6\\
HD   177374  & -91 &  -15 &  -20 &   79.97 $\pm$ 0.14 & 5.78 $\pm$ 0.34 & 11.40 $\pm$ 0.72 & 0.42 $\pm$ 0.05 & 0.33 $\pm$ 0.01  & 82 & 18\\
HD   181433  & -56 &   -1 &    8 &   60.69 $\pm$ 0.17 & 6.79 $\pm$ 0.38 & 10.47 $\pm$ 0.63 & 0.03 $\pm$ 0.02 & 0.21 $\pm$ 0.01  & 97 &  3\\ 
HD   224230  & -78 &  -16 &   -6 &   59.61 $\pm$ 0.06 & 5.88 $\pm$ 0.35 & 10.80 $\pm$ 0.67 & 0.18 $\pm$ 0.04 & 0.29 $\pm$ 0.01  & 92 &  8\\

\\
\multicolumn{11}{c}{Thick disk} \\ 
G 161-029    &-149 &  -89 &   40 &   23.00 $\pm$ 0.16 &            ...  &             ...  &            ...  &              ... &  0 & 97\\
BD-02	180  &  -1 &  -82 &  -46 &   16.32 $\pm$ 0.19 & 3.82 $\pm$ 0.22 &  8.56 $\pm$ 0.48 & 0.92 $\pm$ 0.11 & 0.38 $\pm$ 0.01  &  0 & 99\\ 
BD-05  5798  &  15 &  -54 &  -53 &    6.51 $\pm$ 0.10 & 5.14 $\pm$ 0.30 &  8.49 $\pm$ 0.47 & 1.10 $\pm$ 0.13 & 0.25 $\pm$ 0.01  &  4 & 96\\
BD-17  6035  &  35 &  -83 &   11 &  -65.48 $\pm$ 0.08 & 3.62 $\pm$ 0.22 &  8.55 $\pm$ 0.51 & 0.06 $\pm$ 0.02 & 0.40 $\pm$ 0.01  &  1 & 99\\
CD-32  0327  &  52 &  -69 &  -10 &   14.50 $\pm$ 0.10 & 4.17 $\pm$ 0.25 &  8.79 $\pm$ 0.54 & 0.22 $\pm$ 0.04 & 0.36 $\pm$ 0.01  &  5 & 95\\
CD-40 15036  &  88 &  -80 &   -9 &  -12.58 $\pm$ 0.06 & 3.58 $\pm$ 0.22 &  9.36 $\pm$ 0.57 & 0.23 $\pm$ 0.04 & 0.45 $\pm$ 0.01  &  0 & 99\\
HD     9424  &  57 &  -96 &   -1 &   43.61 $\pm$ 0.18 & 3.09 $\pm$ 0.17 &  8.75 $\pm$ 0.50 & 0.11 $\pm$ 0.03 & 0.48 $\pm$ 0.01  &  0 & 99\\
HD    10576  &  56 &  -93 &  -19 &   54.86 $\pm$ 0.06 & 3.23 $\pm$ 0.20 &  8.73 $\pm$ 0.51 & 0.33 $\pm$ 0.05 & 0.46 $\pm$ 0.01  &  0 & 99\\
HD    13386  &  17 &  -68 &  -13 &   32.39 $\pm$ 0.10 & 4.33 $\pm$ 0.24 &  8.57 $\pm$ 0.49 & 0.25 $\pm$ 0.04 & 0.33 $\pm$ 0.01  & 10 & 90\\
HD    15133  &  45 &  -75 &   -8 &   38.43 $\pm$ 0.13 & 3.96 $\pm$ 0.23 &  8.72 $\pm$ 0.50 & 0.19 $\pm$ 0.03 & 0.38 $\pm$ 0.01  &  3 & 97\\
HD    15555  &  74 &  -56 &    0 &   36.18 $\pm$ 0.12 & 4.61 $\pm$ 0.25 &  9.30 $\pm$ 0.51 & 0.10 $\pm$ 0.03 & 0.34 $\pm$ 0.01  & 18 & 82\\
HD    16905  & -31 &  -59 &  -31 &   64.77 $\pm$ 0.14 & 4.66 $\pm$ 0.28 &  8.82 $\pm$ 0.51 & 0.58 $\pm$ 0.07 & 0.31 $\pm$ 0.01  & 13 & 87\\
HD    25061  &  88 &  -51 &   -7 &   47.55 $\pm$ 0.13 & 4.71 $\pm$ 0.27 &  9.64 $\pm$ 0.59 & 0.19 $\pm$ 0.04 & 0.34 $\pm$ 0.01  & 17 & 83\\
HD    27894  &  63 &  -73 &  -41 &   82.85 $\pm$ 0.14 & 4.05 $\pm$ 0.25 &  8.94 $\pm$ 0.54 & 0.82 $\pm$ 0.10 & 0.38 $\pm$ 0.01  &  0 & 99\\
HD    31827  & -17 &  -69 &    9 &   44.75 $\pm$ 0.28 & 4.21 $\pm$ 0.25 &  8.63 $\pm$ 0.53 & 0.04 $\pm$ 0.01 & 0.34 $\pm$ 0.01  & 13 & 86\\
HD    39213  &  10 &  -70 &   27 &   49.57 $\pm$ 0.09 & 4.23 $\pm$ 0.25 &  8.51 $\pm$ 0.50 & 0.25 $\pm$ 0.04 & 0.34 $\pm$ 0.01  &  8 & 92\\
HD    81767  &  50 &  -77 &   -2 &   81.81 $\pm$ 0.14 & 3.85 $\pm$ 0.22 &  8.74 $\pm$ 0.49 & 0.11 $\pm$ 0.03 & 0.39 $\pm$ 0.01  &  2 & 98\\
HD    90054  &   5 &  -99 &  -17 &   48.93 $\pm$ 0.36 & 3.06 $\pm$ 0.19 &  8.52 $\pm$ 0.50 & 0.31 $\pm$ 0.05 & 0.47 $\pm$ 0.01  &  0 & 99\\
HD    94374  &  38 & -108 &  -16 &   75.05 $\pm$ 0.17 & 2.71 $\pm$ 0.17 &  8.59 $\pm$ 0.51 & 0.29 $\pm$ 0.05 & 0.52 $\pm$ 0.01  &  0 & 99\\
HD    95338  &  24 & -111 &  -44 &   97.05 $\pm$ 0.08 & 2.68 $\pm$ 0.17 &  8.55 $\pm$ 0.50 & 0.88 $\pm$ 0.08 & 0.52 $\pm$ 0.01  &  0 & 98\\
HD   104212  &  30 &  -96 &  -32 &   66.04 $\pm$ 0.12 & 3.18 $\pm$ 0.19 &  8.51 $\pm$ 0.49 & 0.60 $\pm$ 0.07 & 0.46 $\pm$ 0.01  &  0 & 99\\
HD   107509  &  33 & -100 &  -24 &   70.43 $\pm$ 0.13 & 3.01 $\pm$ 0.18 &  8.56 $\pm$ 0.51 & 0.44 $\pm$ 0.06 & 0.48 $\pm$ 0.01  &  0 & 99\\
HD   120329  &  71 & -103 &  -78 &   24.85 $\pm$ 0.05 & 3.04 $\pm$ 0.19 &  8.95 $\pm$ 0.53 & 2.02 $\pm$ 0.19 & 0.49 $\pm$ 0.01  &  0 & 92\\
HD   143102  &  53 &  -92 &  -20 &    7.95 $\pm$ 0.02 & 3.26 $\pm$ 0.21 &  8.68 $\pm$ 0.53 & 0.34 $\pm$ 0.06 & 0.45 $\pm$ 0.01  &  0 & 99\\
HD   148530  & -81 &  -80 &  -34 &   25.77 $\pm$ 0.22 & 3.56 $\pm$ 0.20 &  9.71 $\pm$ 0.56 & 0.68 $\pm$ 0.09 & 0.46 $\pm$ 0.01  &  0 & 99\\
HD   149256  &  -7 &  -98 &  -48 &   25.42 $\pm$ 0.08 & 3.13 $\pm$ 0.20 &  8.46 $\pm$ 0.50 & 0.95 $\pm$ 0.12 & 0.46 $\pm$ 0.01  &  0 & 99\\
HD   149606  &  28 &  -77 &    4 &   -2.27 $\pm$ 0.22 & 3.88 $\pm$ 0.23 &  8.53 $\pm$ 0.50 & 0.04 $\pm$ 0.02 & 0.37 $\pm$ 0.01  &  3 & 97\\
HD   149933  &  22 &  -63 &   -9 &  -13.95 $\pm$ 0.06 & 4.50 $\pm$ 0.25 &  8.49 $\pm$ 0.47 & 0.20 $\pm$ 0.03 & 0.31 $\pm$ 0.01  & 19 & 81\\
HD   165920  & -66 &  -49 &  -42 &   61.25 $\pm$ 0.12 & 4.86 $\pm$ 0.31 &  9.64 $\pm$ 0.60 & 0.87 $\pm$ 0.11 & 0.33 $\pm$ 0.01  & 11 & 89\\
HD   168714  &  60 &  -74 &   17 &   -9.76 $\pm$ 0.26 & 3.91 $\pm$ 0.22 &  8.82 $\pm$ 0.51 & 0.12 $\pm$ 0.03 & 0.39 $\pm$ 0.01  &  2 & 97\\
HD   171999  &  13 &  -87 &  -11 &  -46.32 $\pm$ 0.28 & 3.49 $\pm$ 0.21 &  8.49 $\pm$ 0.49 & 0.22 $\pm$ 0.04 & 0.42 $\pm$ 0.01  &  1 & 99\\
HD   179764  &  21 & -106 &    6 &  -66.04 $\pm$ 0.17 & 2.76 $\pm$ 0.17 &  8.46 $\pm$ 0.49 & 0.03 $\pm$ 0.02 & 0.51 $\pm$ 0.01  &  0 & 99\\
HD   180865  &  27 &  -72 &  -14 &   18.41 $\pm$ 0.15 & 4.11 $\pm$ 0.23 &  8.52 $\pm$ 0.49 & 0.26 $\pm$ 0.04 & 0.35 $\pm$ 0.01  &  5 & 95\\
HD   181234  &   5 &  -92 &    2 &  -46.33 $\pm$ 0.14 & 3.28 $\pm$ 0.20 &  8.44 $\pm$ 0.51 & 0.06 $\pm$ 0.02 & 0.44 $\pm$ 0.01  &  0 & 99\\
HD   196397  &  -1 &  -69 &   16 &  -16.94 $\pm$ 0.07 & 4.24 $\pm$ 0.24 &  8.50 $\pm$ 0.50 & 0.11 $\pm$ 0.03 & 0.33 $\pm$ 0.01  & 13 & 87\\
HD   201237  & -86 &  -79 &   -1 &   30.57 $\pm$ 0.34 & 3.51 $\pm$ 0.21 &  9.74 $\pm$ 0.60 & 0.13 $\pm$ 0.03 & 0.47 $\pm$ 0.01  &  1 & 99\\
HD   209721  &  14 &  -76 &   31 &    7.53 $\pm$ 0.11 & 3.93 $\pm$ 0.23 &  8.43 $\pm$ 0.50 & 0.30 $\pm$ 0.04 & 0.36 $\pm$ 0.01  &  3 & 97\\
HD   211706  &  80 & -106 &  -13 &  -62.79 $\pm$ 0.06 & 2.68 $\pm$ 0.16 &  9.08 $\pm$ 0.53 & 0.27 $\pm$ 0.04 & 0.54 $\pm$ 0.01  &  0 & 98\\
HD   218566  &  77 &  -61 &   -8 &  -37.21 $\pm$ 0.25 & 4.37 $\pm$ 0.25 &  9.28 $\pm$ 0.54 & 0.19 $\pm$ 0.04 & 0.36 $\pm$ 0.01  &  8 & 92\\
HD   218750  & -25 &  -74 &   13 &   17.13 $\pm$ 0.10 & 3.98 $\pm$ 0.23 &  8.69 $\pm$ 0.51 & 0.08 $\pm$ 0.02 & 0.37 $\pm$ 0.01  &  6 & 94\\
HD   221313  &   5 &  -65 &  -47 &   41.60 $\pm$ 0.21 & 4.56 $\pm$ 0.30 &  8.48 $\pm$ 0.53 & 0.94 $\pm$ 0.12 & 0.30 $\pm$ 0.01  &  2 & 98\\
HD   221974  &  95 &  -48 &   -3 &  -25.65 $\pm$ 0.17 & 4.74 $\pm$ 0.29 &  9.91 $\pm$ 0.62 & 0.14 $\pm$ 0.03 & 0.35 $\pm$ 0.01  & 19 & 81\\
HD   224383  &  74 &  -84 &   -1 &  -30.47 $\pm$ 0.10 & 3.46 $\pm$ 0.20 &  9.03 $\pm$ 0.54 & 0.11 $\pm$ 0.03 & 0.45 $\pm$ 0.01  &  0 & 99\\

\\
\multicolumn{11}{c}{Intermediate sample} \\
HD     8389  &  45 &  -28 &  -25 &   35.87 $\pm$ 0.08 & 6.19 $\pm$ 0.34 &  8.90 $\pm$ 0.51 & 0.45 $\pm$ 0.06 & 0.18 $\pm$ 0.01  & 76 & 24\\
HD     9174  & -22 &  -56 &  -29 &   24.51 $\pm$ 0.05 & 4.85 $\pm$ 0.31 &  8.71 $\pm$ 0.54 & 0.54 $\pm$ 0.07 & 0.28 $\pm$ 0.01  & 21 & 78\\
HD    12789  &  74 &  -35 &  -12 &   26.83 $\pm$ 0.20 & 5.52 $\pm$ 0.34 &  9.50 $\pm$ 0.61 & 0.27 $\pm$ 0.04 & 0.26 $\pm$ 0.01  & 60 & 40\\
HD    30295  &  65 &  -30 &  -39 &   46.49 $\pm$ 0.11 & 5.93 $\pm$ 0.34 &  9.35 $\pm$ 0.55 & 0.77 $\pm$ 0.09 & 0.22 $\pm$ 0.01  & 37 & 63\\
HD    31452  &  -6 &  -62 &  -10 &   14.69 $\pm$ 0.03 & 4.57 $\pm$ 0.27 &  8.58 $\pm$ 0.48 & 0.21 $\pm$ 0.04 & 0.31 $\pm$ 0.01  & 25 & 75\\
HD    37986  &  28 &  -59 &    1 &   59.45 $\pm$ 0.28 & 4.70 $\pm$ 0.28 &  8.60 $\pm$ 0.51 & 0.07 $\pm$ 0.03 & 0.29 $\pm$ 0.01  & 32 & 68\\
HD    39715  & -74 &  -50 &  -21 &  -33.66 $\pm$ 0.17 & 4.71 $\pm$ 0.29 &  9.89 $\pm$ 0.62 & 0.41 $\pm$ 0.06 & 0.35 $\pm$ 0.01  & 29 & 71\\
HD    43848  &  53 &  -58 &    1 &   44.92 $\pm$ 0.12 & 4.66 $\pm$ 0.27 &  8.85 $\pm$ 0.52 & 0.07 $\pm$ 0.03 & 0.31 $\pm$ 0.01  & 24 & 76\\
HD    86065  &  69 &  -44 &    2 &   55.30 $\pm$ 0.12 & 5.18 $\pm$ 0.29 &  9.27 $\pm$ 0.52 & 0.07 $\pm$ 0.03 & 0.28 $\pm$ 0.01  & 51 & 49\\
HD    87007  &  40 &  -47 &  -13 &   30.30 $\pm$ 0.09 & 5.22 $\pm$ 0.30 &  8.66 $\pm$ 0.50 & 0.25 $\pm$ 0.04 & 0.25 $\pm$ 0.01  & 53 & 47\\
HD    91585  & 100 &  -41 &    5 &   44.48 $\pm$ 0.12 & 5.00 $\pm$ 0.29 & 10.20 $\pm$ 0.63 & 0.04 $\pm$ 0.02 & 0.34 $\pm$ 0.01  & 30 & 70\\
HD    91669  & -76 &  -18 &  -28 &  -12.40 $\pm$ 0.12 & 5.88 $\pm$ 0.34 & 10.81 $\pm$ 0.67 & 0.57 $\pm$ 0.08 & 0.29 $\pm$ 0.01  & 79 & 21\\
HD   182572  & 116 &  -31 &  -19 &  -99.86 $\pm$ 0.09 & 5.15 $\pm$ 0.30 & 11.08 $\pm$ 0.69 & 0.38 $\pm$ 0.04 & 0.37 $\pm$ 0.01  & 21 & 79\\
HD   196794  &  57 &  -37 &  -14 &  -52.78 $\pm$ 0.07 & 5.58 $\pm$ 0.33 &  9.06 $\pm$ 0.54 & 0.28 $\pm$ 0.04 & 0.24 $\pm$ 0.01  & 66 & 34\\
HD   197921  &  64 &  -42 &    3 &  -38.57 $\pm$ 0.13 & 5.29 $\pm$ 0.31 &  9.16 $\pm$ 0.55 & 0.06 $\pm$ 0.02 & 0.27 $\pm$ 0.01  & 60 & 40\\
HD   213996  & -90 &  -38 &    5 &  -17.99 $\pm$ 0.12 & 4.99 $\pm$ 0.28 & 10.62 $\pm$ 0.63 & 0.05 $\pm$ 0.02 & 0.36 $\pm$ 0.01  & 65 & 35\\
HD   214463  &  10 &  -56 &  -17 &    3.26 $\pm$ 0.25 & 4.87 $\pm$ 0.29 &  8.49 $\pm$ 0.50 & 0.31 $\pm$ 0.04 & 0.27 $\pm$ 0.01  & 34 & 66\\

\hline 
\end{longtable}
\normalsize

\scriptsize
\begin{longtable}{cccccrrcc}

\caption{\label{Tab_final}  Photometric and adopted spectroscopic stellar parameters} \\

\hline\hline

Star & $T_{\rm eff}$ & $\log g$ & $\xi$ & [\ion{Fe}{I}/\ion{H}]$_{Gen}$ &  \multicolumn{1}{c}{[\ion{Fe}{I}/\ion{H}]} &  \multicolumn{1}{c}{[\ion{Fe}{II}/H]} & Mass & Age \\ 
     & K & dex & km s$^{-1}$ & dex & \multicolumn{1}{c}{dex} & \multicolumn{1}{c}{dex} & ($M_{\odot}$) & (Gyr)\\
(1) & (2) & (3) & (4) & (5) & \multicolumn{1}{c}{(6)} & \multicolumn{1}{c}{(7)}   \\

\hline
\endfirsthead
\caption{continued.}\\
\hline\hline

Star & $T_{\rm eff}$ & $\log g$ & $\xi$ & [\ion{Fe}{I}/\ion{H}]$_{Gen}$ &  \multicolumn{1}{c}{[\ion{Fe}{I}/\ion{H}]} &  \multicolumn{1}{c}{[\ion{Fe}{II}/H]} & Mass & Age \\ 
     & K & dex & km s$^{-1}$ & dex & \multicolumn{1}{c}{dex} & \multicolumn{1}{c}{dex} & ($M_{\odot}$) & (Gyr)\\
(1) & (2) & (3) & (4) & (5) & \multicolumn{1}{c}{(6)} & \multicolumn{1}{c}{(7)}   \\

\hline
\endhead
\hline
 \\

 {\bf Notes.} & \multicolumn{8}{l}{$^{(a)}$ Spectroscopic gravity. Other $\log g$s were derived using HIPPARCOS parallaxes.}\\
\endfoot
 
\hline

\\
\multicolumn{9}{c}{Thin disk} \\
HD    11608  & 4966 $\pm$ 47 & 4.57 $\pm$ 0.06 &  0.30 $\pm$ 0.31  & 0.60  &   0.39 $\pm$ 0.05 &  0.42 $\pm$ 0.13  &  0.88 $\pm$ 0.06 &   5.6 $\pm$ 2.9 \\   
HD    26151  & 5383 $\pm$ 47 & 4.41 $\pm$ 0.05 &  0.67 $\pm$ 0.03  & 0.42  &   0.33 $\pm$ 0.05 &  0.34 $\pm$ 0.08  &  0.97 $\pm$ 0.07 &   8.2 $\pm$ 2.4 \\   
HD    26794  & 4920 $\pm$ 52 & 4.49 $\pm$ 0.06 &  0.30 $\pm$ 0.12  & 0.30  &   0.07 $\pm$ 0.04 &  0.10 $\pm$ 0.12  &  0.78 $\pm$ 0.05 &  11.8 $\pm$ 2.2 \\   
HD    35854  & 4901 $\pm$ 37 & 4.57 $\pm$ 0.03 &  0.30 $\pm$ 0.09  & 0.33  &  -0.04 $\pm$ 0.03 & -0.00 $\pm$ 0.12  &  0.76 $\pm$ 0.05 &   9.0 $\pm$ 0.8 \\   
HD    77338  & 5346 $\pm$ 42 & 4.55 $\pm$ 0.05 &  0.44 $\pm$ 0.08  & 0.43  &   0.41 $\pm$ 0.05 &  0.49 $\pm$ 0.08  &  0.98 $\pm$ 0.07 &   2.0 $\pm$ 1.2 \\   
HD    82943  & 5929 $\pm$ 45 & 4.35 $\pm$ 0.04 &  1.22 $\pm$ 0.03  & 0.34  &   0.23 $\pm$ 0.05 &  0.26 $\pm$ 0.07  &  1.14 $\pm$ 0.08 &   3.0 $\pm$ 0.7 \\   
HD    86249  & 4957 $\pm$ 42 & 4.59 $\pm$ 0.05 &  0.30 $\pm$ 0.11  & 0.32  &   0.12 $\pm$ 0.04 &  0.20 $\pm$ 0.10  &  0.82 $\pm$ 0.06 &   7.3 $\pm$ 2.1 \\   
HD    93800  & 5181 $\pm$ 43 & 4.44 $\pm$ 0.07 &  0.30 $\pm$ 0.09  & 0.43  &   0.49 $\pm$ 0.04 &  0.57 $\pm$ 0.09  &  0.92 $\pm$ 0.07 &   8.8 $\pm$ 3.5 \\   
HD   177374  & 5044 $\pm$ 45 & 4.38 $\pm$ 0.08 &  0.30 $\pm$ 0.10  & 0.40  &  -0.08 $\pm$ 0.04 &  0.12 $\pm$ 0.10  &  0.79 $\pm$ 0.05 &  13.7 $\pm$ 1.0 \\   
HD   181433  & 4902 $\pm$ 41 & 4.57 $\pm$ 0.04 &  0.30 $\pm$ 0.21  & 0.65  &   0.41 $\pm$ 0.04 &  0.52 $\pm$ 0.12  &  0.86 $\pm$ 0.06 &   6.7 $\pm$ 1.8 \\   
HD   224230  & 4873 $\pm$ 54 & 4.58 $\pm$ 0.06 &  0.30 $\pm$ 0.17  & 0.34  &  -0.08 $\pm$ 0.04 &  0.11 $\pm$ 0.13  &  0.75 $\pm$ 0.05 &   9.4 $\pm$ 2.4 \\   

\\
\multicolumn{9}{c}{Thick disk} \\
G 161-029    & 4798 $\pm$ 34 & 4.60$^a\pm$0.04 &  0.30 $\pm$ 0.63  & 0.42  &   0.01 $\pm$ 0.09 &  0.12 $\pm$ 0.13  &         ...      &        ...      \\
BD-02   180  & 5004 $\pm$ 57 & 4.46 $\pm$ 0.10 &  0.30 $\pm$ 0.17  & 0.45  &   0.33 $\pm$ 0.04 &  0.45 $\pm$ 0.08  &  0.86 $\pm$ 0.06 &  10.3 $\pm$ 4.0 \\   
BD-05  5798  & 4902 $\pm$ 65 & 4.44 $\pm$ 0.13 &  0.64 $\pm$ 0.05  & 0.57  &   0.20 $\pm$ 0.05 &  0.27 $\pm$ 0.13  &  0.82 $\pm$ 0.06 &  11.4 $\pm$ 3.7 \\   
BD-17  6035  & 4892 $\pm$ 73 & 4.43 $\pm$ 0.15 &  0.85 $\pm$ 0.03  & 0.60  &   0.09 $\pm$ 0.05 &  0.17 $\pm$ 0.14  &  0.79 $\pm$ 0.06 &  11.5 $\pm$ 3.2 \\   
CD-32  0327  & 4957 $\pm$ 66 & 4.52 $\pm$ 0.11 &  0.34 $\pm$ 0.06  & 0.49  &  -0.01 $\pm$ 0.06 &  0.19 $\pm$ 0.10  &  0.79 $\pm$ 0.05 &  10.3 $\pm$ 3.5 \\   
CD-40 15036  & 5429 $\pm$ 66 & 4.41 $\pm$ 0.14 &  0.58 $\pm$ 0.04  & 0.07  &  -0.03 $\pm$ 0.07 & -0.03 $\pm$ 0.09  &  0.89 $\pm$ 0.07 &   9.8 $\pm$ 4.6 \\   
HD     9424  & 5449 $\pm$ 48 & 4.48 $\pm$ 0.06 &  0.66 $\pm$ 0.03  & 0.12  &   0.12 $\pm$ 0.06 &  0.10 $\pm$ 0.08  &  0.95 $\pm$ 0.07 &   6.1 $\pm$ 3.2 \\   
HD    10576  & 5929 $\pm$ 67 & 4.14 $\pm$ 0.06 &  1.36 $\pm$ 0.06  & 0.11  &   0.02 $\pm$ 0.07 &  0.02 $\pm$ 0.08  &  1.12 $\pm$ 0.08 &   5.8 $\pm$ 0.9 \\   
HD    13386  & 5269 $\pm$ 43 & 4.54 $\pm$ 0.05 &  0.46 $\pm$ 0.04  & 0.35  &   0.36 $\pm$ 0.05 &  0.34 $\pm$ 0.09  &  0.96 $\pm$ 0.07 &   3.6 $\pm$ 2.4 \\   
HD    15133  & 5223 $\pm$ 46 & 4.47 $\pm$ 0.08 &  0.30 $\pm$ 0.05  & 0.52  &   0.46 $\pm$ 0.04 &  0.51 $\pm$ 0.09  &  0.94 $\pm$ 0.07 &   6.9 $\pm$ 3.7 \\   
HD    15555  & 4867 $\pm$ 40 & 3.69 $\pm$ 0.06 &  1.04 $\pm$ 0.03  & 0.32  &   0.37 $\pm$ 0.05 &  0.38 $\pm$ 0.11  &  1.22 $\pm$ 0.10 &   6.3 $\pm$ 0.9 \\   
HD    16905  & 4866 $\pm$ 42 & 4.58 $\pm$ 0.05 &  0.30 $\pm$ 0.14  & 0.53  &   0.27 $\pm$ 0.04 &  0.44 $\pm$ 0.12  &  0.83 $\pm$ 0.06 &   7.4 $\pm$ 1.8 \\   
HD    25061  & 5307 $\pm$ 49 & 4.49 $\pm$ 0.05 &  0.77 $\pm$ 0.03  & 0.40  &   0.18 $\pm$ 0.05 &  0.19 $\pm$ 0.09  &  0.92 $\pm$ 0.06 &   6.7 $\pm$ 3.3 \\   
HD    27894  & 4920 $\pm$ 45 & 4.54 $\pm$ 0.05 &  0.30 $\pm$ 0.12  & 0.50  &   0.37 $\pm$ 0.04 &  0.51 $\pm$ 0.10  &  0.86 $\pm$ 0.06 &   7.7 $\pm$ 2.3 \\   
HD    31827  & 5608 $\pm$ 49 & 4.35 $\pm$ 0.04 &  0.82 $\pm$ 0.03  & 0.40  &   0.48 $\pm$ 0.05 &  0.49 $\pm$ 0.08  &  1.08 $\pm$ 0.07 &   4.8 $\pm$ 0.9 \\   
HD    39213  & 5473 $\pm$ 48 & 4.38 $\pm$ 0.06 &  0.74 $\pm$ 0.04  & 0.39  &   0.45 $\pm$ 0.06 &  0.44 $\pm$ 0.08  &  1.02 $\pm$ 0.07 &   6.3 $\pm$ 2.0 \\   
HD    81767  & 4966 $\pm$ 52 & 4.49 $\pm$ 0.07 &  0.30 $\pm$ 0.11  & 0.49  &   0.22 $\pm$ 0.05 &  0.25 $\pm$ 0.13  &  0.83 $\pm$ 0.05 &  10.4 $\pm$ 3.0 \\   
HD    90054  & 6047 $\pm$ 52 & 4.18 $\pm$ 0.06 &  1.52 $\pm$ 0.05  & 0.39  &   0.29 $\pm$ 0.06 &  0.33 $\pm$ 0.07  &  1.28 $\pm$ 0.08 &   3.2 $\pm$ 0.4 \\   
HD    94374  & 5000 $\pm$ 38 & 4.63 $\pm$ 0.05 &  0.30 $\pm$ 0.08  & 0.28  &  -0.10 $\pm$ 0.03 & -0.18 $\pm$ 0.13  &  0.79 $\pm$ 0.05 &   4.8 $\pm$ 2.4 \\   
HD    95338  & 5175 $\pm$ 42 & 4.52 $\pm$ 0.04 &  0.43 $\pm$ 0.05  & 0.34  &   0.21 $\pm$ 0.04 &  0.22 $\pm$ 0.09  &  0.90 $\pm$ 0.06 &   6.3 $\pm$ 2.6 \\   
HD   104212  & 5833 $\pm$ 53 & 4.07 $\pm$ 0.09 &  1.21 $\pm$ 0.03  & 0.09  &   0.13 $\pm$ 0.06 &  0.07 $\pm$ 0.07  &  1.14 $\pm$ 0.08 &   5.9 $\pm$ 0.8 \\   
HD   107509  & 6102 $\pm$ 60 & 4.18 $\pm$ 0.06 &  1.80 $\pm$ 0.06  & 0.08  &   0.03 $\pm$ 0.06 & -0.06 $\pm$ 0.08  &  1.21 $\pm$ 0.09 &   3.9 $\pm$ 0.5 \\   
HD   120329  & 5617 $\pm$ 48 & 4.14 $\pm$ 0.07 &  1.22 $\pm$ 0.02  & 0.27  &   0.31 $\pm$ 0.06 &  0.39 $\pm$ 0.07  &  1.11 $\pm$ 0.08 &   7.3 $\pm$ 1.0 \\   
HD   143102  & 5547 $\pm$ 48 & 3.94 $\pm$ 0.07 &  1.13 $\pm$ 0.03  & 0.01  &   0.16 $\pm$ 0.06 &  0.21 $\pm$ 0.08  &  1.16 $\pm$ 0.08 &   6.8 $\pm$ 1.0 \\   
HD   148530  & 5392 $\pm$ 53 & 4.49 $\pm$ 0.06 &  0.72 $\pm$ 0.04  & 0.11  &   0.03 $\pm$ 0.06 &  0.02 $\pm$ 0.09  &  0.90 $\pm$ 0.08 &   7.1 $\pm$ 3.6 \\   
HD   149256  & 5406 $\pm$ 47 & 4.01 $\pm$ 0.08 &  1.15 $\pm$ 0.02  & 0.27  &   0.34 $\pm$ 0.06 &  0.38 $\pm$ 0.08  &  1.08 $\pm$ 0.08 &   8.9 $\pm$ 1.5 \\   
HD   149606  & 4976 $\pm$ 50 & 4.63 $\pm$ 0.05 &  0.30 $\pm$ 0.08  & 0.35  &   0.20 $\pm$ 0.05 &  0.41 $\pm$ 0.11  &  0.85 $\pm$ 0.10 &   4.1 $\pm$ 2.6 \\   
HD   149933  & 5486 $\pm$ 49 & 4.44 $\pm$ 0.05 &  1.11 $\pm$ 0.02  & 0.31  &   0.13 $\pm$ 0.06 &  0.17 $\pm$ 0.08  &  0.95 $\pm$ 0.06 &   7.2 $\pm$ 2.9 \\   
HD   165920  & 5336 $\pm$ 44 & 4.47 $\pm$ 0.04 &  0.64 $\pm$ 0.03  & 0.40  &   0.36 $\pm$ 0.05 &  0.40 $\pm$ 0.08  &  0.97 $\pm$ 0.07 &   5.2 $\pm$ 2.5 \\   
HD   168714  & 5686 $\pm$ 48 & 4.30 $\pm$ 0.11 &  0.79 $\pm$ 0.03  & 0.47  &   0.48 $\pm$ 0.06 &  0.47 $\pm$ 0.08  &  1.12 $\pm$ 0.07 &   4.0 $\pm$ 1.1 \\   
HD   171999  & 5304 $\pm$ 45 & 4.49 $\pm$ 0.05 &  0.58 $\pm$ 0.03  & 0.33  &   0.29 $\pm$ 0.05 &  0.35 $\pm$ 0.08  &  0.96 $\pm$ 0.09 &   6.0 $\pm$ 2.8 \\   
HD   179764  & 5323 $\pm$ 48 & 4.28 $\pm$ 0.08 &  0.92 $\pm$ 0.04  & 0.16  &  -0.05 $\pm$ 0.05 &  0.03 $\pm$ 0.09  &  0.85 $\pm$ 0.06 &  14.6 $\pm$ 0.6 \\   
HD   180865  & 5218 $\pm$ 44 & 4.53 $\pm$ 0.06 &  0.48 $\pm$ 0.06  & 0.31  &   0.21 $\pm$ 0.05 &  0.27 $\pm$ 0.09  &  0.91 $\pm$ 0.06 &   5.9 $\pm$ 3.1 \\   
HD   181234  & 5311 $\pm$ 45 & 4.37 $\pm$ 0.06 &  0.30 $\pm$ 0.07  & 0.30  &   0.45 $\pm$ 0.04 &  0.52 $\pm$ 0.09  &  0.96 $\pm$ 0.06 &   9.4 $\pm$ 2.2 \\   
HD   196397  & 5404 $\pm$ 54 & 4.49 $\pm$ 0.08 &  0.55 $\pm$ 0.07  & 0.36  &   0.38 $\pm$ 0.06 &  0.43 $\pm$ 0.09  &  0.99 $\pm$ 0.07 &   3.7 $\pm$ 2.8 \\   
HD   201237  & 4829 $\pm$ 82 & 4.14 $\pm$ 0.16 &  0.50 $\pm$ 0.05  & 0.50  &   0.00 $\pm$ 0.04 & -0.06 $\pm$ 0.18  &  0.77 $\pm$ 0.05 &  13.8 $\pm$ 0.9 \\   
HD   209721  & 5503 $\pm$ 51 & 4.30 $\pm$ 0.10 &  1.05 $\pm$ 0.03  & 0.35  &   0.28 $\pm$ 0.05 &  0.23 $\pm$ 0.08  &  1.00 $\pm$ 0.08 &   8.8 $\pm$ 2.1 \\   
HD   211706  & 6017 $\pm$ 69 & 4.33 $\pm$ 0.11 &  1.44 $\pm$ 0.08  & 0.09  &   0.09 $\pm$ 0.07 &  0.07 $\pm$ 0.08  &  1.14 $\pm$ 0.08 &   3.0 $\pm$ 1.5 \\   
HD   218566  & 4849 $\pm$ 42 & 4.48 $\pm$ 0.04 &  0.30 $\pm$ 0.68  & 0.46  &   0.28 $\pm$ 0.14 &  0.43 $\pm$ 0.22  &  0.81 $\pm$ 0.06 &  11.5 $\pm$ 1.9 \\   
HD   218750  & 5134 $\pm$ 49 & 4.41 $\pm$ 0.08 &  0.37 $\pm$ 0.08  & 0.31  &   0.17 $\pm$ 0.05 &  0.23 $\pm$ 0.09  &  0.85 $\pm$ 0.06 &  12.8 $\pm$ 2.4 \\   
HD   221313  & 5153 $\pm$ 55 & 4.36 $\pm$ 0.13 &  0.62 $\pm$ 0.04  & 0.50  &   0.31 $\pm$ 0.05 &  0.34 $\pm$ 0.10  &  0.90 $\pm$ 0.06 &  12.0 $\pm$ 3.6 \\   
HD   221974  & 5213 $\pm$ 52 & 4.60 $\pm$ 0.07 &  0.30 $\pm$ 0.11  & 0.49  &   0.46 $\pm$ 0.05 &  0.56 $\pm$ 0.11  &  0.95 $\pm$ 0.07 &   2.2 $\pm$ 1.8 \\   
HD   224383  & 5760 $\pm$ 53 & 4.28 $\pm$ 0.06 &  1.16 $\pm$ 0.09  & 0.00  &  -0.10 $\pm$ 0.06 & -0.15 $\pm$ 0.08  &  0.98 $\pm$ 0.07 &   9.0 $\pm$ 1.3 \\ 
\\
\multicolumn{9}{c}{Intermediate population} \\
HD     8389  & 5274 $\pm$ 42 & 4.47 $\pm$ 0.04 &  0.33 $\pm$ 0.12  & 0.47  &   0.58 $\pm$ 0.04 &  0.58 $\pm$ 0.08  &  0.96 $\pm$ 0.07 &   5.4 $\pm$ 1.8 \\   
HD     9174  & 5599 $\pm$ 55 & 4.15 $\pm$ 0.08 &  0.95 $\pm$ 0.02  & 0.36  &   0.41 $\pm$ 0.07 &  0.35 $\pm$ 0.08  &  1.13 $\pm$ 0.08 &   6.7 $\pm$ 1.0 \\   
HD    12789  & 5810 $\pm$ 47 & 4.21 $\pm$ 0.09 &  1.15 $\pm$ 0.03  & 0.31  &   0.27 $\pm$ 0.06 &  0.36 $\pm$ 0.07  &  1.17 $\pm$ 0.08 &   4.9 $\pm$ 0.9 \\   
HD    30295  & 5406 $\pm$ 45 & 4.36 $\pm$ 0.05 &  0.72 $\pm$ 0.03  & 0.41  &   0.32 $\pm$ 0.05 &  0.33 $\pm$ 0.08  &  0.97 $\pm$ 0.07 &   9.1 $\pm$ 1.8 \\   
HD    31452  & 5250 $\pm$ 45 & 4.44 $\pm$ 0.05 &  0.61 $\pm$ 0.05  & 0.30  &   0.23 $\pm$ 0.05 &  0.19 $\pm$ 0.09  &  0.90 $\pm$ 0.06 &  10.0 $\pm$ 3.1 \\   
HD    37986  & 5503 $\pm$ 44 & 4.47 $\pm$ 0.04 &  0.95 $\pm$ 0.02  & 0.47  &   0.30 $\pm$ 0.05 &  0.33 $\pm$ 0.08  &  1.02 $\pm$ 0.07 &   4.0 $\pm$ 1.6 \\   
HD    39715  & 4741 $\pm$ 63 & 4.57 $\pm$ 0.04 &  0.30 $\pm$ 0.15  & 0.33  &  -0.10 $\pm$ 0.03 &  0.07 $\pm$ 0.15  &  0.72 $\pm$ 0.05 &   9.6 $\pm$ 1.6 \\   
HD    43848  & 5161 $\pm$ 41 & 4.54 $\pm$ 0.04 &  0.30 $\pm$ 0.08  & 0.52  &   0.43 $\pm$ 0.04 &  0.46 $\pm$ 0.09  &  0.94 $\pm$ 0.06 &   3.7 $\pm$ 1.7 \\   
HD    86065  & 4938 $\pm$ 48 & 4.62 $\pm$ 0.05 &  0.30 $\pm$ 0.10  & 0.36  &   0.09 $\pm$ 0.04 &  0.21 $\pm$ 0.12  &  0.82 $\pm$ 0.06 &   6.0 $\pm$ 2.6 \\   
HD    87007  & 5282 $\pm$ 59 & 4.54 $\pm$ 0.06 &  0.61 $\pm$ 0.04  & 0.44  &   0.29 $\pm$ 0.06 &  0.45 $\pm$ 0.09  &  0.95 $\pm$ 0.06 &   4.0 $\pm$ 2.9 \\   
HD    91585  & 5144 $\pm$ 50 & 4.55 $\pm$ 0.07 &  0.48 $\pm$ 0.07  & 0.30  &   0.25 $\pm$ 0.05 &  0.33 $\pm$ 0.10  &  0.91 $\pm$ 0.09 &   5.4 $\pm$ 3.5 \\   
HD    91669  & 5278 $\pm$ 57 & 4.34 $\pm$ 0.11 &  0.64 $\pm$ 0.04  & 0.42  &   0.44 $\pm$ 0.05 &  0.45 $\pm$ 0.09  &  0.95 $\pm$ 0.06 &  10.1 $\pm$ 3.4 \\   
HD   182572  & 5700 $\pm$ 32 & 4.18 $\pm$ 0.03 &  1.00 $\pm$ 0.02  & 0.31  &   0.48 $\pm$ 0.04 &  0.39 $\pm$ 0.07  &  1.16 $\pm$ 0.07 &   4.5 $\pm$ 0.2 \\   
HD   196794  & 5094 $\pm$ 44 & 4.64 $\pm$ 0.04 &  0.30 $\pm$ 0.05  & 0.33  &   0.06 $\pm$ 0.05 &  0.16 $\pm$ 0.09  &  0.84 $\pm$ 0.06 &   3.7 $\pm$ 2.0 \\   
HD   197921  & 4866 $\pm$ 45 & 4.48 $\pm$ 0.06 &  0.30 $\pm$ 0.15  & 0.39  &   0.22 $\pm$ 0.04 &  0.33 $\pm$ 0.10  &  0.80 $\pm$ 0.05 &  11.3 $\pm$ 2.4 \\   
HD   213996  & 5314 $\pm$ 53 & 4.49 $\pm$ 0.05 &  0.75 $\pm$ 0.04  & 0.43  &   0.33 $\pm$ 0.05 &  0.35 $\pm$ 0.10  &  0.96 $\pm$ 0.09 &   5.1 $\pm$ 3.0 \\   
HD   214463  & 5122 $\pm$ 47 & 4.40 $\pm$ 0.11 &  0.80 $\pm$ 0.04  & 0.33  &   0.34 $\pm$ 0.05 &  0.30 $\pm$ 0.09  &  0.90 $\pm$ 0.06 &  11.8 $\pm$ 3.7 \\

\hline
\end{longtable}


\scriptsize
\begin{longtable}{ccccccccc}

\caption{\label{Tab_abonds1}  Final abundances} \\

\hline\hline

Star & [C/H] & [C/Fe] & [Ni/H] & [Ni/Fe] & [O/H] & [O/Fe] & [Mg/H] & [Mg/Fe] \\

\hline
\endfirsthead
\caption{continued.}\\
\hline\hline

Star & [C/H] & [C/Fe] & [Ni/H] & [Ni/Fe] & [O/H] & [O/Fe] & [Mg/H] & [Mg/Fe] \\

\hline
\endhead
\hline
\\
\multicolumn{9}{c}{Thin disk} \\
  HD	11608 &  0.26 & -0.13 &  0.52 $\pm$ 0.06 &  0.13  &   0.24 &  -0.15 &	0.36 & -0.03  \\
  HD	26151 &  0.24 & -0.09 &  0.42 $\pm$ 0.05 &  0.09  &   ...  &   ...  &	0.31 & -0.02  \\
  HD	26794 &  0.07 & -0.00 &  0.10 $\pm$ 0.03 &  0.03  &   0.25 &   0.18 &	0.15 &  0.08  \\
  HD	35854 & -0.00 &  0.04 & -0.03 $\pm$ 0.03 &  0.01  &   0.05 &   0.09 &  -0.07 & -0.03  \\
  HD	77338 &  0.44 &  0.03 &  0.52 $\pm$ 0.04 &  0.11  &   ...  &   ...  &	0.33 & -0.08  \\
  HD	82943 &  0.18 & -0.05 &  0.25 $\pm$ 0.06 &  0.02  &   0.45 &   0.22 &	0.16 & -0.07  \\
  HD	86249 &  0.15 &  0.03 &  0.12 $\pm$ 0.03 &  0.00  &   ...  &   ...  &	0.02 & -0.10  \\
  HD	93800 &  0.44 & -0.05 &  0.64 $\pm$ 0.04 &  0.15  &  -0.03 &  -0.52 &	0.46 & -0.03  \\
  HD   177374 &  0.32 &  0.40 & -0.01 $\pm$ 0.03 &  0.07  &   ...  &   ...   &	0.19 &  0.27  \\
  HD   181433 &  ...  &  ...  &  0.56 $\pm$ 0.04 &  0.15  &   0.42 &   0.01 &	0.44 &  0.03  \\
  HD   224230 &  0.10 &  0.18 & -0.05 $\pm$ 0.03 &  0.02  &  -0.01 &   0.07 &  -0.06 &  0.03  \\

\\
\multicolumn{9}{c}{Thick disk} \\
  G 161-029   &  0.04 &  0.03 &  0.10 $\pm$ 0.09 &  0.08  &   ...  &   ...  &	0.10 &  0.09  \\
  BD-02   180 &  0.35 &  0.02 &  0.41 $\pm$ 0.04 &  0.08  &   0.60 &   0.27 &	0.30 & -0.03  \\
  BD-05  5798 &  0.28 &  0.08 &  0.19 $\pm$ 0.04 & -0.01  &   ...  &   ...  &	0.25 &  0.05  \\
  BD-17  6035 &  0.31 &  0.22 &  0.12 $\pm$ 0.04 &  0.03  &   ...  &   ...  &	0.26 &  0.17  \\
  CD-32  0327 &  ...  &  ...  &  0.09 $\pm$ 0.04 &  0.11  &   ...  &   ...  &	0.11 &  0.19  \\
  CD-40 15036 & -0.01 &  0.02 & -0.02 $\pm$ 0.07 &  0.00  &   ...  &   ...  &  -0.14 & -0.01  \\
  HD	 9424 &  0.17 &  0.05 &  0.17 $\pm$ 0.05 &  0.05  &   0.09 &  -0.03 &	0.10 & -0.02  \\
  HD	10576 &  ...  &  ...  & -0.01 $\pm$ 0.07 & -0.03  &   0.25 &   0.23 &	0.00 & -0.02  \\
  HD	13386 &  0.25 & -0.11 &  0.41 $\pm$ 0.04 &  0.05  &   0.07 &  -0.29 &	0.30 & -0.16  \\
  HD	15133 &  0.25 & -0.21 &  0.59 $\pm$ 0.04 &  0.13  &   0.27 &  -0.19 &	0.37 & -0.09  \\
  HD	15555 & -0.11 & -0.48 &  0.51 $\pm$ 0.04 &  0.14  &   0.35 &  -0.02 &	0.33 & -0.04  \\
  HD	16905 &  0.23 & -0.04 &  0.38 $\pm$ 0.04 &  0.11  &   0.28 &   0.01 &	0.26 & -0.01  \\
  HD	25061 &  0.13 & -0.05 &  0.23 $\pm$ 0.04 &  0.05  &   ...  &   ...  &	0.12 & -0.06  \\
  HD	27894 &  0.25 & -0.12 &  0.43 $\pm$ 0.04 &  0.06  &   0.32 &  -0.05 &	0.34 & -0.03  \\
  HD	31827 &  0.46 & -0.02 &  0.59 $\pm$ 0.05 &  0.11  &   0.02 &  -0.46 &	0.47 & -0.01  \\
  HD	39213 &  ...  &  ...  &  0.59 $\pm$ 0.05 &  0.14  &   ...  &   ...  &	0.39 & -0.06  \\
  HD	81767 & -0.01 & -0.23 &  0.22 $\pm$ 0.03 &  0.00  &   0.28 &   0.06 &	0.16 & -0.06  \\
  HD	90054 &  0.31 &  0.02 &  0.31 $\pm$ 0.06 &  0.02  &   0.17 &  -0.12 &	0.22 & -0.07  \\
  HD	94374 &  0.25 &  0.35 & -0.08 $\pm$ 0.03 &  0.02  &   ...  &   ...  &  -0.04 &  0.06  \\
  HD	95338 &  0.06 & -0.15 &  0.25 $\pm$ 0.04 &  0.04  &   0.06 &  -0.15 &	0.18 & -0.03  \\
  HD   104212 &  0.11 & -0.02 &  0.13 $\pm$ 0.07 &  0.00  &   0.35 &   0.22 &	0.07 & -0.06  \\
  HD   107509 & -0.05 & -0.08 &  0.00 $\pm$ 0.07 & -0.02  &   0.13 &   0.10 &	0.01 &  0.05  \\
  HD   120329 &  0.39 &  0.08 &  0.33 $\pm$ 0.06 &  0.02  &   ...  &   ...  &	0.29 & -0.02  \\
  HD   143102 &   ... &  ...  &  0.17 $\pm$ 0.06 &  0.01  &   0.17 &   0.01 &	0.10 & -0.06  \\
  HD   148530 &  0.06 &  0.03 &  0.04 $\pm$ 0.05 &  0.01  &   0.32 &   0.29 &	0.07 &  0.04  \\
  HD   149256 &  ...  &  ...  &  0.40 $\pm$ 0.05 &  0.06  &   0.61 &   0.27 &	0.37 &  0.03  \\
  HD   149606 &  0.24 &  0.04 &  0.22 $\pm$ 0.04 &  0.02  &   0.40 &   0.20 &	0.17 & -0.03  \\
  HD   149933 &  0.21 &  0.08 &  0.23 $\pm$ 0.05 &  0.10  &   0.10 &  -0.03 &	0.21 &  0.05  \\
  HD   165920 &  0.34 & -0.02 &  0.42 $\pm$ 0.04 &  0.06  &   0.17 &  -0.19 &	0.29 & -0.07  \\
  HD   168714 &  0.35 & -0.13 &  0.62 $\pm$ 0.05 &  0.14  &   ...  &   ...  &	0.44 & -0.04  \\
  HD   171999 &  0.30 &  0.01 &  0.33 $\pm$ 0.04 &  0.04  &   0.27 &  -0.02 &	0.24 & -0.05  \\
  HD   179764 &  ...  &  ...  & -0.02 $\pm$ 0.05 &  0.03  &   0.04 &   0.09 &	0.02 &  0.07  \\
  HD   180865 &  0.20 & -0.01 &  0.28 $\pm$ 0.04 &  0.07  &   0.39 &   0.18 &	0.26 &  0.05  \\
  HD   181234 &  0.30 & -0.15 &  0.55 $\pm$ 0.04 &  0.10  &   0.37 &  -0.08 &	0.44 & -0.01  \\
  HD   196397 &  0.37 & -0.01 &  0.46 $\pm$ 0.05 &  0.08  &    ... &   ...  &	0.32 & -0.06  \\
  HD   201237 & -0.19 & -0.19 &  0.03 $\pm$ 0.04 &  0.03  &   0.12 &   0.12 &	0.01 &  0.01  \\
  HD   209721 &  0.35 &  0.07 &  0.36 $\pm$ 0.05 &  0.08  &   0.17 &  -0.11 &	0.32 &  0.04  \\
  HD   211706 &  0.09 &  0.00 &  0.11 $\pm$ 0.08 &  0.02  &   ...  &   ...  &	0.02 & -0.07  \\
  HD   218566 &  0.21 & -0.07 &  0.39 $\pm$ 0.15 &  0.11  &   0.17 &  -0.11 &	0.12 &  0.10  \\
  HD   218750 &  0.22 &  0.05 &  0.21 $\pm$ 0.04 &  0.04  &   ...  &   ...  &	0.24 &  0.07  \\
  HD   221313 &  ...  &  ...  &  0.41 $\pm$ 0.04 &  0.10  &   0.52 &   0.21 &	0.33 &  0.02  \\
  HD   221974 &  0.39 & -0.07 &  0.59 $\pm$ 0.04 &  0.13  &   0.40 &  -0.06 &	0.37 & -0.09  \\
  HD   224383 & -0.03 &  0.07 & -0.10 $\pm$ 0.06 & -0.01  &   ...  &   ...  &  -0.04 &  0.06  \\ 
\\
\multicolumn{9}{c}{Intermediate population} \\
  HD	 8389 &  0.46 & -0.12 &  0.71 $\pm$ 0.04 &  0.22  &   0.35 &  -0.23 &	0.45 & -0.13  \\
  HD	 9174 &  ...  &  ...  &  0.51 $\pm$ 0.06 &  0.10  &   0.55 &   0.14 &	0.33 & -0.08  \\
  HD	12789 &  ...  &  ...  &  0.34 $\pm$ 0.06 &  0.07  &   0.44 &   0.17 &	0.14 & -0.13  \\
  HD	30295 &  0.33 &  0.01 &  0.37 $\pm$ 0.04 &  0.05  &   0.34 &   0.02 &	0.33 &  0.01  \\
  HD	31452 &  ...  &  ...  &  0.25 $\pm$ 0.04 &  0.02  &   ...  &   ...  &	0.15 & -0.08  \\
  HD	37986 &  0.34 &  0.04 &  0.35 $\pm$ 0.04 &  0.05  &   0.22 &  -0.08 &	0.26 & -0.04  \\
  HD	39715 &  ...  &  ...  & -0.10 $\pm$ 0.03 &  0.01  &   ...  &   ...  &  -0.13 & -0.03  \\
  HD	43848 &  0.29 & -0.14 &  0.53 $\pm$ 0.04 &  0.10  &   0.36 &  -0.07 &	0.38 & -0.15  \\
  HD	86065 &  0.11 &  0.02 &  0.13 $\pm$ 0.04 &  0.04  &   0.17 &   0.08 &	0.09 & -0.00  \\
  HD	87007 &  0.23 & -0.06 &  0.39 $\pm$ 0.04 &  0.10  &   0.14 &  -0.15 &	0.31 &  0.02  \\
  HD	91585 &  0.28 &  0.03 &  0.34 $\pm$ 0.04 &  0.09  &   0.44 &   0.19 &	0.33 &  0.08  \\
  HD	91669 &  0.32 & -0.12 &  0.54 $\pm$ 0.04 &  0.10  &   0.31 &  -0.13 &	0.38 & -0.06  \\
  HD   182572 &  0.29 & -0.19 &  0.52 $\pm$ 0.04 &  0.04  &   0.44 &  -0.04 &	0.44 & -0.04  \\
  HD   196794 & -0.02 & -0.08 &  0.08 $\pm$ 0.04 &  0.02  &   0.14 &   0.08 &	0.00 & -0.06  \\
  HD   197921 &  0.12 & -0.10 &  0.30 $\pm$ 0.04 &  0.08  &   0.27 &   0.05 &	0.29 &  0.07  \\
  HD   213996 &  0.27 & -0.06 &  0.39 $\pm$ 0.04 &  0.06  &   0.22 &  -0.11 &	0.30 & -0.03  \\
  HD   214463 &  0.22 & -0.12 &  0.43 $\pm$ 0.03 &  0.09  &   0.27 &  -0.07 &	0.28 & -0.06  \\

\hline
\end{longtable}

\newpage
\scriptsize
\centering
\begin{longtable}{ccccccc}

\caption{\label{Tab_abonds2}  Final abundances} \\

\hline\hline

Star &  [Ca/H] & [Ca/Fe] & [Si/H] & [Si/Fe] & [Ti/H] & [Ti/Fe] \\

\hline
\endfirsthead
\caption{continued.}\\
\hline\hline

Star &  [Ca/H] & [Ca/Fe] & [Si/H] & [Si/Fe] & [Ti/H] & [Ti/Fe] \\

\hline
\endhead
\hline
\\
\multicolumn{7}{c}{Thin disk} \\
HD    11608  &   0.33 $\pm$ 0.10 &  -0.06 &  0.40 $\pm$ 0.06 & 0.01 &  0.35  $\pm$ 0.09 & -0.04  \\
HD    26151  &   0.34 $\pm$ 0.04 &   0.01 &  0.35 $\pm$ 0.04 & 0.02 &  0.35  $\pm$ 0.09 &  0.02  \\
HD    26794  &   0.11 $\pm$ 0.09 &   0.04 &  0.11 $\pm$ 0.05 & 0.04 &  0.15  $\pm$ 0.11 &  0.08  \\
HD    35854  &  -0.05 $\pm$ 0.11 &  -0.01 & -0.04 $\pm$ 0.05 &-0.00 & -0.10  $\pm$ 0.08 & -0.06  \\
HD    77338  &   0.34 $\pm$ 0.06 &  -0.07 &  0.47 $\pm$ 0.05 & 0.06 &  0.32  $\pm$ 0.08 & -0.09  \\
HD    82943  &   0.30 $\pm$ 0.05 &   0.07 &  0.26 $\pm$ 0.05 & 0.03 &  0.20  $\pm$ 0.07 & -0.03  \\
HD    86249  &   0.11 $\pm$ 0.10 &  -0.01 &  0.10 $\pm$ 0.05 &-0.02 &  0.04  $\pm$ 0.09 & -0.08  \\
HD    93800  &   0.38 $\pm$ 0.10 &  -0.11 &  0.51 $\pm$ 0.05 & 0.02 &  0.42  $\pm$ 0.08 & -0.07  \\
HD   177374  &   0.04 $\pm$ 0.14 &   0.12 &  0.13 $\pm$ 0.05 & 0.21 & -0.24  $\pm$ 0.09 & -0.16  \\
HD   181433  &   0.30 $\pm$ 0.13 &  -0.11 &  0.52 $\pm$ 0.05 & 0.11 &  0.37  $\pm$ 0.08 & -0.04  \\
HD   224230  &  -0.18 $\pm$ 0.20 &  -0.10 & -0.14 $\pm$ 0.06 &-0.07 & -0.09  $\pm$ 0.11 & -0.01  \\
\\
\multicolumn{7}{c}{Thick disk} \\
G 161-029    &   0.00 $\pm$ 0.12 &  -0.01 &  0.05 $\pm$ 0.06 & 0.04 &  0.04  $\pm$ 0.12 &  0.02  \\
BD-02	180  &   0.35 $\pm$ 0.04 &   0.02 &  0.34 $\pm$ 0.06 & 0.01 &  0.41  $\pm$ 0.11 &  0.08  \\
BD-05  5798  &   0.16 $\pm$ 0.15 &  -0.04 &  0.14 $\pm$ 0.06 &-0.06 &  0.19  $\pm$ 0.13 & -0.01  \\
BD-17  6035  &   0.06 $\pm$ 0.18 &  -0.03 &  0.10 $\pm$ 0.06 & 0.01 &  0.05  $\pm$ 0.17 & -0.04  \\
CD-32  0327  &  -0.02 $\pm$ 0.19 &  -0.01 &  0.03 $\pm$ 0.05 & 0.04 & -0.01  $\pm$ 0.15 &  0.01  \\
CD-40 15036  &  -0.02 $\pm$ 0.11 &   0.01 & -0.03 $\pm$ 0.03 & 0.00 &  0.00  $\pm$ 0.12 &  0.03  \\
HD     9424  &   0.12 $\pm$ 0.10 &  -0.00 &  0.16 $\pm$ 0.04 & 0.04 &  0.20  $\pm$ 0.09 &  0.08  \\
HD    10576  &  -0.00 $\pm$ 0.11 &  -0.02 &  0.03 $\pm$ 0.05 & 0.00 &  0.04  $\pm$ 0.10 &  0.02  \\
HD    13386  &   0.34 $\pm$ 0.08 &  -0.02 &  0.32 $\pm$ 0.05 &-0.04 &  0.35  $\pm$ 0.09 & -0.01  \\
HD    15133  &   0.48 $\pm$ 0.10 &   0.02 &  0.44 $\pm$ 0.05 &-0.02 &  0.51  $\pm$ 0.09 &  0.05  \\
HD    15555  &   0.17 $\pm$ 0.08 &  -0.20 &  0.45 $\pm$ 0.05 & 0.08 &  0.27  $\pm$ 0.09 & -0.10  \\
HD    16905  &   0.24 $\pm$ 0.08 &  -0.03 &  0.36 $\pm$ 0.05 & 0.09 &  0.25  $\pm$ 0.09 & -0.02  \\
HD    25061  &   0.20 $\pm$ 0.09 &   0.02 &  0.17 $\pm$ 0.04 &-0.01 &  0.18  $\pm$ 0.09 & -0.00  \\
HD    27894  &   0.32 $\pm$ 0.10 &  -0.05 &  0.41 $\pm$ 0.05 & 0.04 &  0.40  $\pm$ 0.10 &  0.03  \\
HD    31827  &   0.45 $\pm$ 0.05 &  -0.03 &  0.55 $\pm$ 0.04 & 0.07 &  0.41  $\pm$ 0.09 & -0.07  \\
HD    39213  &   0.45 $\pm$ 0.05 &  -0.00 &  0.48 $\pm$ 0.04 & 0.03 &  0.50  $\pm$ 0.08 &  0.05  \\
HD    81767  &   0.21 $\pm$ 0.09 &  -0.01 &  0.21 $\pm$ 0.05 &-0.01 &  0.18  $\pm$ 0.11 & -0.04  \\
HD    90054  &   0.24 $\pm$ 0.15 &  -0.05 &  0.26 $\pm$ 0.05 &-0.03 &  0.22  $\pm$ 0.08 & -0.07  \\
HD    94374  &   0.18 $\pm$ 0.09 &   0.28 & -0.23 $\pm$ 0.04 &-0.13 &  0.18  $\pm$ 0.08 &  0.28  \\
HD    95338  &   0.14 $\pm$ 0.05 &  -0.07 &  0.19 $\pm$ 0.04 &-0.02 &  0.17  $\pm$ 0.09 & -0.04  \\
HD   104212  &   0.13 $\pm$ 0.07 &   0.00 &  0.14 $\pm$ 0.04 & 0.01 &  0.14  $\pm$ 0.08 &  0.01  \\
HD   107509  &  -0.03 $\pm$ 0.23 &  -0.06 &  0.03 $\pm$ 0.05 &-0.00 &  0.02  $\pm$ 0.08 & -0.00  \\
HD   120329  &   0.26 $\pm$ 0.06 &  -0.05 &  0.34 $\pm$ 0.04 & 0.03 &  0.29  $\pm$ 0.09 & -0.02  \\
HD   143102  &   0.09 $\pm$ 0.06 &  -0.07 &  0.20 $\pm$ 0.04 & 0.04 &  0.13  $\pm$ 0.08 & -0.03  \\
HD   148530  &   0.04 $\pm$ 0.11 &   0.01 &  0.06 $\pm$ 0.04 & 0.03 &  0.04  $\pm$ 0.10 &  0.01  \\
HD   149256  &   0.24 $\pm$ 0.08 &  -0.10 &  0.40 $\pm$ 0.05 & 0.06 &  0.34  $\pm$ 0.09 &  0.00  \\
HD   149606  &   0.12 $\pm$ 0.14 &  -0.08 &  0.16 $\pm$ 0.05 &-0.04 &  0.13  $\pm$ 0.10 & -0.07  \\
HD   149933  &   0.11 $\pm$ 0.12 &  -0.02 &  0.19 $\pm$ 0.04 & 0.06 &  0.13  $\pm$ 0.09 & -0.00  \\
HD   165920  &   0.33 $\pm$ 0.06 &  -0.03 &  0.40 $\pm$ 0.04 & 0.04 &  0.30  $\pm$ 0.08 & -0.06  \\
HD   168714  &   0.46 $\pm$ 0.12 &  -0.02 &  0.41 $\pm$ 0.04 &-0.07 &  0.37  $\pm$ 0.09 & -0.11  \\
HD   171999  &   0.23 $\pm$ 0.10 &  -0.06 &  0.32 $\pm$ 0.04 & 0.03 &  0.20  $\pm$ 0.09 & -0.09  \\
HD   179764  &  -0.04 $\pm$ 0.06 &   0.01 &  0.08 $\pm$ 0.04 & 0.13 & -0.10  $\pm$ 0.09 & -0.05  \\
HD   180865  &   0.21 $\pm$ 0.07 &  -0.00 &  0.24 $\pm$ 0.04 & 0.03 &  0.28  $\pm$ 0.10 &  0.07  \\
HD   181234  &   0.48 $\pm$ 0.10 &   0.03 &  0.52 $\pm$ 0.04 & 0.07 &  0.49  $\pm$ 0.08 &  0.04  \\
HD   196397  &   0.33 $\pm$ 0.09 &  -0.05 &  0.41 $\pm$ 0.05 & 0.03 &  0.36  $\pm$ 0.10 & -0.02  \\
HD   201237  &   0.10 $\pm$ 0.17 &   0.10 &  0.05 $\pm$ 0.07 & 0.05 &  0.07  $\pm$ 0.17 &  0.07  \\
HD   209721  &   0.25 $\pm$ 0.12 &  -0.03 &  0.29 $\pm$ 0.04 & 0.01 &  0.26  $\pm$ 0.09 & -0.02  \\
HD   211706  &   0.09 $\pm$ 0.15 &  -0.00 &  0.09 $\pm$ 0.05 &-0.00 &  0.10  $\pm$ 0.10 &  0.01  \\
HD   218566  &   0.26 $\pm$ 0.09 &  -0.02 &  0.34 $\pm$ 0.08 & 0.06 &  0.24  $\pm$ 0.15 & -0.04  \\
HD   218750  &   0.14 $\pm$ 0.12 &  -0.03 &  0.23 $\pm$ 0.05 & 0.06 &  0.17  $\pm$ 0.09 & -0.00  \\
HD   221313  &   0.34 $\pm$ 0.13 &   0.03 &  0.31 $\pm$ 0.05 & 0.00 &  0.35  $\pm$ 0.11 &  0.04  \\
HD   221974  &   0.40 $\pm$ 0.09 &  -0.06 &  0.47 $\pm$ 0.05 & 0.01 &  0.48  $\pm$ 0.10 &  0.02  \\
HD   224383  &  -0.10 $\pm$ 0.13 &   0.00 & -0.14 $\pm$ 0.04 &-0.05 & -0.02  $\pm$ 0.08 &  0.07  \\ 
 \\
\multicolumn{7}{c}{Intermediate population} \\
HD     8389  &   0.54 $\pm$ 0.10 &  -0.04 &  0.60 $\pm$ 0.05 & 0.02 &  0.57  $\pm$ 0.08 & -0.01  \\
HD     9174  &   0.36 $\pm$ 0.09 &  -0.05 &  0.50 $\pm$ 0.05 & 0.09 &  0.38  $\pm$ 0.10 & -0.03  \\
HD    12789  &   0.28 $\pm$ 0.05 &   0.01 &  0.31 $\pm$ 0.05 & 0.04 &  0.27  $\pm$ 0.08 &  0.00  \\
HD    30295  &   0.33 $\pm$ 0.07 &   0.01 &  0.32 $\pm$ 0.04 & 0.00 &  0.30  $\pm$ 0.09 & -0.02  \\
HD    31452  &   0.19 $\pm$ 0.08 &  -0.04 &  0.23 $\pm$ 0.04 & 0.00 &  0.19  $\pm$ 0.09 & -0.04  \\
HD    37986  &   0.25 $\pm$ 0.04 &  -0.05 &  0.36 $\pm$ 0.04 & 0.06 &  0.27  $\pm$ 0.08 & -0.03  \\
HD    39715  &  -0.09 $\pm$ 0.11 &   0.01 & -0.13 $\pm$ 0.06 &-0.03 & -0.14  $\pm$ 0.13 & -0.04  \\
HD    43848  &   0.35 $\pm$ 0.08 &  -0.08 &  0.40 $\pm$ 0.05 &-0.03 &  0.37  $\pm$ 0.08 & -0.06  \\
HD    86065  &   0.03 $\pm$ 0.09 &  -0.06 &  0.11 $\pm$ 0.05 & 0.02 &  0.02  $\pm$ 0.10 & -0.08  \\
HD    87007  &   0.24 $\pm$ 0.09 &  -0.05 &  0.38 $\pm$ 0.05 & 0.09 &  0.23  $\pm$ 0.12 & -0.06  \\
HD    91585  &   0.23 $\pm$ 0.10 &  -0.02 &  0.29 $\pm$ 0.05 & 0.04 &  0.27  $\pm$ 0.10 &  0.02  \\
HD    91669  &   0.41 $\pm$ 0.09 &  -0.03 &  0.48 $\pm$ 0.05 & 0.04 &  0.38  $\pm$ 0.12 & -0.06  \\
HD   182572  &   0.50 $\pm$ 0.05 &   0.02 &  0.46 $\pm$ 0.04 &-0.02 &  0.50  $\pm$ 0.06 &  0.02  \\
HD   196794  &   0.10 $\pm$ 0.07 &   0.04 &  0.04 $\pm$ 0.04 &-0.03 &  0.08  $\pm$ 0.10 &  0.01  \\
HD   197921  &   0.22 $\pm$ 0.11 &   0.00 &  0.24 $\pm$ 0.05 & 0.02 &  0.24  $\pm$ 0.10 &  0.02  \\
HD   213996  &   0.34 $\pm$ 0.07 &   0.01 &  0.35 $\pm$ 0.05 & 0.02 &  0.35  $\pm$ 0.10 &  0.02  \\
HD   214463  &   0.38 $\pm$ 0.11 &   0.04 &  0.32 $\pm$ 0.05 &-0.02 &  0.32  $\pm$ 0.10 & -0.02  \\

\hline
\end{longtable}
\newpage
\begin{appendix}

\section{Online tables}

\begin{table*}[ht]
\caption[1]{Log of spectroscopic observations.} 
\label{Tab_logbook}
\centering
\begin{tabular}{cccccccc}
\hline \hline
{\rm star} & $\alpha_{\rm J2000}$   & $\delta_{\rm J2000}$   &  date  & UT & exp         &  Airmass & (S/N)  \\  
           &  [h m s]               &  [d m s]               &        &    & [s]         &          &        \\
\hline
G 161-029    & 09:25:41.84  & -06:46:05.80 & 2001 Jan 15 & 04:33:58 & 4500 & 1.15 & 57   \\
BD-02   180  &  01:21:20.93 & -01:43:45.62 & 2001 Jan 15 & 00:35:33 & 3600 & 1.54 & 64   \\ 
BD-05  5798  &  22:20:19.62 & -04:50:06.72 & 1999 Sep 27 & 04:54:58 & 2800 & 1.37 & 62   \\
BD-17  6035  &  20:53:53.32 & -16:45:54.98 & 1999 Sep 24 & 02:28:19 & 3600 & 1.14 & 65   \\
CD-32  0327  &  00:15:14.75 & -31:19:57.97 & 1999 Sep 26 & 06:50:42 & 3600 & 1.61 & 90   \\
  $\vdots$   &  $\vdots$    &     $\vdots$ &   $\vdots$  &  $\vdots$ & $\vdots$ & $\vdots$ & $\vdots$  \\ 
\hline 
\end{tabular}
\end{table*}

\begin{table*}[ht]
\caption[1]{Data from the PASTEL catalogue.} 
\label{Tab_pastel}
\centering
\begin{tabular}{lcccr}
\hline \hline
Star    & \multicolumn{1}{c}{$T_{\rm eff}$} & \multicolumn{1}{c}{$\log g$} & \multicolumn{1}{c}{[Fe/H]} & Reference \\
\hline
HD     8389 &  5283 $\pm$  64 &  4.37 $\pm$  0.12     &   0.34 $\pm$ 0.05     & \citet{Sousa.etal:2008} \\
            &  5378 $\pm$  84 &  4.50 $\pm$  0.12     &   0.47 $\pm$ 0.08     & \citet{Sousa.etal:2006} \\ 
HD     9424 &  5420 $\pm$  43 &\multicolumn{1}{c}{...}&\multicolumn{1}{c}{...}& \citet{Masana.etal:2006} \\
HD    10576 &  5882 $\pm$  56 &\multicolumn{1}{c}{...}&\multicolumn{1}{c}{...}& \citet{Masana.etal:2006} \\
HD    13386 &  5226 $\pm$  56 &  4.28 $\pm$  0.09     &   0.26 $\pm$ 0.06     & \citet{Sousa.etal:2006} \\
            &  5361 $\pm$  43 &\multicolumn{1}{c}{...}&\multicolumn{1}{c}{...}& \citet{Masana.etal:2006} \\
HD    15555 &  4820 $\pm$  43 &\multicolumn{1}{c}{...}&\multicolumn{1}{c}{...}& \citet{GonzalezHernandez.Bonifacio:2009} \\
            &  4855 $\pm$  53 &\multicolumn{1}{c}{...}&\multicolumn{1}{c}{...}& \citet{Ramirez.Melendez:2005} \\
            &  4855 $\pm$  67 &\multicolumn{1}{c}{...}&\multicolumn{1}{c}{...}& \citet{Alonso.etal:1999a} \\
$\vdots$ & $\vdots$ & $\vdots$ & $\vdots$ & $\vdots$ \\
\hline 
\end{tabular}
\end{table*}


\begin{table*}[ht]
\caption{ Si, Ca, and Ti line list}
 \label{Tab_lines_Si_Ca_Ti}
\centering
\begin{tabular}{cccccccc}
\hline\hline
Species   & $\lambda$ & $\chi_{exc}$  & $\log gf$ & $\log gf$  & $\log gf$ &  $\log gf$ &  $\log gf$  \\
          &   (\AA)   &    (eV)       & (Sun)     & (BFL04)    & (NIST)    & (VALD)     &  (BZO+09)  \\ 
\hline

Si I & 5665.56 & 4.92 & -2.01 & -1.94 & -2.04 & -1.75 &   ... \\ 
Si I & 5684.48 & 4.95 & -1.63 & -1.55 & -1.42 & -1.73 &   ... \\ 
Si I & 5690.43 & 4.93 & -1.81 & -1.77 & -1.87 & -1.77 &   ... \\ 
Si I & 5701.10 & 4.93 & -2.00 & -1.95 & -2.05 & -1.58 &   ... \\ 
Si I & 5708.40 & 4.95 & -1.40 &   ... & -1.47 & -1.03 &   ... \\ 
$\vdots$ & $\vdots$ & $\vdots$ & $\vdots$ & $\vdots$ & $\vdots$ & $\vdots$ & $\vdots$\\
\hline 
\end{tabular}
\tablefoot{
 BFL04: \citet{Bensby.etal:2004}; BZO+09: \citet{Barbuy.etal:2009}.
}
\end{table*}


\begin{table*}[ht]
\caption{\ion{Fe}{I} and \ion{Fe}{II} line list} 
\label{Tab_ironlinelist}
\centering
\begin{tabular}{cccccccc}
\hline \hline
\hbox{Ion} & \hbox{$\lambda$ ({\rm \AA})} & \hbox{$\chi_{ex}$ (eV)} & C6 & \hbox{log $gf$} & \hbox{log $gf$} 
& \hbox{log $gf$} & \hbox{log $gf$} \\ 
           &            &       &    & (Sun) & (VALD) & (FW06) & (MAGY09) \\

\hline

\ion{Fe}{I} &  5522.45 &  4.21 & 3.0200e-31 & -1.49  &  -1.55  & -1.52  &    ...\\
\ion{Fe}{I} &  5546.51 &  4.37 & 3.9100e-31 & -1.18  &  -1.31  & -1.28  &    ...\\ 
\ion{Fe}{I} &  5560.21 &  4.43 & 4.7900e-31 & -1.14  &  -1.19  & -1.16  &    ...\\
\ion{Fe}{I} &  5577.02 &  5.03 & 1.0000e-32 & -1.61  &  -1.55  &   ...  &    ...\\
\ion{Fe}{I} &  5618.63 &  4.21 & 2.9000e-31 & -1.39  &  -1.28  & -1.28  &    ...\\
$\vdots$ & $\vdots$ & $\vdots$ & $\vdots$ & $\vdots$ & $\vdots$ & $\vdots$ & $\vdots$ \\
\hline 
\end{tabular}
\tablefoot{
FW06: \citet{Fuhr.Wiese:2006}; MAGY09: \citet{Melendez.etal:2009}; MB09: \citet{Melendez.Barbuy:2009}.
}
\end{table*}


\end{appendix}

\end{document}